\documentclass[11pt]{article}
\usepackage{amsmath,amssymb,amsthm,amsxtra,overpic,bbm,bm,epsfig}
\usepackage{color}
\def\thefootnote{\fnsymbol{footnote}}

\begin{document}

\vspace{0.2cm}

\begin{center}
{\Large\bf Majorana CP-violating phases in neutrino-antineutrino
oscillations and other lepton-number-violating processes}
\end{center}

\vspace{0.2cm}

\begin{center}
{\bf Zhi-zhong Xing} \footnote{E-mail: xingzz@ihep.ac.cn}
\quad
{\bf Ye-Ling Zhou} \footnote{E-mail: zhouyeling@ihep.ac.cn}
\\
{Institute of High Energy Physics, Chinese Academy of
Sciences, P.O. Box 918, Beijing 100049, China \\
Theoretical Physics Center for Science Facilities, Chinese Academy
of Sciences, Beijing 100049, China }
\end{center}

\vspace{1.5cm}

\begin{abstract}
If the massive neutrinos are identified to be the Majorana particles
via a convincing measurement of the neutrinoless double beta
($0\nu\beta\beta$) decay, how to determine the Majorana CP-violating
phases in the $3\times 3$ lepton flavor mixing matrix $U$ will
become a desirable experimental question. The answer to this
question is to explore all the possible lepton-number-violating
(LNV) processes in which the Majorana phases really matter. In this
paper we carry out a systematic study of CP violation in
neutrino-antineutrino oscillations, whose CP-conserving parts
involve six independent $0\nu\beta\beta$-like mass terms $\langle
m\rangle^{}_{\alpha\beta}$ and CP-violating parts are associated
with nine independent Jarlskog-like parameters ${\cal
V}^{ij}_{\alpha\beta}$ (for $\alpha, \beta = e, \mu, \tau$ and $i, j
= 1, 2 ,3$). With the help of current neutrino oscillation data, we
analyze the sensitivities of $|\langle m\rangle^{}_{\alpha\beta}|$
and ${\cal V}^{ij}_{\alpha\beta}$ to the three CP-violating phases
of $U$, and illustrate the salient features of six independent
CP-violating asymmetries between $\nu^{}_\alpha \to
\overline{\nu}^{}_\beta$ and $\overline{\nu}^{}_\alpha \to
\nu^{}_\beta$ oscillations. As a by-product, the effects of the
CP-violating phases on the LNV decays of doubly- and singly-charged
Higgs bosons are reexamined by taking account of the unsuppressed
value of $\theta^{}_{13}$. Such CP-conserving LNV processes can be
complementary to the possible measurements of neutrino-antineutrino
oscillations in the distant future.
\end{abstract}

\begin{flushleft}
\hspace{0.8cm} PACS number(s): 14.60.Pq, 13.10.+q, 25.30.Pt \\
\hspace{0.8cm} Keywords: Majorana phases, neutrino-antineutrino
oscillations, CP violation
\end{flushleft}

\def\thefootnote{\arabic{footnote}}
\setcounter{footnote}{0}

\newpage

\section{Introduction}

Neutrinos are the most elusive fermions in the standard electroweak
model, partly because they are electrically neutral and their masses
are too small as compared with those charged leptons and quarks. The
neutrality and smallness of neutrinos make it experimentally
difficult to identify whether they are the Dirac or Majorana
particles, but most theorists believe that massive neutrinos should
have the Majorana nature (i.e., they are their own antiparticles
\cite{Majorana}). To verify the Majorana nature of massive
neutrinos, the most feasible way up to our current experimental
techniques is to detect the neutrinoless double beta
($0\nu\beta\beta$) decay of some even-even nuclei \cite{2B}: $A(Z,N)
\to A(Z+2, N-2) + 2 e^-$, in which the lepton number is violated by
two units. However, the $0\nu\beta\beta$ decay is a CP-conserving
process and cannot directly be used to probe the Majorana
CP-violating phases. Hence one has to consider other possible ways
out of such a situation.

Given three massive neutrinos of the Majorana nature, the $3\times
3$ Pontecorvo-Maki-Nakagawa-Sakata (PMNS) matrix $U$ \cite{PMNS} can
be parametrized in terms of three flavor mixing angles
($\theta^{}_{12}, \theta^{}_{13}, \theta^{}_{23}$) and three
CP-violating phases ($\delta, \rho, \sigma$) as follows:
\begin{eqnarray}
U \hspace{-0.2cm} & = & \hspace{-0.2cm} \left( \begin{matrix}
c^{}_{12} c^{}_{13} & s^{}_{12} c^{}_{13} & s^{}_{13} e^{-{\rm i}
\delta} \cr -s^{}_{12} c^{}_{23} - c^{}_{12} s^{}_{13} s^{}_{23}
e^{{\rm i} \delta} & c^{}_{12} c^{}_{23} - s^{}_{12} s^{}_{13}
s^{}_{23} e^{{\rm i} \delta} & c^{}_{13} s^{}_{23} \cr s^{}_{12}
s^{}_{23} - c^{}_{12} s^{}_{13} c^{}_{23} e^{{\rm i} \delta} &
-c^{}_{12} s^{}_{23} - s^{}_{12} s^{}_{13} c^{}_{23} e^{{\rm i}
\delta} & c^{}_{13} c^{}_{23} \cr
\end{matrix} \right) \left(
\begin{matrix} e^{{\rm i} \rho} & 0 & 0 \cr 0 & e^{{\rm i} \sigma}
& 0 \cr 0 & 0 & 1 \cr \end{matrix} \right) \; ,
\end{eqnarray}
where $c^{}_{ij} \equiv \cos\theta^{}_{ij}$ and $s^{}_{ij} \equiv
\sin\theta^{}_{ij}$ (for $ij = 12, 13, 23$). Although $\delta$ is
usually referred to as the ``Dirac" CP-violating phase which
naturally appears in those lepton-number-conserving processes such
as neutrino-neutrino and antineutrino-antineutrino oscillations, one
should keep in mind that it is actually a Majorana phase like $\rho$
or $\sigma$ and can also show up in those lepton-number-violating
(LNV) processes such as the $0\nu\beta\beta$ decay and
neutrino-antineutrino oscillations. This point will soon become
clear. So far all the three neutrino mixing angles have been
measured to a good degree of accuracy in a number of solar,
atmospheric, reactor and accelerator neutrino oscillation
experiments \cite{PDG}. A determination of the phase parameter
$\delta$ via a measurement of the Jarlskog invariant ${\cal J} =
c^{}_{12} s^{}_{12} c^2_{13} s^{}_{13} c^{}_{23} s^{}_{23}
\sin\delta$ \cite{J} will be one of the major goals of the
next-generation long-baseline neutrino oscillation experiments. The
most challenging task is to detect the Majorana phases $\rho$ and
$\sigma$, which can only emerge in the LNV processes. As formulated
by one of us in Ref. \cite{Xing13}, it is {\it in principle}
possible to determine all the three phases from the CP-violating
asymmetries ${\cal A}^{}_{\alpha\beta}$ between $\nu^{}_\alpha \to
\overline{\nu}^{}_\beta$ and $\overline{\nu}^{}_\alpha \to
\nu^{}_\beta$ oscillations. Nevertheless, a systematic study of this
problem has been lacking.

The present work aims to go beyond Ref. \cite{Xing13} by carrying
out a systematic analysis of the Majorana CP-violating phases in
both neutrino-antineutrino oscillations and LNV decays of doubly-
and singly-charged Higgs bosons based on the type-II seesaw
mechanism \cite{SS2}
\footnote{As the $0\nu\beta\beta$ decay has been extensively
discussed in the literature \cite{Rode}, here we shall not pay
particular attention to it.},
in order to reveal their distinct properties which might be more or
less associated with the observed matter-antimatter asymmetry of the
Universe \cite{LEP}. Our study is different from the previous ones
at least in the following aspects:
\begin{itemize}
\item     All the $0\nu\beta\beta$-like mass terms $\langle
m\rangle^{}_{\alpha\beta}$ and the Jarlskog-like parameters ${\cal
V}^{ij}_{\alpha\beta}$ (for $\alpha, \beta = e, \mu, \tau$ and $i, j
= 1, 2 ,3$), which measure the CP-conserving and CP-violating
properties of Majorana neutrinos respectively, are analyzed in
detail.

\item     The sensitivities of all the CP-violating asymmetries
${\cal A}^{}_{\alpha\beta}$ to the phase parameters and the neutrino
mass spectrum are discussed in a systematic way, and the
``pseudo-Dirac" case with vanishing $\rho$ and $\sigma$ is also
explored to illustrate why $\delta$ is of the Majorana nature.

\item     The CP-conserving LNV decays of $H^{\pm\pm}$ and $H^{\pm}$
bosons are reexamined by taking account of the unsuppressed value of
$\theta^{}_{13}$ reported by the Daya Bay \cite{DYB} and RENO
\cite{RENO} Collaborations, and the dependence of their branching
ratios on $\delta$, $\rho$ and $\sigma$ is investigated.
\end{itemize}
Such a comprehensive analysis of the Majorana phases in CP-violating
and CP-conserving LNV processes should be useful to illustrate how
important they are in both lepton flavor mixing and CP violation and
how difficult they are to be measured in reality.

The remaining parts of this paper are organized as follows. In
section 2 we briefly review the salient features of three-flavor
neutrino-antineutrino oscillations, including a concise discussion
about the CP- and T-violating asymmetries. Section 3 is devoted to a
detailed analysis of six independent $0\nu\beta\beta$-like mass
terms $\langle m\rangle^{}_{\alpha\beta}$ and nine independent
Jalskog-like parameters ${\cal V}^{ij}_{\alpha\beta}$ (for $\alpha,
\beta = e, \mu, \tau$ and $i, j = 1, 2 ,3$), which appear in the
probabilities of $\nu^{}_\alpha \to \overline{\nu}^{}_\beta$
oscillations and their CP- or T-conjugate processes. A comparison
between ${\cal V}^{ij}_{\alpha\beta}$ and ${\cal J}$ is made by
switching off the Majorana phases $\rho$ and $\sigma$. As a
by-product, the effects of three CP-violating phases on the LNV
decays of doubly- and singly-charged Higgs bosons are also
reexamined by taking account of the unsuppressed value of
$\theta^{}_{13}$. In section 4 we carry out a systematic study of
the sensitivities of six possible CP-violating asymmetries ${\cal
A}^{}_{\alpha\beta}$ to the three phase parameters, the absolute
scale and hierarchies of three neutrino masses, and the ratio of the
neutrino beam energy $E$ to the baseline length $L$. Our numerical
results illustrate the distinct roles of $\delta$, $\rho$ and
$\sigma$ or their combinations in neutrino-antineutrino
oscillations. Section 5 is devoted to a summary of this work with
some main conclusions.
\begin{figure}[t]
\centering \vspace{-0.15cm}
\includegraphics[width=0.28\textwidth]{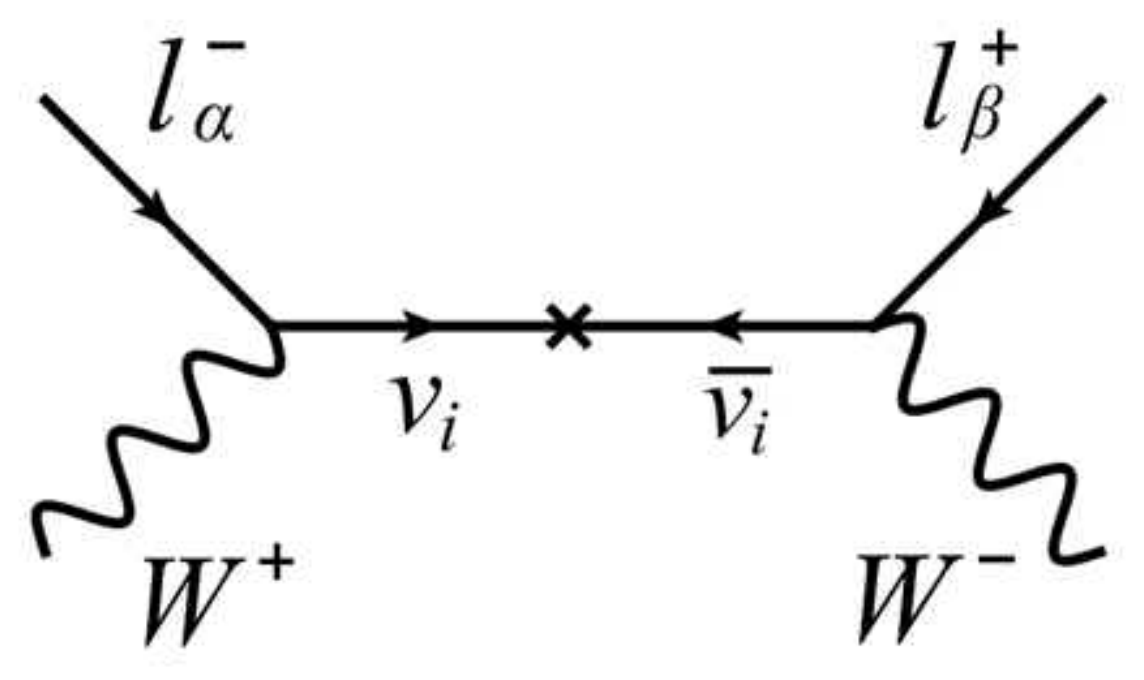}
\vspace{-0.2cm} \caption{The Feynman diagram for $\nu^{}_\alpha \to
\overline{\nu}^{}_\beta$ oscillations, where ``$\times$" stands for
the chirality flip in the neutrino propagator which is proportional
to the mass $m^{}_i$ of the Majorana neutrino $\nu^{}_i =
\overline{\nu}^{}_i$.}
\end{figure}

\section{Salient features of neutrino-antineutrino oscillations}

Let us consider $\nu^{}_\alpha \to \overline{\nu}^{}_\beta$
oscillations (for $\alpha, \beta = e, \mu, \tau$), as schematically
illustrated in Figure 1, where the production of $\nu^{}_\alpha$ and
the detection of $\overline{\nu}^{}_\beta$ are both governed by the
standard weak charged-current interactions. The amplitudes of
$\nu^{}_\alpha \to \overline{\nu}^{}_\beta$ transitions and their
CP-conjugate processes $\overline{\nu}^{}_\alpha \to \nu^{}_\beta$
can be written as \cite{Xing13,Valle}
\begin{eqnarray}
A(\nu^{}_\alpha \to \overline{\nu}^{}_\beta) \hspace{-0.2cm} & = &
\hspace{-0.2cm} \sum_i \left[ U^*_{\alpha i} U^*_{\beta i}
\frac{m^{}_i}{E} \exp\left(-{\rm
i}\frac{m^2_i}{2E} L\right) \right] K \; , \nonumber \\
A(\overline{\nu}^{}_\alpha \to \nu^{}_\beta) \hspace{-0.2cm} & = &
\hspace{-0.2cm} \sum_i \left[ U^{}_{\alpha i} U^{}_{\beta i}
\frac{m^{}_i}{E} \exp\left(-{\rm i}\frac{m^2_i}{2E} L\right) \right]
\overline{K} \; ,
\end{eqnarray}
where $m^{}_i$ denotes the mass of $\nu^{}_i$, $E$ is the neutrino
(or antineutrino) beam energy, $L$ stands for the baseline length,
$K$ and $\overline{K}$ are the kinematical factors independent of
the index $i$ (and satisfying $|K| = |\overline{K}|$). The helicity
suppression in the transition between $\nu^{}_i$ and
$\overline{\nu}^{}_i$ is characterized by $m^{}_i/E$. The
neutrino-antineutrino oscillation probabilities $P(\nu^{}_\alpha \to
\overline{\nu}^{}_\beta) \equiv |A(\nu^{}_\alpha \to
\overline{\nu}^{}_\beta)|^2$ and $P(\overline{\nu}^{}_\alpha \to
\nu^{}_\beta) \equiv |A(\overline{\nu}^{}_\alpha \to
\nu^{}_\beta)|^2$ turn out to be \cite{Xing13}
\begin{eqnarray}
P(\nu^{}_\alpha \to \overline{\nu}^{}_\beta) \hspace{-0.2cm} & = &
\hspace{-0.2cm} \frac{|K|^2}{E^2} \left[ \left| \langle m
\rangle^{}_{\alpha \beta} \right|^2 - 4 \sum_{i<j} m^{}_i m^{}_j
{\cal C}^{ij}_{\alpha\beta} \sin^2 \phi^{}_{ji} + 2 \sum_{i<j}
m^{}_i m^{}_j {\cal V}^{ij}_{\alpha\beta} \sin 2\phi^{}_{ji} \right]
\; ,
\nonumber \\
P(\overline{\nu}^{}_\alpha \to \nu^{}_\beta) \hspace{-0.2cm} & = &
\hspace{-0.2cm} \frac{|\overline{K}|^2}{E^2} \left[ \left| \langle
m \rangle^{}_{\alpha \beta} \right|^2 - 4 \sum_{i<j} m^{}_i m^{}_j
{\cal C}^{ij}_{\alpha\beta} \sin^2 \phi^{}_{ji} - 2 \sum_{i<j}
m^{}_i m^{}_j {\cal V}^{ij}_{\alpha\beta} \sin 2\phi^{}_{ji} \right]
\; ,
\end{eqnarray}
in which $\phi^{}_{ji} \equiv \Delta m^2_{ji} L/(4 E)$ with $\Delta
m^2_{ji} \equiv m^2_j - m^2_i$, the effective mass term $\langle m
\rangle^{}_{\alpha \beta}$ is defined as
\begin{eqnarray}
\langle m \rangle^{}_{\alpha \beta} \hspace{-0.2cm} & \equiv &
\hspace{-0.2cm} \sum_i m^{}_i U^{}_{\alpha i} U^{}_{\beta i} \; ,
\end{eqnarray}
and the CP-conserving and CP-violating contributions of the PMNS
flavor mixing matrix elements are described by
\begin{eqnarray}
{\cal C}^{ij}_{\alpha\beta} \hspace{-0.2cm} & \equiv &
\hspace{-0.2cm} {\rm Re}\left(U^{}_{\alpha i} U^{}_{\beta i}
U^*_{\alpha j} U^*_{\beta j} \right) \; ,
\nonumber \\
{\cal V}^{ij}_{\alpha\beta} \hspace{-0.2cm} & \equiv &
\hspace{-0.2cm} {\rm Im}\left(U^{}_{\alpha i} U^{}_{\beta i}
U^*_{\alpha j} U^*_{\beta j} \right) \; ,
\end{eqnarray}
with the Greek and Latin subscripts running over $(e, \mu, \tau)$
and $(1, 2, 3)$, respectively. Note that $\langle m
\rangle^{}_{\alpha \beta}$ is the $(\alpha, \beta)$ element of the
Majorana neutrino mass matrix $M^{}_\nu = U \widehat{M}^{}_\nu U^T$
with $\widehat{M}^{}_\nu \equiv {\rm Diag}\{m^{}_1, m^{}_2,
m^{}_3\}$ in the flavor basis where the charged-lepton mass matrix
is diagonal, and thus $\langle m \rangle^{}_{\alpha \beta} = \langle
m \rangle^{}_{\beta\alpha}$ holds as a result of the symmetry of
$M^{}_\nu$. Because $\langle m \rangle^{}_{ee}$ is simply the
effective mass term of the $0\nu\beta\beta$ decay, we refer to
$\langle m \rangle^{}_{\alpha \beta}$ as the $0\nu\beta\beta$-like
mass terms. Similarly, the CP- and T-violating quantities ${\cal
V}^{ij}_{\alpha\beta}$ are referred to as the Jarlskog-like
parameters.

By definition, the CP-conserving quantities ${\cal
C}^{ij}_{\alpha\beta}$ satisfy ${\cal C}^{ij}_{\alpha\beta} = {\cal
C}^{ij}_{\beta\alpha} = {\cal C}^{ji}_{\alpha\beta} = {\cal
C}^{ji}_{\beta\alpha}$. In addition, ${\cal C}^{ij}_{\alpha\beta}$
and $\langle m\rangle^{}_{\alpha\beta}$ are related to each other
through
\begin{eqnarray}
\left|\langle m\rangle^{}_{\alpha\beta}\right|^2 \hspace{-0.2cm} &
= & \hspace{-0.2cm} \sum^{}_{i} m^2_i {\cal C}^{ii}_{\alpha\beta} +
2\sum^{}_{i<j} m^{}_i m^{}_j {\cal C}^{ij}_{\alpha\beta} \; .
\end{eqnarray}
This relation allows us to rewrite Eq. (3) as
\begin{eqnarray}
P(\nu^{}_\alpha \to \overline{\nu}^{}_\beta) \hspace{-0.2cm} & = &
\hspace{-0.2cm} \frac{|K|^2}{E^2} \left[\sum^{}_{i} m^2_i {\cal
C}^{ii}_{\alpha\beta} + 2\sum^{}_{i<j} m^{}_i m^{}_j \left({\cal
C}^{ij}_{\alpha\beta} \cos 2\phi_{ji} + {\cal V}^{ij}_{\alpha\beta}
\sin 2\phi_{ji} \right) \right] \; ,
\nonumber\\
P(\overline{\nu}^{}_\alpha \to \nu^{}_\beta) \hspace{-0.2cm} & = &
\hspace{-0.2cm} \frac{|\overline{K}|^2}{E^2} \left[ \sum^{}_{i}
m^2_i {\cal C}^{ii}_{\alpha\beta} + 2\sum^{}_{i<j} m^{}_i m^{}_j
\left({\cal C}^{ij}_{\alpha\beta} \cos 2\phi_{ji} - {\cal
V}^{ij}_{\alpha\beta} \sin 2\phi_{ji} \right) \right] \; .
\end{eqnarray}
The unitarity of the PMNS matrix $U$ leads us to the relations
\begin{eqnarray}
\sum_\alpha {\cal C}^{ij}_{\alpha\beta} \hspace{-0.2cm} & = &
\hspace{-0.2cm} \sum_\beta {\cal C}^{ij}_{\alpha\beta} = 0 \; ,
\nonumber \\
\sum_\alpha {\cal V}^{ij}_{\alpha\beta} \hspace{-0.2cm} & = &
\hspace{-0.2cm} \sum_\beta {\cal V}^{ij}_{\alpha\beta} = 0 \; ,
\end{eqnarray}
for $i \neq j$. Then we arrive at the following sum rule for the
probabilities of $\nu^{}_\alpha \to \overline{\nu}^{}_\beta$ and
$\overline{\nu}^{}_\alpha \to \nu^{}_\beta$ oscillations:
\begin{eqnarray}
\sum_\beta P(\nu^{}_\alpha \to \overline{\nu}^{}_\beta)
\hspace{-0.2cm} & = & \hspace{-0.2cm} \sum_\beta
P(\overline{\nu}^{}_\alpha \to \nu^{}_\beta) = \frac{|K|^2}{E^2}
\left|\langle m \rangle^{}_{\alpha}\right|^2 \; ,
\end{eqnarray}
where
\begin{eqnarray}
\left|\langle m \rangle^{}_{\alpha}\right|^2 \hspace{-0.2cm} &
\equiv & \hspace{-0.2cm} \sum_i m^2_i \left|U^{}_{\alpha i}\right|^2
\; ,
\end{eqnarray}
which is actually the $(\alpha,\alpha)$ element of $M^{}_\nu
M^\dag_\nu$. In particular, $\langle m \rangle^{}_e$ is just the
effective mass term appearing in the rate of the tritium beta decay
${}^3_1 {\rm H} \to {}^3_2 {\rm He} + e^- + \overline{\nu}^{}_e$. In
comparison with Eq. (9), the so-called {\it zero-distance effect} of
neutrino-antineutrino oscillations at $L=0$ is given by
\begin{eqnarray}
P(\nu^{}_\alpha \to \overline{\nu}^{}_\beta) \hspace{-0.2cm} & = &
\hspace{-0.2cm} P(\overline{\nu}^{}_\alpha \to \nu^{}_\beta) =
\frac{|K|^2}{E^2} \left| \langle m \rangle^{}_{\alpha \beta}
\right|^2 \; .
\end{eqnarray}
Because of $m^{}_i \ll E$, the effects in both Eqs. (9) and (11)
are extremely suppressed.

Thanks to CPT invariance, it is easy to check that $P(\nu^{}_\alpha
\to \overline{\nu}^{}_\beta) = P(\nu^{}_\beta \to
\overline{\nu}^{}_\alpha)$ and $P(\overline{\nu}^{}_\alpha \to
\nu^{}_\beta) = P(\overline{\nu}^{}_\beta \to \nu^{}_\alpha)$ hold.
Hence the T-violating asymmetry between $\nu^{}_\alpha \to
\overline{\nu}^{}_\beta$ and $\overline{\nu}^{}_\beta \to
\nu^{}_\alpha$ oscillations must be exactly equal to the
CP-violating asymmetry between $\nu^{}_\alpha \to
\overline{\nu}^{}_\beta$ and $\overline{\nu}^{}_\alpha \to
\nu^{}_\beta$ oscillations. To eliminate the $|K|^2/E^2$ and
$|\overline{K}|^2/E^2$ factors, we define the CP-violating asymmetry
between $\nu^{}_\alpha \to \overline{\nu}^{}_\beta$ and
$\overline{\nu}^{}_\alpha \to \nu^{}_\beta$ oscillations as the
ratio of the difference $P(\nu^{}_\alpha \to
\overline{\nu}^{}_\beta) - P(\overline{\nu}^{}_\alpha \to
\nu^{}_\beta)$ to the sum $P(\nu^{}_\alpha \to
\overline{\nu}^{}_\beta) + P(\overline{\nu}^{}_\alpha \to
\nu^{}_\beta)$, denoted by ${\cal A}^{}_{\alpha\beta}$
\cite{Xing13}. Therefore,
\begin{eqnarray}
{\cal A}^{}_{\alpha\beta} \hspace{-0.2cm} & = & \hspace{-0.2cm}
\frac{\displaystyle 2\sum^{}_{i<j} m^{}_i m^{}_j {\cal
V}^{ij}_{\alpha\beta} \sin 2\phi^{}_{ji}}{\displaystyle \sum^{}_{i}
m^2_i {\cal C}^{ii}_{\alpha\beta} + 2\sum^{}_{i<j} m^{}_i m^{}_j
{\cal C}^{ij}_{\alpha\beta} \cos 2\phi^{}_{ji}} \hspace{0.3cm}
\nonumber \\
\hspace{-0.2cm} & = & \hspace{-0.2cm} \frac{\displaystyle
2\sum^{}_{i<j} m^{}_i m^{}_j {\cal V}^{ij}_{\alpha\beta} \sin
2\phi^{}_{ji}}{\displaystyle \left|\langle m
\rangle^{}_{\alpha\beta}\right|^2 - 4\sum^{}_{i<j} m^{}_i m^{}_j
{\cal C}^{ij}_{\alpha\beta} \sin^2\phi^{}_{ji}} \; .
\end{eqnarray}
We see that ${\cal A}^{}_{\alpha\beta} = {\cal A}^{}_{\beta\alpha}$
holds, so only six of the nine CP-violating asymmetries are
independent and nontrivial. As pointed out in Ref. \cite{Xing13},
Eq. (12) will not be much simplified even if $\alpha = \beta$ is
taken. Namely, the $\nu^{}_\alpha \to \overline{\nu}^{}_\alpha$
oscillation is actually a kind of ``appearance" process and thus it
can accommodate the CP- and T-violating effects.

It is absolutely true that a measurement of neutrino-antineutrino
oscillations is far beyond the capability of nowadays experimental
technology. The main problem arises from the helicity suppression
proportional to $m^{}_i/E$. Given the fact that the neutrino masses
are constrained to be below the eV scale but those currently
available neutrino sources all have $E \gtrsim {\cal
O}\left(1\right)$ MeV, the neutrino-antineutrino oscillation
probabilities are formidably suppressed by the factor $m^2_i/E^2
\lesssim {\cal O}\left(10^{-12}\right)$. A naive suggestion is to
lower $E$ and hence enhance $m^{}_i/E$ in a thought experiment
\cite{Xing13}, implying that the baseline length of such an
experiment must be very short. This point can be more clearly seen
from an estimate of the typical oscillation lengths by taking $E
\sim {\cal O} \left(10\right)$ keV for example
\footnote{For example, the M$\rm\ddot{o}$ssbauer electron
antineutrinos are the $E = 18.6$ keV $\overline{\nu}^{}_e$ events
which could be used to do an oscillation experiment \cite{M}. In
this case we have $L^{\text{osc}}_{31} \simeq 18$ m and
$L^{\text{osc}}_{21}\simeq 600$ m, and the size of the detector
could be as small as $\mathcal{O}(10^{-2})$ m by using metal
crystals.}:
\begin{eqnarray}
(1) ~~~~ L^{\text{osc}}_{31} \hspace{-0.2cm} & \simeq &
\hspace{-0.2cm} L^{\text{osc}}_{32} \simeq \frac{E}{10 ~\text{keV}}
\times 10 ~\text{m} \; ,
\nonumber\\
(2) ~~~~ L^{\text{osc}}_{21} \hspace{-0.2cm} & \simeq &
\hspace{-0.2cm} \frac{E}{10 ~\text{keV}} \times 330 ~\text{m} \; ,
\end{eqnarray}
corresponding to $\Delta m^2_{21} \simeq 7.5 \times 10^{-5} ~{\rm
eV}^2$ and $|\Delta m^2_{31}| \simeq |\Delta m^2_{32}| \simeq 2.4
\times 10^{-3} ~{\rm eV}^2$, respectively. In this case, however,
the sizes of the neutrino (or antineutrino) source and the detector
must be much smaller than the ones characterized by
$L^{\text{osc}}_{21}$ and (or) $L^{\text{osc}}_{31} \simeq
L^{\text{osc}}_{32}$. Note that the result in Eq. (11) is
essentially equivalent to the $L^{\text{osc}}_{ji} \gg L$ case. If
$L^{\text{osc}}_{ji} \ll L$, instead, the $\Delta
m^2_{ji}$-dependent oscillation terms will be averaged out and then
the probabilities will be simplified to
\begin{eqnarray}
P(\nu^{}_\alpha \to \overline{\nu}^{}_\beta) \hspace{-0.2cm} & = &
\hspace{-0.2cm} P(\overline{\nu}^{}_\alpha \to \nu^{}_\beta) =
\frac{|K|^2}{E^2} \sum_{i} m^2_i {\cal C}^{ii}_{\alpha\beta} \; .
\end{eqnarray}
This CP-conserving result can be compared with the ones in Eqs. (9)
and (11).

\section {Properties and profiles of ${\cal V}^{ij}_{\alpha\beta}$ and
$\langle m\rangle^{}_{\alpha\beta}$}

As shown in section 2, neutrino-antineutrino oscillations are
closely associated with the effective mass terms $\langle
m\rangle^{}_{\alpha\beta}$ and the CP-violating quantities ${\cal
V}^{ij}_{\alpha\beta}$. The former may also appear in some other LNV
processes in which CP and T symmetries are conserved. Let us explore
the analytical properties and numerical profiles of ${\cal
V}^{ij}_{\alpha\beta}$ and $\langle m\rangle^{}_{\alpha\beta}$ in
some detail in this section.

\subsection{The Jarlskog-like parameters ${\cal V}^{ij}_{\alpha\beta}$}

It is well known that the strength of CP and T violation in normal
neutrino-neutrino and antineutrino-antineutrino oscillations is
measured by a single rephasing-invariant quantity, the so-called
Jarlskog parameter ${\cal J}$ \cite{J}, defined through
\begin{eqnarray}
{\rm Im} \left(U^{}_{\alpha i} U^{}_{\beta j} U^*_{\alpha j}
U^*_{\beta i} \right) \hspace{-0.2cm} & = & \hspace{-0.2cm} {\cal J}
\sum_{\gamma} \epsilon^{}_{\alpha\beta\gamma} \sum^{}_k
\epsilon^{}_{ijk} \; ,
\end{eqnarray}
where $U$ is the PMNS matrix. In terms of the standard
parametrization of $U$ given in Eq. (1), we have
\begin{eqnarray}
{\cal J} \hspace{-0.2cm} & = & \hspace{-0.2cm} c^{}_{12} s^{}_{12}
c^2_{13} s^{}_{13} c^{}_{23} s^{}_{23} \sin\delta \; .
\end{eqnarray}
Therefore, a measurement of the CP-violating asymmetry between
$P(\nu^{}_\alpha \to \nu^{}_\beta)$ and $P(\overline{\nu}^{}_\alpha
\to \overline{\nu}^{}_\beta)$ or the T-violating asymmetry between
$P(\nu^{}_\alpha \to \nu^{}_\beta)$ and $P(\nu^{}_\beta \to
\nu^{}_\alpha)$ can only probe the ``Dirac" phase $\delta$
\cite{X13}. In contrast, the other two phases of $U$ (i.e., $\rho$
and $\sigma$) may contribute to the Jarlskog-like quantities ${\cal
V}^{ij}_{\alpha\beta}$ defined in Eq. (5), and thus they can in
principle be measured in neutrino-antineutrino oscillations.

By definition, the Jarlskog-like parameters ${\cal
V}^{ij}_{\alpha\beta}$ satisfy the relations
\begin{eqnarray}
{\cal V}^{ij}_{\alpha\beta} \hspace{-0.2cm} & = & \hspace{-0.2cm}
{\cal V}^{ij}_{\beta\alpha} = -{\cal V}^{ji}_{\alpha\beta} = -{\cal
V}^{ji}_{\beta\alpha} \; ,
\end{eqnarray}
and ${\cal V}^{ii}_{\alpha\beta}=0$; but ${\cal
V}^{ij}_{\alpha\alpha} \neq 0$ for $i\neq j$. With the help of Eq.
(17), one may express ${\cal V}^{ij}_{\alpha\beta}$ in terms of
three different ${\cal V}^{ij}_{\alpha\alpha}$ as follows:
\begin{eqnarray}
{\cal V}^{ij}_{e\mu} \hspace{-0.2cm} & = & \hspace{-0.2cm}
\frac{1}{2}\left({\cal V}^{ij}_{\tau\tau} - {\cal V}^{ij}_{ee} -
{\cal V}^{ij}_{\mu\mu}\right) \; ,
\nonumber \\
{\cal V}^{ij}_{e\tau} \hspace{-0.2cm} & = & \hspace{-0.2cm}
\frac{1}{2}\left({\cal V}^{ij}_{\mu\mu} - {\cal V}^{ij}_{ee} - {\cal
V}^{ij}_{\tau\tau}\right) \; ,
\nonumber \\
{\cal V}^{ij}_{\mu\tau} \hspace{-0.2cm} & = & \hspace{-0.2cm}
\frac{1}{2}\left({\cal V}^{ij}_{ee} - {\cal V}^{ij}_{\mu\mu} -
{\cal V}^{ij}_{\tau\tau}\right) \; .
\end{eqnarray}
This result implies that only nine ${\cal V}^{ij}_{\alpha\beta}$
are independent.

To see the explicit dependence of each ${\cal V}^{ij}_{\alpha\beta}$
on the CP-violating phases, let us calculate ${\cal
V}^{ij}_{\alpha\alpha}$ in the standard parametrization of $U$ given
by Eq. (1). We obtain
\begin{eqnarray}
{\cal V}_{ee}^{12} \hspace{-0.2cm} & = & \hspace{-0.2cm} c_{12}^2
s_{12}^2 c_{13}^4 \sin 2\left(\rho - \sigma\right) \; ,
\nonumber \\
{\cal V}_{ee}^{13} \hspace{-0.2cm} & = & \hspace{-0.2cm} c_{12}^2
c_{13}^2 s_{13}^2 \sin 2\left(\delta + \rho\right) \; ,
\nonumber \\
{\cal V}_{ee}^{23} \hspace{-0.2cm} & = & \hspace{-0.2cm} s_{12}^2
c_{13}^2 s_{13}^2 \sin 2\left(\delta + \sigma\right) \; ;
\end{eqnarray}
and
\begin{eqnarray}
{\cal V}_{\mu \mu }^{12} \hspace{-0.2cm} & = & \hspace{-0.2cm}
c_{12}^2 s_{12}^2 \left(c_{23}^4 - 4 s_{13}^2 c_{23}^2 s_{23}^2 +
s_{13}^4 s_{23}^4\right) \sin 2\left(\rho -\sigma\right)
\nonumber \\
\hspace{-0.2cm} & & \hspace{-0.2cm} + 2c_{12}^{} s_{12}^{} s_{13}^{}
c_{23}^{} s_{23} ^{} \left(c_{23}^2 - s_{13}^2 s_{23}^2\right)
\left[c_{12}^2 \sin \left(2 \rho -2 \sigma +\delta\right) - s_{12}^2
\sin \left(2 \rho - 2\sigma -\delta\right) \right]
\nonumber \\
\hspace{-0.2cm} & & \hspace{-0.2cm} + s_{13}^2 c_{23}^2 s_{23}^2
\left[c_{12}^4 \sin 2\left(\rho -\sigma + \delta\right) +
s_{12}^4 \sin 2\left(\rho - \sigma - \delta\right) \right] \; ,
\nonumber \\
{\cal V}_{\mu \mu }^{13} \hspace{-0.2cm} & = & \hspace{-0.2cm}
c_{13}^2 s_{23}^2 \left[ s_{12}^2 c_{23}^2 \sin 2\rho + 2 c_{12}^{}
s_{12}^{} s_{13}^{} c_{23}^{} s_{23}^{} \sin \left(\delta +2
\rho\right) + c_{12}^2 s_{13}^2 s_{23}^2 \sin 2\left(\delta +
\rho\right) \right] \; ,
\nonumber \\
{\cal V}_{\mu \mu }^{23} \hspace{-0.2cm} & = & \hspace{-0.2cm}
c_{13}^2 s_{23}^2 \left[ c_{12}^2 c_{23}^2 \sin 2\sigma - 2
c_{12}^{} s_{12}^{} s_{13}^{} c_{23}^{} s_{23}^{} \sin \left(\delta
+ 2\sigma\right) + s_{12}^2 s_{13}^2 s_{23}^2 \sin 2\left(\delta +
\sigma\right) \right] \; ;
\end{eqnarray}
and
\begin{eqnarray}
{\cal V}_{\tau \tau }^{12} \hspace{-0.2cm} & = & \hspace{-0.2cm}
c_{12}^2 s_{12}^2 \left(s_{23}^4 - 4 s_{13}^2c_{23}^2 s_{23}^2 +
s_{13}^4c_{23}^4 \right) \sin 2\left(\rho - \sigma\right)
\nonumber \\
\hspace{-0.2cm} & & \hspace{-0.2cm} - 2c_{12}^{} s_{12}^{} s_{13}^{}
c_{23}^{} s_{23}^{} \left(s_{23}^2 - s_{13}^2 c_{23}^2\right)
\left[c_{12}^2 \sin \left(2 \rho - 2\sigma + \delta\right) -
s_{12}^2 \sin \left(2 \rho - 2\sigma - \delta\right) \right]
\nonumber \\
\hspace{-0.2cm} & & \hspace{-0.2cm} + s_{13}^2 c_{23}^2 s_{23}^2
\left[c_{12}^4 \sin 2\left(\rho - \sigma + \delta\right) + s_{12}^4
\sin 2\left(\rho - \sigma - \delta\right) \right] \; ,
\nonumber \\
{\cal V}_{\tau \tau }^{13} \hspace{-0.2cm} & = & \hspace{-0.2cm}
c_{13}^2 c_{23}^2 \left[ s_{12}^2 s_{23}^2 \sin 2\rho - 2 c^{}_{12}
s^{}_{12} s^{}_{13} c^{}_{23} s^{}_{23} \sin \left(\delta +
2\rho\right) + c_{12}^2 s_{13}^2 c_{23}^2 \sin 2\left(\delta +
\rho\right) \right] \; ,
\nonumber \\
{\cal V}_{\tau \tau }^{23} \hspace{-0.2cm} & = & \hspace{-0.2cm}
c_{13}^2 c_{23}^2 \left[ c_{12}^2 s_{23}^2 \sin 2\sigma + 2
c^{}_{12} s^{}_{12} s^{}_{13} c^{}_{23} s^{}_{23} \sin \left(\delta
+ 2\sigma\right) + s_{12}^2 s_{13}^2 c_{23}^2 \sin 2\left(\delta +
\sigma\right) \right] \; .
\end{eqnarray}
Taking account of Eq. (18), we can immediately write out the
explicit expressions of ${\cal V}^{ij}_{\alpha\beta}$ (for $\alpha
\neq \beta$) with the help of Eqs. (19)---(21):
\begin{eqnarray}
{\cal V}_{e\mu}^{12} \hspace{-0.2cm} & = & \hspace{-0.2cm} -c_{12}^2
s_{12}^2 c_{13}^2 \left(c_{23}^2 - s_{13}^2 s_{23}^2\right) \sin
2\left(\rho - \sigma\right)
\nonumber \\
\hspace{-0.2cm} & & \hspace{-0.2cm} - c^{}_{12} s^{}_{12} c_{13}^2
s^{}_{13} c^{}_{23} s^{}_{23} \left[c_{12}^2 \sin \left(2\rho -
2\sigma + \delta\right) - s_{12}^2 \sin \left(2\rho - 2\sigma -
\delta\right) \right] \; ,
\nonumber \\
{\cal V}_{e\mu}^{13} \hspace{-0.2cm} & = & \hspace{-0.2cm} -
c^{}_{12} c_{13}^2 s^{}_{13} s^{}_{23} \left[s^{}_{12} c^{}_{23}
\sin \left(\delta + 2\rho\right) - c^{}_{12} s^{}_{13} s^{}_{23}
\sin 2\left(\delta + \rho\right) \right] \; ,
\nonumber \\
{\cal V}_{e\mu}^{23} \hspace{-0.2cm} & = & \hspace{-0.2cm}
+s^{}_{12} c_{13}^2 s^{}_{13} s^{}_{23} \left[c^{}_{12} c^{}_{23}
\sin \left(\delta + 2\sigma\right) - s^{}_{12} s^{}_{13} s^{}_{23}
\sin 2\left(\delta + \sigma\right) \right] \; ;
\end{eqnarray}
and
\begin{eqnarray}
{\cal V}_{e\tau}^{12} \hspace{-0.2cm} & = & \hspace{-0.2cm} c_{12}^2
s_{12}^2 c_{13}^2 \left(c_{23}^2 s_{13}^2 - s_{23}^2\right) \sin
2\left(\rho - \sigma\right)
\nonumber \\
\hspace{-0.2cm} & & \hspace{-0.2cm} + c^{}_{12} s^{}_{12} c_{13}^2
s^{}_{13} c^{}_{23} s^{}_{23} \left[c_{12}^2 \sin \left(2\rho -
2\sigma + \delta\right) - s_{12}^2 \sin \left(2\rho - 2\sigma -
\delta\right) \right] \; ,
\nonumber \\
{\cal V}_{e\tau}^{13} \hspace{-0.2cm} & = & \hspace{-0.2cm}
+c^{}_{12} c_{13}^2 s^{}_{13} c^{}_{23} \left[s^{}_{12} s^{}_{23}
\sin \left(\delta + 2\rho\right) - c^{}_{12} s^{}_{13} c^{}_{23}
\sin 2\left(\delta + \rho\right) \right] \; ,
\nonumber \\
{\cal V}_{e\tau}^{23} \hspace{-0.2cm} & = & \hspace{-0.2cm}
-s^{}_{12} c_{13}^2 s^{}_{13} c^{}_{23} \left[c^{}_{12} s^{}_{23}
\sin \left(\delta + 2\sigma\right) + s^{}_{12} s^{}_{13} c^{}_{23}
\sin 2\left(\delta + \sigma\right) \right] \; ;
\end{eqnarray}
and
\begin{eqnarray}
{\cal V}_{\mu \tau }^{12} \hspace{-0.2cm} & = & \hspace{-0.2cm}
-c_{12}^2 s_{12}^2 \left[c_{23}^4 s_{13}^2 - \left(1 +
s_{13}^2\right)^2 c_{23}^2 s_{23}^2 + s_{13}^2 s_{23}^4\right] \sin
2\left(\rho - \sigma\right)
\nonumber \\
\hspace{-0.2cm} & & \hspace{-0.2cm} - c^{}_{12} s^{}_{12} s^{}_{13}
c^{}_{23} s^{}_{23} \left(1 + s_{13}^2\right) \left(c_{23}^2 -
s_{23}^2\right) \left[c_{12}^2 \sin \left(2\rho - 2\sigma +
\delta\right) - s_{12}^2 \sin \left(2\rho - 2\sigma - \delta\right)
\right]
\nonumber \\
\hspace{-0.2cm} & & \hspace{-0.2cm} -s_{13}^2 c_{23}^2 s_{23}^2
\left[c_{12}^4 \sin 2\left(\rho - \sigma + \delta\right) + s_{12}^4
\sin 2\left(\rho - \sigma - \delta\right)\right] \; ,
\nonumber \\
{\cal V}_{\mu \tau }^{13} \hspace{-0.2cm} & = & \hspace{-0.2cm}
c_{13}^2 c^{}_{23} s^{}_{23} \left[- s_{12}^2 c^{}_{23} s^{}_{23}
\sin 2\rho + c^{}_{12} s^{}_{12} s^{}_{13} \left(c_{23}^2 -
s_{23}^2\right) \sin \left(\delta + 2\rho\right) + c_{12}^2 s_{13}^2
c^{}_{23} s^{}_{23} \sin 2\left(\delta + \rho\right)\right] \; ,
\nonumber \\
{\cal V}_{\mu \tau }^{23} \hspace{-0.2cm} & = & \hspace{-0.2cm}
c_{13}^2 c^{}_{23} s^{}_{23} \left[- c_{12}^2c^{}_{23} s^{}_{23}
\sin 2\sigma - c^{}_{12} s^{}_{12} s^{}_{13} \left(c_{23}^2 -
s_{23}^2\right) \sin \left(\delta + 2\sigma\right) + s_{12}^2
s_{13}^2 c^{}_{23} s^{}_{23} \sin 2\left(\delta +
\sigma\right)\right] \; . \hspace{0.9cm}
\end{eqnarray}
Similar expressions for ${\cal C}^{ij}_{\alpha\beta}$ have been
listed in Appendix A. These results clearly tell us how the
CP-violating quantities ${\cal V}^{ij}_{\alpha\beta}$ depend on the
CP-violating phases $\delta$, $\rho$ and $\sigma$: (a) each ${\cal
V}^{12}_{\alpha\beta}$ is a function of $\rho-\sigma$ and $\delta$
(the only exception is ${\cal V}^{12}_{ee}$, which only involves
$\rho-\sigma$); (b) each ${\cal V}^{13}_{\alpha\beta}$ is a function
of $\rho$ and $\delta$; and (c) each ${\cal V}^{23}_{\alpha\beta}$
is a function of $\sigma$ and $\delta$. The following extreme cases
are particularly interesting.
\begin{table}[t]
\vspace{-0.3cm} \caption{The simplified expressions of the
Jarlskog-like parameters ${\cal V}^{ij}_{\alpha\beta}$ and their
relations with the Jarlskog parameter ${\cal J}$ in the $\rho =
\sigma = 0$ limit. Their typical numerical results are obtained by
inputting $\theta^{}_{12} \simeq 33.4^\circ$, $\theta^{}_{13} \simeq
8.66^\circ$ and $\theta^{}_{23} \simeq 40.0^\circ$ \cite{FIT} in the
$\delta = 45^\circ$ and $\delta = 90^\circ$ cases.}
\begin{center}
\begin{tabular}{l|ll}
\hline\hline Jarlskog-like parameter & $\delta = 45^\circ$
\hspace{1cm} & $\delta = 90^\circ$
\hspace{1cm} \\
\hline\hline
${\cal V}^{12}_{ee} = 0$ & $\hspace{0.3cm}0$ & $\hspace{0.3cm}0$ \\
\hline
${\cal V}^{13}_{ee} = c^2_{12} c^2_{13} s^2_{13} \sin 2\delta$ &
$+0.016$ & $\hspace{0.3cm}0$ \\
\hline
${\cal V}^{23}_{ee} = s^2_{12} c^2_{13} s^2_{13} \sin 2\delta$ &
$+0.0067$ & $\hspace{0.3cm}0$ \\
\hline
${\cal V}^{12}_{\mu\mu} = \left[2{\cal J} \left(c^2_{23} - s^2_{13}
s^2_{23}\right) + \left({\cal V}^{13}_{ee} - {\cal V}^{23}_{ee}
\right) c^2_{23} s^2_{23}\right]/c^2_{13}$ &
$+0.030$ & $+0.039$ \\
\hline
${\cal V}^{13}_{\mu\mu} = 2{\cal J} s^2_{23} + {\cal V}^{13}_{ee}
s^4_{23}$ &
$+0.022$ & $+0.028$ \\
\hline
${\cal V}^{23}_{\mu\mu} = -2{\cal J} s^2_{23} + {\cal V}^{23}_{ee}
s^4_{23}$ &
$-0.018$ & $-0.028$ \\
\hline
${\cal V}^{12}_{\tau\tau} = \left[2{\cal J} \left(s^2_{13} c^2_{23}
- s^2_{23}\right) + \left({\cal V}^{13}_{ee} - {\cal V}^{23}_{ee}
\right) c^2_{23} s^2_{23}\right]/c^2_{13}$ &
$-0.017$ & $-0.027$ \\
\hline
${\cal V}^{13}_{\tau\tau} = -2{\cal J} c^2_{23} + {\cal V}^{13}_{ee}
c^4_{23}$ &
$-0.022$ & $-0.039$ \\
\hline
${\cal V}^{23}_{\tau\tau} = 2{\cal J} c^2_{23} + {\cal V}^{23}_{ee}
c^4_{23}$ &
$+0.030$ & $+0.039$ \\
\hline\hline
${\cal V}^{12}_{e\mu} = -{\cal J}$ & $-0.024$ & $-0.033$ \\
\hline
${\cal V}^{13}_{e\mu} = -{\cal J} + {\cal V}^{13}_{ee} s^2_{23}$ &
$-0.030$ & $-0.033$ \\
\hline
${\cal V}^{23}_{e\mu} = {\cal J} - {\cal V}^{13}_{ee} s^2_{23}$ &
$+0.021$ & $+0.033$ \\
\hline
${\cal V}^{12}_{e\tau} = {\cal J}$ & $+0.024$ & $+0.033$ \\
\hline
${\cal V}^{13}_{e\tau} = {\cal J} - {\cal V}^{13}_{ee} c^2_{23}$ &
$+0.014$ & $+0.033$ \\
\hline
${\cal V}^{23}_{e\tau} = -{\cal J} - {\cal V}^{23}_{ee} c^2_{23}$ &
$-0.028$ & $-0.033$ \\
\hline
${\cal V}^{12}_{\mu\tau} = \left[{\cal J} \left(1 + s^2_{13}\right)
\left(s^2_{23} - c^2_{23}\right) - \left({\cal V}^{13}_{ee} - {\cal
V}^{23}_{ee} \right) c^2_{23} s^2_{23}\right]/c^2_{13}$ &
$-0.0064$ & $-0.0060$ \\
\hline
${\cal V}^{13}_{\mu\tau} = {\cal J} \left(c^2_{23} - s^2_{23}\right)
+ {\cal V}^{13}_{ee} c^2_{23} s^2_{23}$ &
$+0.0078$ & $+0.0058$ \\
\hline
${\cal V}^{23}_{\mu\tau} = {\cal J} \left(s^2_{23} - c^2_{23}\right)
+ {\cal V}^{23}_{ee} c^2_{23} s^2_{23}$ &
$-0.0025$ & $-0.0058$ \\
\hline\hline
\end{tabular}
\end{center}
\end{table}
\begin{itemize}
\item     In the $\delta =0$ (or $\pi$) limit, ${\cal J} =0$ holds, but
all the ${\cal V}^{ij}_{\alpha\beta}$ are in general nonvanishing.
In this special case there will be no CP or T violation in normal
neutrino-neutrino and antineutrino-antineutrino oscillations, but
large CP or T violation in neutrino-antineutrino oscillations is
possible.

\item     In the $\theta^{}_{13} = 0$ limit, ${\cal J} =0$ holds, so
does
\begin{eqnarray}
{\cal V}^{13}_{ee} \hspace{-0.2cm} & = & \hspace{-0.2cm} {\cal
V}^{23}_{ee} = {\cal V}^{13}_{e\mu} = {\cal V}^{23}_{e\mu} = {\cal
V}^{13}_{e\tau} = {\cal V}^{23}_{e\tau} = 0 \; ,
\end{eqnarray}
simply because all of them involve the $U^{}_{e 3} = s^{}_{13}
e^{-{\rm i}\delta}$ element. Those nonvanishing Jarlskog-like
parameters depend on either $\rho$ or $\sigma$, or their difference
$\rho -\sigma$. However, such an extreme case is not favored by the
recent reactor antineutrino oscillation data (i.e., $\theta^{}_{13}
\simeq 9^\circ$ \cite{DYB,RENO}).

\item     In the $\rho = \sigma = 0$ limit, which looks like a
``pseudo-Dirac" case with a single CP-violating parameter $\delta$,
we obtain ${\cal V}^{12}_{ee} = 0$, ${\cal V}^{13}_{ee} = c^2_{12}
c^2_{13} s^2_{13} \sin 2\delta$ and ${\cal V}^{23}_{ee} = s^2_{12}
c^2_{13} s^2_{13} \sin 2\delta$. The other fifteen ${\cal
V}^{ij}_{\alpha\beta}$ can all be given in terms of ${\cal J}$,
${\cal V}^{13}_{ee}$ and ${\cal V}^{23}_{ee}$, as listed in Table 1.
We see ${\cal V}^{12}_{e\tau} = -{\cal V}^{12}_{e\mu} = {\cal J}$,
and the other nonvanishing ${\cal V}^{ij}_{\alpha\beta}$ may also
receive the higher-order contributions proportional to $s^2_{13}
\sin 2\delta$ (i.e., the ${\cal V}^{13}_{ee}$ and ${\cal
V}^{23}_{ee}$ terms). In this case the Jarlskog parameter ${\cal J}$
governs CP and T violation in both normal neutrino-neutrino (or
antineutrino-antineutrino) oscillations and neutrino-antineutrino
oscillations.
\end{itemize}
Of course, it is possible to relate ${\cal V}^{ij}_{\alpha\beta}$ to
$\mathcal{J}$ in some other special cases. For example, $\rho =
\sigma = -\delta$ leads to ${\cal V}^{12}_{e\tau} = -{\cal
V}^{13}_{e\tau} = {\cal V}^{23}_{e\tau} = -{\cal V}^{12}_{e\mu} =
{\cal V}^{13}_{e\mu} = -{\cal V}^{23}_{e\mu} = {\cal J}$.
\begin{figure}[t]
\center
\begin{overpic}[width=14.3cm]{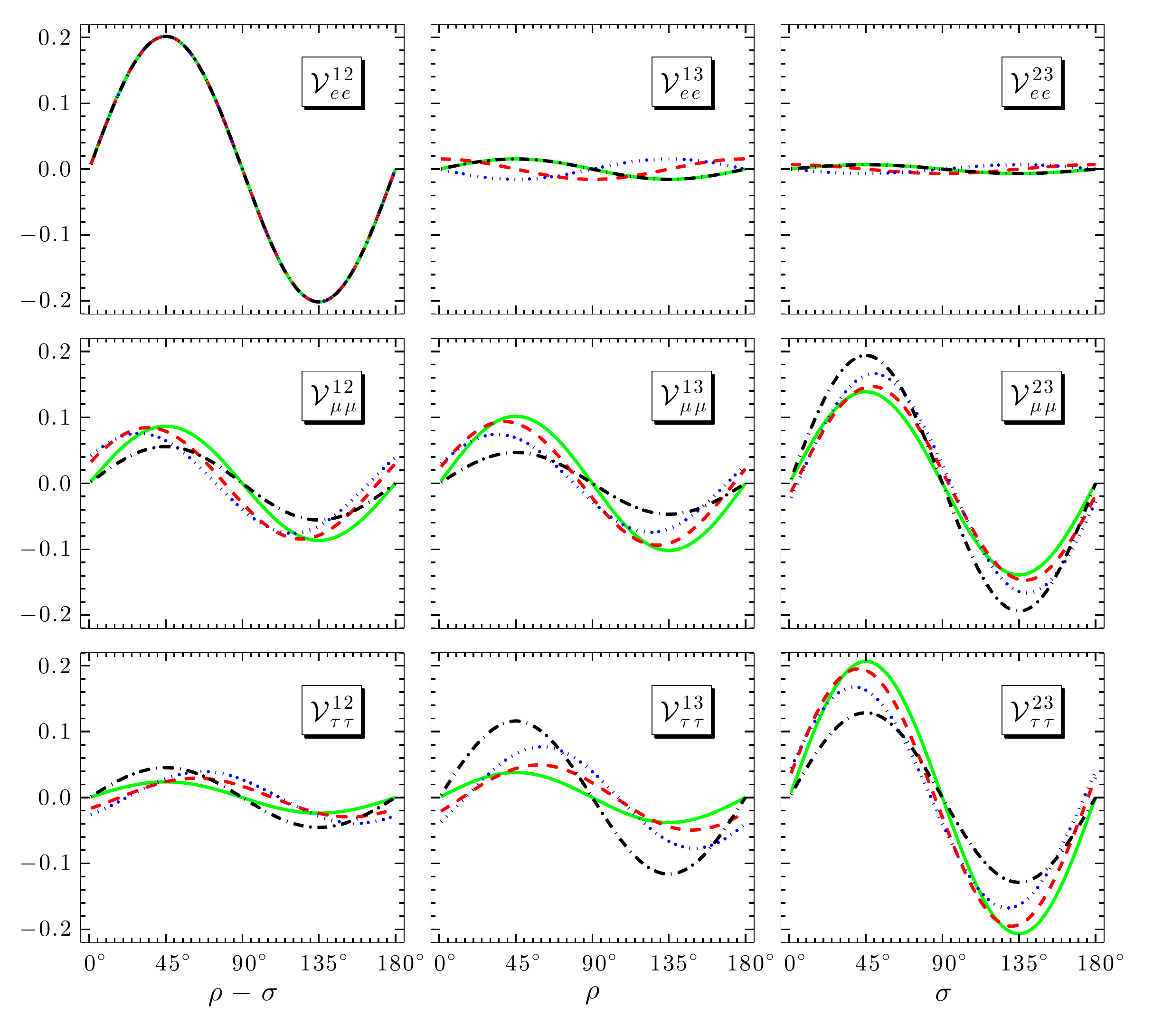}
\end{overpic}
\vspace{-0.4cm} \caption{The Jarlskog-like parameters ${\cal
V}^{ij}_{\alpha\alpha}$ changing with the CP-violating phases. The
green solid, red dashed, blue dotted and black dashed-dotted lines
correspond to $\delta=0^\circ$, $45^\circ$, $90^\circ$ and
$180^\circ$, respectively. The typical inputs are $\theta^{}_{12}
\simeq 33.4^\circ$, $\theta^{}_{13} \simeq 8.66^\circ$ and
$\theta^{}_{23} \simeq 40.0^\circ$\cite{FIT}.}
\end{figure}

We proceed to illustrate the numerical dependence of ${\cal
V}^{ij}_{\alpha\beta}$ on $\rho$, $\sigma$ and $\delta$ by taking
$\theta^{}_{12} \simeq 33.4^\circ$, $\theta^{}_{13} \simeq
8.66^\circ$ and $\theta^{}_{23} \simeq 40.0^\circ$ as the typical
inputs \cite{FIT}. As the ``Dirac" phase $\delta$ is expected to be
determined earlier than the Majorana phases $\rho$ and $\sigma$, one
may fix the value of $\delta$ (for example, $\delta = 0^\circ$,
$45^\circ$, $90^\circ$ or $180^\circ$) to show how ${\cal
V}^{ij}_{\alpha\beta}$ can change with $\rho-\sigma$, $\rho$ or
$\sigma$. Our numerical results of ${\cal V}^{ij}_{\alpha\beta}$ for
the $\alpha=\beta$ and $\alpha\neq\beta$ cases are shown in Figures
2 and 3, respectively. In addition, the magnitude of each ${\cal
V}^{ij}_{\alpha\beta}$ in the $\rho =\sigma =0$ limit is illustrated
in Table 1. Some comments are in order.
\begin{figure}[t]
\center
\begin{overpic}[width=14.3cm]{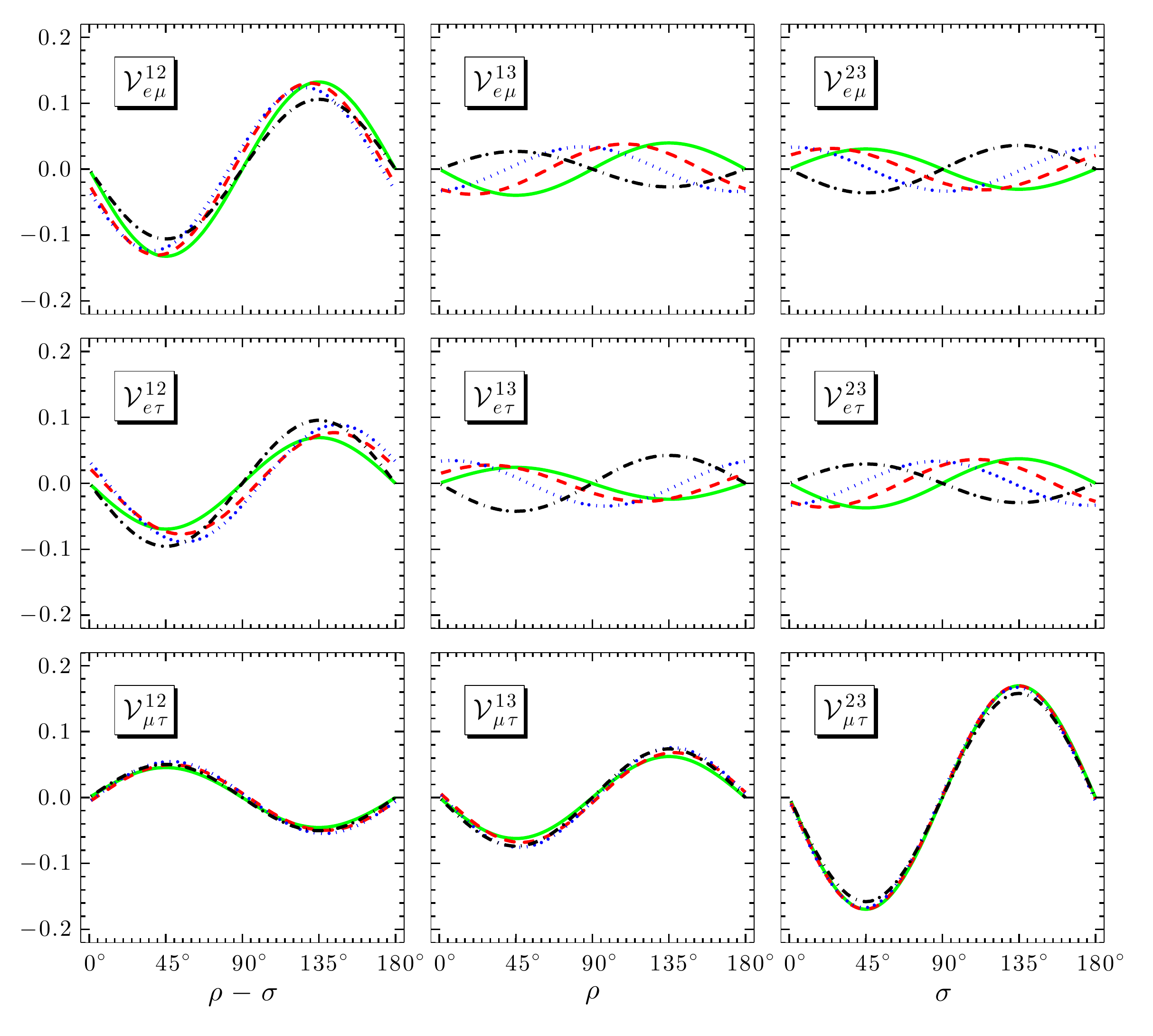}
\end{overpic}
\vspace{-0.4cm} \caption{The Jarlskog-like parameters ${\cal
V}^{ij}_{\alpha\beta}$ (for $\alpha \neq \beta$) changing with the
CP-violating phases. The green solid, red dashed, blue dotted and
black dashed-dotted lines correspond to $\delta=0^\circ$,
$45^\circ$, $90^\circ$ and $180^\circ$, respectively. The typical
inputs are $\theta^{}_{12} \simeq 33.4^\circ$, $\theta^{}_{13}
\simeq 8.66^\circ$ and $\theta^{}_{23} \simeq
40.0^\circ$\cite{FIT}.}
\end{figure}
\begin{itemize}
\item     It is amazing that ${\cal V}^{12}_{ee}$, ${\cal
V}^{23}_{\mu\mu}$, ${\cal V}^{23}_{\tau\tau}$ and ${\cal
V}^{23}_{\mu\tau}$ can maximally reach about $20\%$ in magnitude. In
comparison, ${\cal J} \leq 1/\left(6\sqrt{3}\right) \simeq 9.6\%$
constrains the strength of CP and T violation in normal
neutrino-neutrino oscillations \cite{Cabibbo}. The reason for
possible largeness of the above four Jarlskog-like parameters is
simply that their leading terms are only slightly suppressed by
$s^2_{12}$ or $s^2_{23}$.

\item     The magnitudes of ${\cal V}^{13}_{ee}$ and ${\cal
V}^{23}_{ee}$ are strongly suppressed, because both of them are
proportional to $s^2_{13} \simeq 2.3\%$. We see that the magnitudes
of ${\cal V}^{13}_{e\mu}$, ${\cal V}^{23}_{e\mu}$, ${\cal
V}^{13}_{e\tau}$ and ${\cal V}^{23}_{e\tau}$ are modest, since their
leading terms are comparable with ${\cal J}$. In other words, they
are essentially constrained to be $\lesssim 10\%$.

\item     ${\cal V}^{12}_{ee}$ has nothing to do with the ``Dirac"
phase $\delta$, as one can see in Eq. (19). The dependence of ${\cal
V}^{12}_{\mu\tau}$, ${\cal V}^{13}_{\mu\tau}$ and ${\cal
V}^{23}_{\mu\tau}$ on $\delta$ is very weak, because this dependence
is suppressed either by the factor $s^{}_{13} \left(c^2_{23} -
s^2_{23}\right) \simeq 2.6\%$ or by the factor $s^2_{13} \simeq
2.3\%$ as shown in Eq. (24).

\item     The ``pseudo-Dirac" case illustrated in Table 1 is
interesting in the sense that appreciable CP- and T-violating
effects are expected to show up in neutrino-antineutrino
oscillations even the Majorana phases $\rho$ and $\sigma$ vanish.
Namely, the Majorana neutrinos with only the ``Dirac" CP-violating
phase behave very differently from the Dirac neutrinos
\footnote{This point can also be seen by examining their distinct
renormalization-group running effects \cite{RGE}.}.
\end{itemize}
Therefore, it is in principle possible to determine all the three
CP-violating phases in neutrino-antineutrino oscillations, in which
the strength of CP and T violation is governed by ${\cal
V}^{ij}_{\alpha\beta}$, whose maximal magnitudes could be larger
than that of ${\cal J}$ by a factor of two or so. We shall come back
to this point in section 4 to analyze the CP-violating asymmetries
between $\nu^{}_\alpha \to \overline{\nu}^{}_\beta$ and
$\overline{\nu}^{}_\alpha \to \nu^{}_\beta$ oscillations.

\subsection{The effective mass terms $\langle m\rangle^{}_{\alpha\beta}$}

The effective mass terms $\langle m\rangle^{}_{\alpha\beta}$ defined
in Eq. (4) are important to understand the origin of neutrino
masses, since they are simply the $(\alpha, \beta)$ elements of the
symmetric Majorana neutrino mass matrix $M^{}_\nu$ in the basis
where the flavor eigenstates of three charged leptons are identified
with their mass eigenstates. Namely,
\begin{eqnarray}
M^{}_\nu \hspace{-0.2cm} & = & \hspace{-0.2cm} \left(\begin{matrix}
\langle m\rangle^{}_{ee} & \langle m\rangle^{}_{e\mu} &
\langle m\rangle^{}_{e\tau} \cr
\langle m\rangle^{}_{e\mu} & \langle m\rangle^{}_{\mu\mu} &
\langle m\rangle^{}_{\mu\tau} \cr
\langle m\rangle^{}_{e\tau} & \langle m\rangle^{}_{\mu\tau} &
\langle m\rangle^{}_{\tau\tau} \end{matrix}\right) \; ,
\end{eqnarray}
where $\langle m\rangle^{}_{\beta\alpha} = \langle
m\rangle^{}_{\alpha\beta}$ has been taken into account. In the
standard parametrization of $U$, we have
\begin{eqnarray}
\langle m\rangle_{ee}^{} \hspace{-0.2cm} & = & \hspace{-0.2cm}
m_1^{} c_{12}^2 c_{13}^2 e^{2 {\rm i} \rho } + m_2^{} s_{12}^2 c_{13}^2
e^{2 {\rm i} \sigma} + m_3^{} s_{13}^2 e^{-2 {\rm i} \delta } \; ,
\nonumber\\
\langle m\rangle_{\mu\mu}^{} \hspace{-0.2cm} & = & \hspace{-0.2cm}
m_1^{} \left(s^{}_{12} c^{}_{23} + c^{}_{12} s^{}_{13} s^{}_{23}
e^{{\rm i} \delta}\right)^2 e^{2 {\rm i} \rho}+m_2^{} \left(c^{}_{12} c^{}_{23}
- s^{}_{12} s^{}_{13} s^{}_{23} e^{{\rm i} \delta}\right)^2 e^{2 {\rm i} \sigma}
+ m_3^{} c_{13}^2 s_{23}^2 \; ,
\nonumber\\
\langle m\rangle_{\tau\tau}^{} \hspace{-0.2cm} & = & \hspace{-0.2cm}
m_1^{} \left(s^{}_{12} s^{}_{23} - c^{}_{12} s^{}_{13} c^{}_{23}
e^{{\rm i} \delta}\right)^2 e^{2 {\rm i} \rho} + m_2^{} \left(c^{}_{12}
s^{}_{23} + s^{}_{12} s^{}_{13} c^{}_{23} e^{{\rm i} \delta}\right)^2 e^{2
{\rm i} \sigma} + m_3^{} c_{13}^2 c_{23}^2 \; ,
\nonumber\\
\langle m\rangle_{e\mu}^{} \hspace{-0.2cm} & = & \hspace{-0.2cm}
-m_1^{} c_{12}^{} c_{13}^{} \left(s_{12}^{} c_{23}^{} + c_{12}^{}
s_{13}^{} s_{23}^{} e^{{\rm i} \delta} \right) e^{2 {\rm i} \rho} + m_2^{}
s_{12}^{} c_{13}^{} \left( c_{23}^{} c_{12}^{} - s_{12}^{} s_{13}^{}
s_{23}^{} e^{i \delta}\right) e^{2 {\rm i} \sigma}
\nonumber\\
\hspace{-0.2cm} & & \hspace{-0.2cm} + m_3^{} c_{13}^{} s_{13}^{}
s_{23}^{} e^{-{\rm i} \delta} \; ,
\nonumber\\
\langle m\rangle_{e\tau}^{} \hspace{-0.2cm} & = & \hspace{-0.2cm}
+m_1^{} c_{12}^{} c_{13}^{} \left(s_{12}^{} s_{23}^{} - c_{12}^{}
s_{13}^{} c_{23}^{} e^{{\rm i} \delta}\right) e^{2 {\rm i} \rho} - m_2^{}
s_{12}^{}c_{13}^{} \left(c_{12}^{} s_{23}^{} + s_{12}^{} s_{13}^{}
c_{23}^{} e^{{\rm i} \delta}\right) e^{2 {\rm i} \sigma}
\nonumber\\
\hspace{-0.2cm} & & \hspace{-0.2cm} + m_3^{} c_{13}^{} s_{13}^{}
c_{23}^{} e^{-{\rm i} \delta} \; ,
\nonumber\\
\langle m\rangle_{\mu\tau}^{} \hspace{-0.2cm} & = & \hspace{-0.2cm}
-m_1^{} \left(s_{12}^{} s_{23}^{} - c_{12}^{} s_{13}^{} c_{23}^{}
e^{{\rm i} \delta}\right) \left(c_{23}^{} s_{12}^{} + c_{12}^{} s_{13}^{}
s_{23}^{} e^{{\rm i} \delta} \right) e^{2 {\rm i} \rho}
\nonumber\\
\hspace{-0.2cm} & & \hspace{-0.2cm} - m_2^{} \left(c_{12}^{}
s_{23}^{} + s_{12}^{} s_{13}^{} c_{23}^{} e^{{\rm i} \delta}\right)
\left(c_{12}^{} c_{23}^{} - s_{12}^{} s_{13}^{} s_{23}^{} e^{{\rm i}
\delta}\right) e^{2 {\rm i} \sigma} + m_3^{} c_{13}^2 c_{23}^{} s_{23}^{}
\; .
\end{eqnarray}
We see that a measurement of the three CP-violating phases is
absolutely necessary in order to fully reconstruct the neutrino mass
matrix $M^{}_\nu$. Without the information on $\rho$ and $\sigma$,
it would be impossible to model-independently look into the
structure of $M^{}_\nu$ via a bottom-up approach. On the other hand,
a predictive model of lepton flavors should be able to specify the
texture of $M^{}_\nu$ via a top-down approach, such that its
predictions can be experimentally tested.

Note that $|\langle m\rangle^{}_\alpha|^2$ in Eq. (10) can be
related to $\langle m\rangle^{}_{\alpha\beta}$ as follows:
\begin{eqnarray}
\sum_{\beta} \left| \langle m \rangle^{}_{\alpha\beta} \right|^2
\hspace{-0.2cm} & = & \hspace{-0.2cm} \left| \langle m
\rangle^{}_{\alpha} \right|^2 = \sum_i m^2_i \left|U^{}_{\alpha
i}\right|^2 \; ,
\end{eqnarray}
It is obvious that all the $|\langle m\rangle^{}_\alpha|^2$ do not
contain any information about the Majorana phases $\rho$ and
$\sigma$, but they may depend on the ``Dirac" phase $\delta$.
Furthermore, we have
\begin{eqnarray}
\sum_{\alpha} \left| \langle m \rangle^{}_\alpha\right|^2
\hspace{-0.2cm} & = & \hspace{-0.2cm} \sum_{i} m^2_i = 3m^2_1 +
\Delta m^2_{21} + \Delta m^2_{31} = 3m^2_3 - \Delta m^2_{21} -
2\Delta m^2_{32} \; ,
\end{eqnarray}
where $\Delta m^2_{ij} \equiv m^2_i - m^2_j$ (for $i, j = 1, 2, 3$).
A global analysis of current neutrino oscillation data has given
$\Delta m^2_{21} \simeq 7.50 \times 10^{-5} ~{\rm eV}^2$ and $\Delta
m^2_{31} \simeq 2.473 \times 10^{-3} ~{\rm eV}^2$ (normal neutrino
mass hierarchy) or $\Delta m^2_{32} \simeq -2.427 \times 10^{-3}
~{\rm eV}^2$ (inverted neutrino mass hierarchy) \cite{FIT}.
Therefore,
\begin{eqnarray}
{\rm Normal ~ hierarchy:~} && \sum_{i} m^2_i \geq \Delta m^2_{21} +
\Delta m^2_{31} \simeq 2.55 \times 10^{-3} ~{\rm eV}^2 \; ,
\nonumber \\
{\rm Inverted ~ hierarchy:} && \sum_{i} m^2_i \geq -\Delta m^2_{21}
- 2\Delta m^2_{32} \simeq 4.78 \times 10^{-3} ~{\rm eV}^2 \; ,
\end{eqnarray}
where the lower bounds correspond to $m^{}_1 =0$ (normal hierarchy)
and $m^{}_3 =0$ (inverted hierarchy), respectively. On the other
hand, the sum of the three neutrino masses can also be written as
\begin{eqnarray}
\sum_{i} m^{}_i \hspace{-0.2cm} & = & \hspace{-0.2cm} m^{}_1 +
\sqrt{m^2_1 + \Delta m^2_{21}} \ + \sqrt{m^2_1 + \Delta m^2_{31}} \
\nonumber \\
\hspace{-0.2cm} & = & \hspace{-0.2cm} m^{}_3 + \sqrt{m^2_3 -\Delta
m^2_{32}} \ + \sqrt{m^2_3 - \Delta m^2_{32} - \Delta m^2_{21}} \ \; .
\end{eqnarray}
This sum has well been constrained thanks to the recent WMAP
\cite{WMAP} and PLANCK \cite{PLANCK} data, and its upper bound is
about $0.23$ eV at the $95\%$ confidence level \cite{PLANCK}. One
may then obtain the allowed range of the lightest neutrino mass by
using the above inputs: $0 \lesssim m^{}_1 \lesssim 0.071$ eV in the
normal hierarchy; or $0 \lesssim m^{}_3 \lesssim 0.065$ eV in the
inverted hierarchy.
\begin{figure}[t]
\center
\begin{overpic}[width=13.3cm]{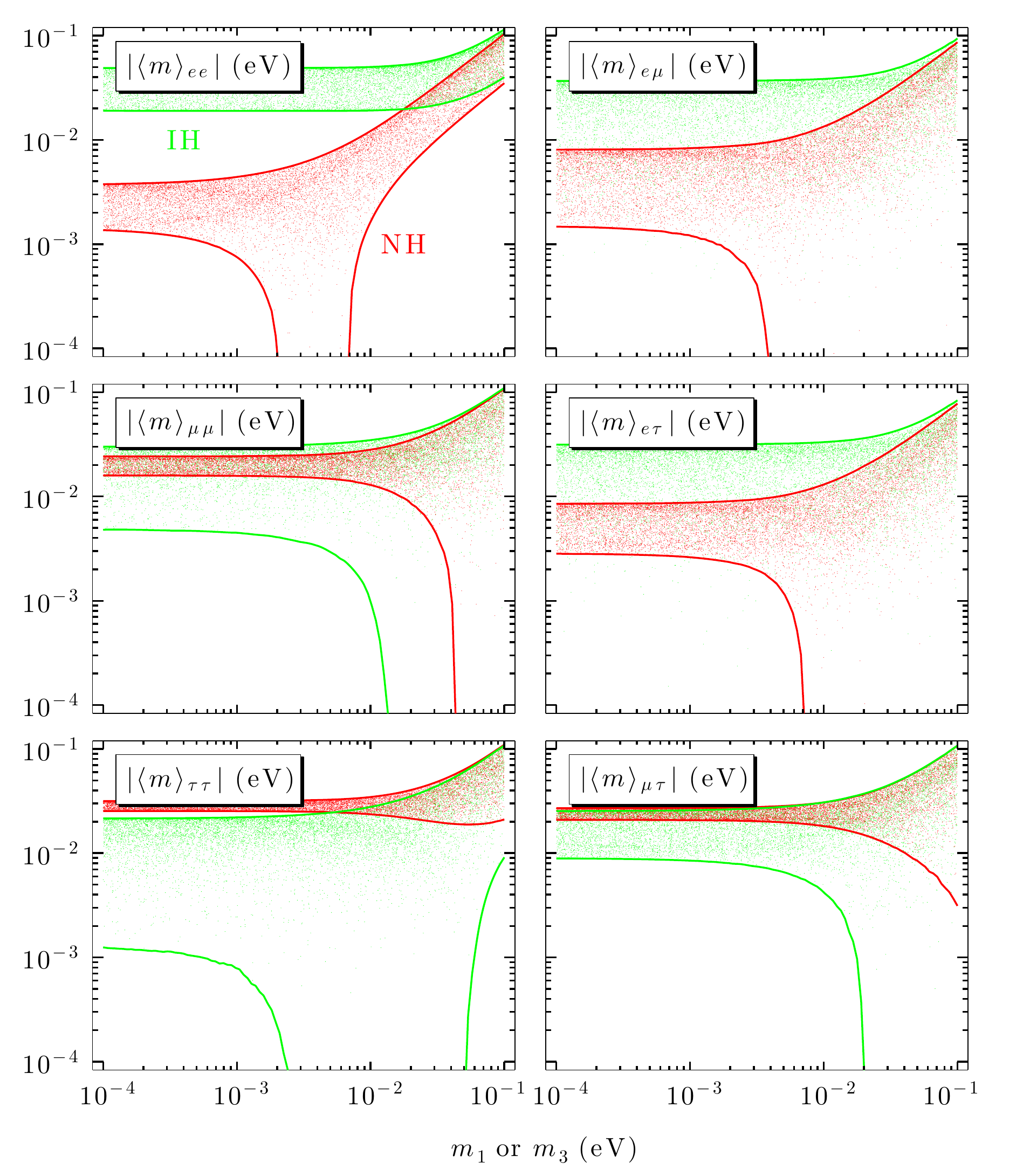}
\end{overpic}
\vspace{-0.2cm} \caption{The profiles of $|\langle
m\rangle^{}_{\alpha\beta}|$ versus the lightest neutrino mass
$m^{}_1$ (normal hierarchy or NH: red region) or $m^{}_3$ (inverted
hierarchy or IH: green region).}
\end{figure}

Figure 4 illustrate the profiles of six $|\langle
m\rangle^{}_{\alpha\beta}|$. Our inputs are $\theta^{}_{12} \simeq
33.4^\circ$, $\theta^{}_{13} \simeq 8.66^\circ$ and $\theta^{}_{23}
\simeq 40.0^\circ$; $\Delta m^2_{21} \simeq 7.50 \times 10^{-5}
~{\rm eV}^2$ and $\Delta m^2_{31} \simeq 2.473 \times 10^{-3} ~{\rm
eV}^2$ (normal hierarchy) or $\Delta m^2_{32} \simeq -2.427 \times
10^{-3} ~{\rm eV}^2$ (inverted hierarchy) \cite{FIT}. As for the
three unknown phase parameters, we allow the ``Dirac" phase $\delta$
to randomly vary between $0^\circ$ and $360^\circ$, and allow the
Majorana phases $\rho$ and $\sigma$ to randomly vary between
$0^\circ$ and $180^\circ$. We plot the results of $|\langle m
\rangle^{}_{\alpha\beta}|$ versus the lightest neutrino mass in
Figure 4 by allowing the latter to vary from $10^{-4}$ eV to
$10^{-1}$ eV, where the upper bound is set by taking account of the
recent PLANCK data \cite{PLANCK}. To understand our numerical
results, we have also made some analytical approximations for
$\langle m \rangle^{}_{\alpha\beta}$ in Appendix B. Some discussions
are in order.
\begin{itemize}
\item     Given the normal neutrino mass hierarchy, most of the
random points of $|\langle m\rangle^{}_{\mu\mu}|$, $|\langle
m\rangle^{}_{\tau\tau}|$, $|\langle m\rangle^{}_{\mu\tau}|$ are
located in the region of $10^{-2}$ eV to $10^{-1}$ eV. This
observation is also true for all the $|\langle
m\rangle^{}_{\alpha\beta}|$ in the inverted hierarchy. Such results
are compatible with the analytical approximations made in Appendix
B. The point is that the relevant $|\langle
m\rangle^{}_{\alpha\beta}|$ are dominated by $\sqrt{|\Delta
m^2_{31}|} \simeq \sqrt{|\Delta m^2_{32}|}\simeq 0.05$ eV when the
lightest neutrino mass is sufficiently small, and all the $|\langle
m\rangle^{}_{\alpha\beta}|$ approach $m^{}_1 \simeq m^{}_2 \simeq
m^{}_3 > 0.05$ eV for a nearly degenerate neutrino mass spectrum.

\item     The random points of $|\langle m\rangle^{}_{ee}|$,
$|\langle m\rangle^{}_{e\mu}|$ and $|\langle m\rangle^{}_{e\tau}|$
in the normal hierarchy are most likely to lie in the region of
$10^{-3}$ eV to $10^{-2}$ eV, especially when $m^2_1 \ll \Delta
m^2_{31}$. Their magnitudes are in general smaller than those in the
inverted hierarchy. The reason is simply that $|\langle
m\rangle^{}_{ee}| \sim \sqrt{\Delta m_{21}^{2}} ~s_{12}^2 \simeq 2.6
\times 10^{-3}$ eV and $|\langle m\rangle^{}_{e\mu}| \sim |\langle
m\rangle^{}_{e\tau}| \sim \sqrt{\Delta m_{31}^{2}} ~s_{13}^{} \simeq
7.5 \times 10^{-3}$ eV hold in the normal hierarchy, while in the
inverted hierarchy the dominant masses $m^{}_1 \simeq m^{}_2 \simeq
\sqrt{-\Delta m^2_{32}}$ do not undergo this $s^{}_{13}$ suppression
(see Appendix B).

\item     In the limit where the lightest neutrino mass approaches
zero or much smaller than $\sqrt{\Delta m^2_{21}}$, the allowed
region of $|\langle m\rangle^{}_{\alpha\beta}|$ in the normal
hierarchy is narrower than that in the inverted hierarchy, as shown
in Figure 4, where the only exception is $|\langle
m\rangle^{}_{ee}|$. The reason can be seen from Eqs. (49)---(52) in
Appendix B: the dominant term of each $|\langle m
\rangle^{}_{\alpha\beta}|$ (for $\alpha\beta \neq ee$) is
proportional to $\sqrt{\Delta m^2_{31}}$ and its uncertainty is
associated with $\sqrt{\Delta m^2_{21}}$ in the normal hierarchy,
while the uncertainty of the same effective mass term in the
inverted hierarchy does not undergo this suppression. Because the
two terms of $\langle m\rangle^{}_{ee}$ in Eq. (49) are almost
comparable in magnitude, its magnitude involves a relatively large
uncertainty in the normal hierarchy as in the inverted hierarchy.

\item     In the $m^{}_1 \simeq m^{}_2 \simeq m^{}_3$ limit,
which is guaranteed if the lightest neutrino mass is larger or much
larger than $\sqrt{|\Delta m^2_{31}|} \simeq \sqrt{|\Delta
m^2_{32}|} \simeq 0.05$ eV, the $|\langle
m\rangle^{}_{\alpha\beta}|$ in both normal and inverted hierarchies
should have the same bounds. This is because the $m^{}_i$ can be
factored out from the expression of each $|\langle
m\rangle^{}_{\alpha\beta}|$, making the latter insensitive to the
ordering of the three masses. Such a feature has essentially been
reflected in Figure 4 (see the limit of $m^{}_1 \to 0.1$ eV or
$m^{}_3 \to 0.1$ eV), and it will become more obvious if $m^{}_1$
(or $m^{}_3$) runs to much larger values, such as $0.2$ eV or even
$0.5$ eV. See also Appendix B for some relevant analytical
approximations in this case.

\item     For some values of the lightest neutrino mass,
$|\langle m\rangle^{}_{\alpha\beta}| =0$ is always allowed, as shown
in Figure 4, either in the normal hierarchy (e.g., $|\langle
m\rangle^{}_{ee}| =0$ \cite{Xing03}) or in the inverted hierarchy
(e.g., $|\langle m\rangle^{}_{\tau\tau}| =0$), or in both of them
(e.g., $|\langle m\rangle^{}_{e\mu}| =0$). This kind of {\it texture
zeros} implies that significant cancellations can happen in
$|\langle m\rangle^{}_{\alpha\beta}|$ due to the unknown
CP-violating phases \cite{Xing03,Zero}. For instance, $\rho =
\sigma$ may lead to $|\langle m\rangle^{}_{e\mu}| \simeq |\langle
m\rangle^{}_{e\tau}| \simeq 0$ when the three neutrino masses are
nearly degenerate, as one can see from Eq. (53) in Appendix B.
Fortunately, it is impossible for all the $|\langle
m\rangle^{}_{\alpha\beta}|$ to be simultaneously vanishing or highly
suppressed, no matter what values $m^{}_1$ or $m^{}_3$ may take.
Unfortunately, current experimental techniques only allow us to
constrain $|\langle m\rangle^{}_{ee}|$ via a careful measurement of
the $0\nu\beta\beta$ decay \cite{Rode}.
\end{itemize}
Taking the upper bound of the sum of three neutrino masses as set by
the recent PLANCK data (i.e., $m^{}_1 + m^{}_2 + m^{}_3 < 0.23$ eV
at the $95\%$ confidence level \cite{PLANCK}), we may obtain the
upper bounds on all the $|\langle m\rangle^{}_{\alpha\beta}|$ from
our numerical calculations:
\begin{eqnarray*}
|\langle m\rangle^{}_{ee}|\lesssim 0.072 \; \text{eV} \; , &\quad
|\langle m\rangle^{}_{\mu\mu}|\lesssim 0.077 \; \text{eV} \;, \quad&
|\langle m\rangle^{}_{\tau\tau}|\lesssim 0.080 \; \text{eV} \; , \\
|\langle m\rangle^{}_{e\mu}|\lesssim 0.060 \; \text{eV} \; , &\quad
|\langle m\rangle^{}_{e\tau}|\lesssim 0.055 \; \text{eV} \; , \quad&
|\langle m\rangle^{}_{\mu\tau}|\lesssim 0.078 \; \text{eV} \; ,
\end{eqnarray*}
for the normal neutrino mass hierarchy; and
\begin{eqnarray*}
|\langle m\rangle^{}_{ee}|\lesssim 0.082 \; \text{eV} \; , &\quad
|\langle m\rangle^{}_{\mu\mu}|\lesssim 0.075 \; \text{eV} \; , \quad&
|\langle m\rangle^{}_{\tau\tau}|\lesssim 0.072 \; \text{eV} \; , \\
|\langle m\rangle^{}_{e\mu}|\lesssim 0.065 \; \text{eV} \; , &\quad
|\langle m\rangle^{}_{e\tau}|\lesssim 0.058 \; \text{eV} \; , \quad&
|\langle m\rangle^{}_{\mu\tau}|\lesssim 0.072 \; \text{eV} \; ,
\end{eqnarray*}
for the inverted neutrino mass hierarchy.
\begin{table}[t]
\vspace{-0.3cm} \caption{The branching ratios of the LNV $H^{+} \to
\ell^+_{\alpha} \overline{\nu}^{}$ decay modes in four different
neutrino mass hierarchies, where $\delta = 90^\circ$ has been
assumed.}
\begin{center}
\begin{tabular}{l|lll}
\hline\hline Normal hierarchy \hspace{3cm} && $m^{}_1 =0$
\hspace{2cm} & $m^{}_1 =0.1$ eV
\hspace{1cm} \\
\hline
${\cal B}(H^{+} \to e^+ \overline{\nu})$ && $0.03$ & $0.31$ \\
\hline
${\cal B}(H^{+} \to \mu^+ \overline{\nu})$ &&
$0.41$ & $0.34$ \\
\hline
${\cal B}(H^{+} \to \tau^+ \overline{\nu})$ &&
$0.56$ & $0.35$ \\
\hline\hline Inverted hierarchy && $m^{}_3 =0$ \hspace{1cm} &
$m^{}_3 =0.1$ eV
\hspace{1cm} \\
\hline
${\cal B}(H^{+} \to e^+ \overline{\nu})$ && $0.21$ & $0.32$ \\
\hline
${\cal B}(H^{+} \to \mu^+ \overline{\nu})$ &&
$0.30$ & $0.33$ \\
\hline
${\cal B}(H^{+} \to \tau^+ \overline{\nu})$ &&
$0.49$ & $0.35$ \\
\hline\hline
\end{tabular}
\end{center}
\end{table}

\subsection{$H^{++} \to \ell^+_\alpha \ell^+_\beta$ and $H^+ \to
\ell^+_\alpha \overline{\nu}$ decays}

There exist a number of viable mechanisms which can explain why the
neutrino masses are naturally tiny \cite{Book}. Among them, the
type-II seesaw mechanism \cite{SS2} is of particular interest
because it can keep the unitarity of the PMNS matrix $U$ unviolated
and lead to very rich collider phenomenology \cite{Han}. The latter
includes the LNV decay modes $H^{++} \to \ell^+_\alpha \ell^+_\beta$
and $H^+ \to \ell^+_\alpha \overline{\nu}$. Their branching ratios
are
\begin{eqnarray}
{\cal B}(H^{++} \to \ell_\alpha^+ \ell_\beta^+) \hspace{-0.2cm} &
\equiv & \hspace{-0.2cm} \frac{\displaystyle \Gamma(H^{++} \to
\ell^+_\alpha \ell^+_\beta)} {\displaystyle\sum_\alpha \sum_\beta
\Gamma(H^{++} \to \ell^+_\alpha \ell^+_\beta)} =
\frac{2}{\displaystyle \left(1+\delta^{}_{\alpha\beta}\right)} \cdot
\frac{\displaystyle \left|\langle m \rangle^{}_{\alpha\beta}
\right|^2} {\displaystyle \sum_i m^2_i} \; ,
\end{eqnarray}
and
\begin{eqnarray}
{\cal B}(H^{+} \to \ell^+_{\alpha} \overline{\nu}^{})
\hspace{-0.2cm} & \equiv & \hspace{-0.2cm} \frac{\displaystyle
\sum^{}_{\beta} \Gamma(H^{+} \to \ell^+_{\alpha}
\overline{\nu}^{}_{\beta})} {\displaystyle \sum_\alpha \sum_\beta
\Gamma(H^{+} \to \ell^+_\alpha \overline{\nu}^{}_\beta)} =
\frac{\displaystyle \left|\langle m \rangle^{}_\alpha
\right|^2}{\displaystyle \sum_i m^2_i} \; ,
\end{eqnarray}
respectively, where the Greek subscripts run over $e$, $\mu$ and
$\tau$. Taking account of Eq. (28), we see that ${\cal B}(H^{+} \to
\ell^+_{\alpha} \overline{\nu}^{})$ only depends on the ``Dirac"
phase $\delta$, while ${\cal B}(H^{++} \to \ell_\alpha^+
\ell_\beta^+)$ is sensitive to all the three CP-violating phases.
These interesting LNV decay modes deserve a reexamination because
the previous works \cite{Han,Ren} were more or less subject to the
assumption of vanishing or very small $\theta^{}_{13}$, making the
role of $\delta$ unimportant. In view of the experimental fact that
$\theta^{}_{13}$ is not that small \cite{DYB,RENO}, we update the
numerical analysis of ${\cal B}(H^{+} \to \ell^+_{\alpha}
\overline{\nu}^{})$ and ${\cal B}(H^{++} \to \ell_\alpha^+
\ell_\beta^+)$ by taking the same inputs as above. Our results are
presented in Table 2 and Figures 5---7, respectively.
\begin{figure}[t]
\center
\begin{overpic}[width=15cm]{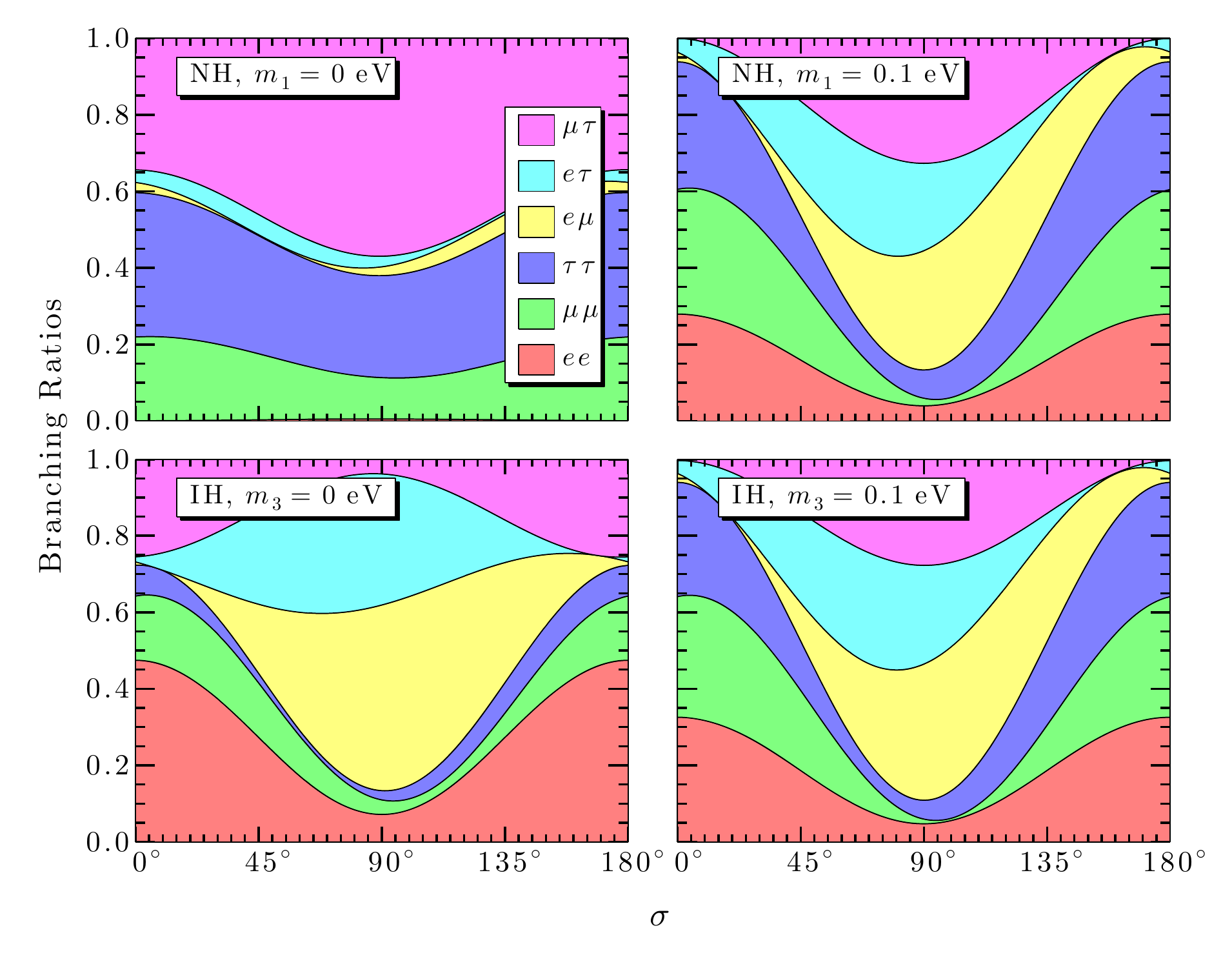}
\end{overpic}
\vspace{-0.4cm} \caption{The branching ratios of the LNV
$H^{++}_{}\to \ell^+_\alpha \ell^+_\beta$ decays as functions of the
Majorana phase $\sigma$, where $\rho = 0^\circ$ and $\delta =
90^\circ$ are taken. Four typical cases of the neutrino mass
spectrum are considered: the normal hierarchy (NH) with $m^{}_1 = 0$
or 0.1 eV; and the inverted hierarchy with $m^{}_3 = 0$ or 0.1 eV.}
\end{figure}
\begin{figure}[t]
\center
\begin{overpic}[width=15cm]{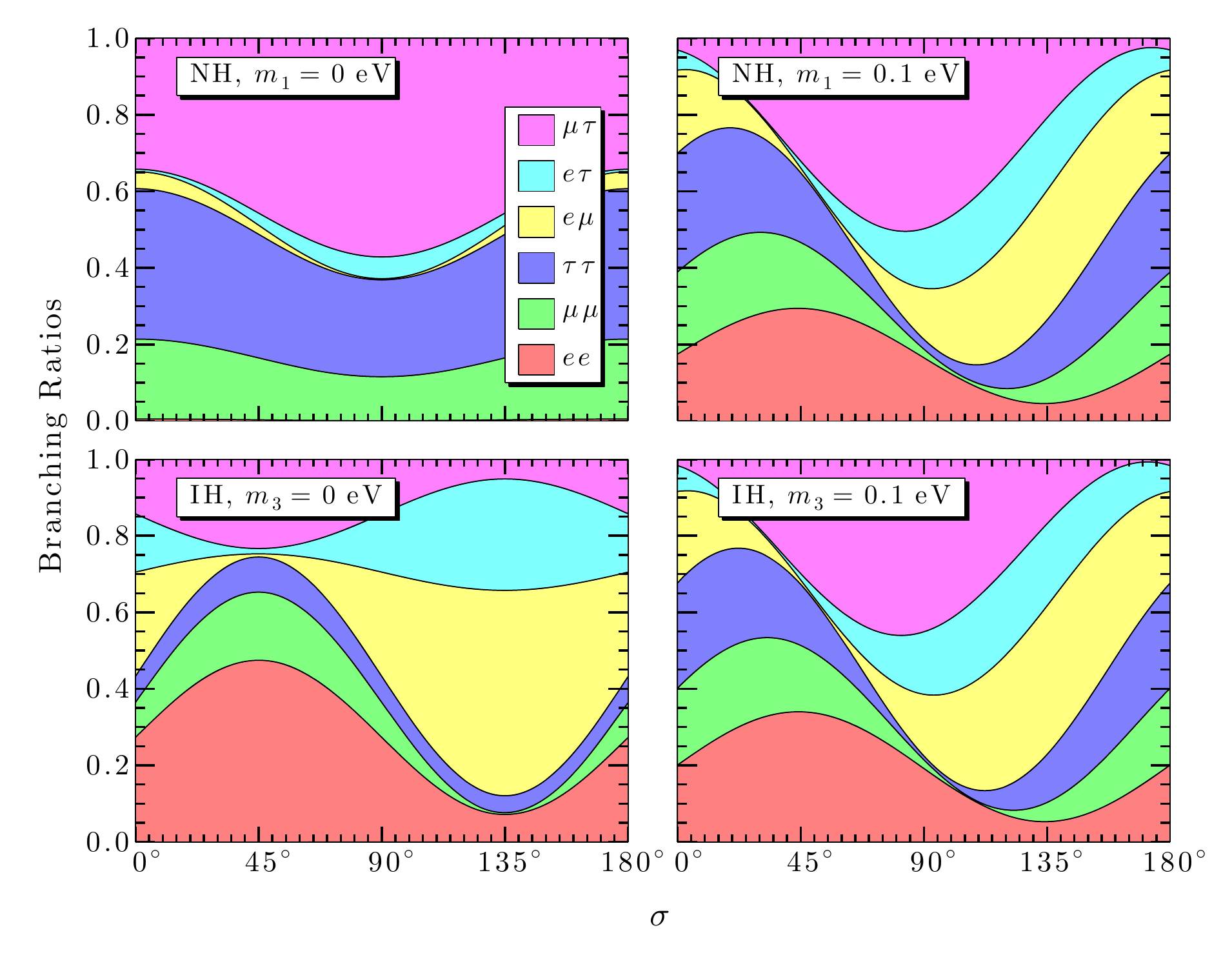}
\end{overpic}
\vspace{-0.4cm} \caption{The branching ratios of the LNV
$H^{++}_{}\to \ell^+_\alpha \ell^+_\beta$ decays as functions of the
Majorana phase $\sigma$, where $\rho = 45^\circ$ and $\delta =
0^\circ$ are taken. Four typical cases of the neutrino mass spectrum
are considered: the normal hierarchy (NH) with $m^{}_1 = 0$ or 0.1
eV; and the inverted hierarchy with $m^{}_3 = 0$ or 0.1 eV.}
\end{figure}
\begin{figure}[t]
\center
\begin{overpic}[width=15cm]{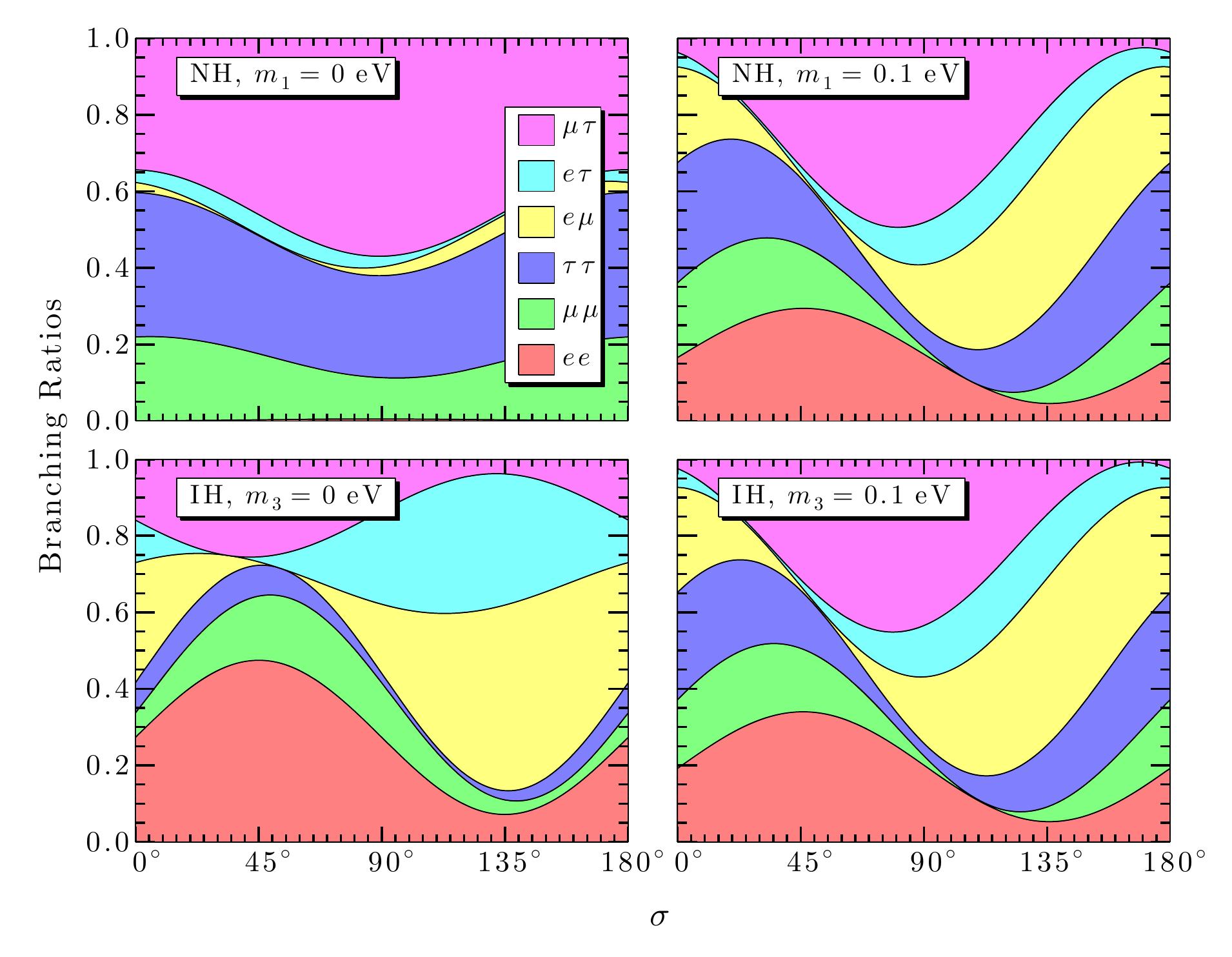}
\end{overpic}
\vspace{-0.4cm} \caption{The branching ratios of the LNV
$H^{++}_{}\to \ell^+_\alpha \ell^+_\beta$ decays as functions of the
Majorana phase $\sigma$, where $\rho = 45^\circ$ and $\delta =
90^\circ$ are taken. Four typical cases of the neutrino mass
spectrum are considered: the normal hierarchy (NH) with $m^{}_1 = 0$
or 0.1 eV; and the inverted hierarchy with $m^{}_3 = 0$ or 0.1 eV.}
\end{figure}

We first look at the branching ratios ${\cal B}(H^{+} \to
\ell^+_{\alpha} \overline{\nu}^{})$, whose magnitudes are governed
by
\begin{eqnarray}
\left|\langle m\rangle^{}_\alpha\right|^2 \hspace{-0.2cm} & = &
\hspace{-0.2cm} \sum^{}_i m_i^2|U^{}_{\alpha i}|^2 =
m_1^2\left(1-|U^{}_{\alpha3}|^2\right) + m_3^2|U^{}_{\alpha3}|^2 +
\Delta m^2_{21}|U^{}_{\alpha2}|^2 \; ,
\end{eqnarray}
in which only the $U^{}_{\alpha2}$ elements (for $\alpha = e, \mu,
\tau$) contain $\delta$, as shown in Eq. (1). Hence the
contributions of $\delta$ to $|\langle m\rangle^{}_\alpha|$ and
${\cal B}(H^{+} \to \ell^+_{\alpha} \overline{\nu}^{})$ are
suppressed not only by the smallness of $\theta^{}_{13}$ but also by
the smallness of $\Delta m^2_{21}$. In particular, ${\cal B}(H^{+}
\to e^+_{} \overline{\nu}^{})$ is exactly independent of $\delta$.
These LNV decay modes are actually not useful to probe the ``Dirac"
phase $\delta$. We use the typical inputs of three neutrino mixing
angles and two neutrino mass-squared differences given above to
calculate ${\cal B}(H^{+} \to \ell^+_{\alpha} \overline{\nu}^{})$,
and list the numerical results in Table 2, where $\delta = 90^\circ$
has been assumed. When varying $\delta$ from $0^\circ$ to
$360^\circ$, we find that the $\delta$-induced uncertainties of all
the branching ratios are lower than $1\%$.

Now let us turn to the branching ratios of the LNV $H^{++} \to
\ell^+_\alpha \ell^+_\beta$ decays. We take $(\rho, \delta) =
(0^\circ, 90^\circ)$, $(45^\circ, 0^\circ)$ and $(45^\circ,
90^\circ)$ to show how ${\cal B}(H^{++} \to \ell_\alpha^+
\ell_\beta^+)$ changes with $\sigma$ in Figures 5---7, respectively.
Both the normal hierarchy with $m^{}_1 =0$ (or $0.1$ eV) and the
inverted hierarchy with $m^{}_3 =0$ (or $0.1$ eV) are considered in
each of the figures. Some discussions are in order.
\begin{itemize}
\item     The sum of the six independent branching ratios is equal to one,
as guaranteed by Eq. (32) and the unitarity of the PMNS matrix $U$.
This point can be clearly seen in each figure, which is exactly
saturated by six different branching ratios.

\item     In the normal neutrino mass hierarchy with $m^{}_1 = 0$,
the magnitude of $\langle m\rangle^{}_{ee}$ is highly suppressed,
and thus ${\cal B}(H^{++} \to e^+e^+) \simeq 0$. In this special
case ${\cal B}(H^{++} \to e^+\mu^+)$ and ${\cal B}(H^{++} \to
e^+\tau^+)$ are also very small. In the inverted neutrino mass
hierarchy with $m^{}_3 = 0$, the $H^{++} \to \tau^+\tau^+$ channel
is strongly suppressed.

\item     The Majorana phases $\rho$ and $\sigma$ play an important
role in all the six LNV decay modes. They may significantly affect
the branching ratio of each process, making themselves easier to be
detected. Given some specific values of $\rho$ and $\delta$, each
${\cal B}(H^{++} \to \ell_\alpha^+ \ell_\beta^+)$ changes as a
simple trigonometric function of $\sigma$. When $\rho$ changes from
one given value to another, the profile of the branching ratio of
each decay mode will more or less shift and deform.

\item     In some cases, the three CP-violating phases may give rise
to large cancellations in $\langle m\rangle^{}_{\alpha\beta}$,
making some of the LNV decay modes significantly suppressed. In the
$(\rho,\sigma)=(0^\circ, 90^\circ_{})$ case, for example, the
$H^{++} \to e^+e^+$, $H^{++} \to \mu^+\mu^+$ and $H^{++} \to
\tau^+\tau^+$ channels are somewhat suppressed when the lightest
neutrino mass is about $0.1$ eV. It is therefore difficult to detect
them.

\item     The ``Dirac" phase $\delta$, whose effect is always
suppressed by the smallness of $\theta^{}_{13}$, has relatively
small influence on the branching ratios of the LNV $H^{++}_{}\to
\ell^+_\alpha \ell^+_\beta$ decays. A comparison between Figures 6
and 7 indicates that the relevant numerical results do not change
much when $\delta$ changes from $0^\circ$ to $90^\circ$. But the
interplay of $\delta$ and $\rho$ (or $\sigma$) is sometimes
important.

\item     The branching ratios in the normal hierarchy with $m^{}_1
= 0.1$ eV and in the inverted hierarchy with $m^{}_3 =0.1$ eV are
almost the same. The reason is simply that these two cases belong to
the nearly degenerate mass hierarchy (i.e., $m^{}_1 \simeq m^{}_2
\simeq m^{}_3$).
\end{itemize}
The behaviors of ${\cal B}(H^{++} \to \ell_\alpha^+ \ell_\beta^+)$
changing with the lightest neutrino mass are essentially similar to
those of $|\langle m\rangle^{}_{\alpha\beta}|$ shown in Figure 4,
and hence we do not go into detail in this connection.

\section{CP violation in neutrino-antineutrino oscillations}

In this section we carry out a detailed analysis of all the possible
CP-violating asymmetries ${\cal A}^{}_{\alpha\beta}$ between
$\nu^{}_\alpha \to \overline{\nu}^{}_\beta$ and
$\overline{\nu}^{}_\alpha \to \nu^{}_\beta$ oscillations. The
generic expression of ${\cal A}^{}_{\alpha\beta}$ has been given in
Eq. (12). Because of the fact $|\Delta m^2_{31}| \simeq |\Delta
m^2_{32}| \simeq 32 \Delta m^2_{21}$, there may exist two
oscillating regions dominated respectively by $\Delta m^2_{21}$ and
$\Delta m^2_{31}$. Let us make some analytical approximations for
each of these two regions.
\begin{figure}[t]
\center
\begin{overpic}[width=12cm]{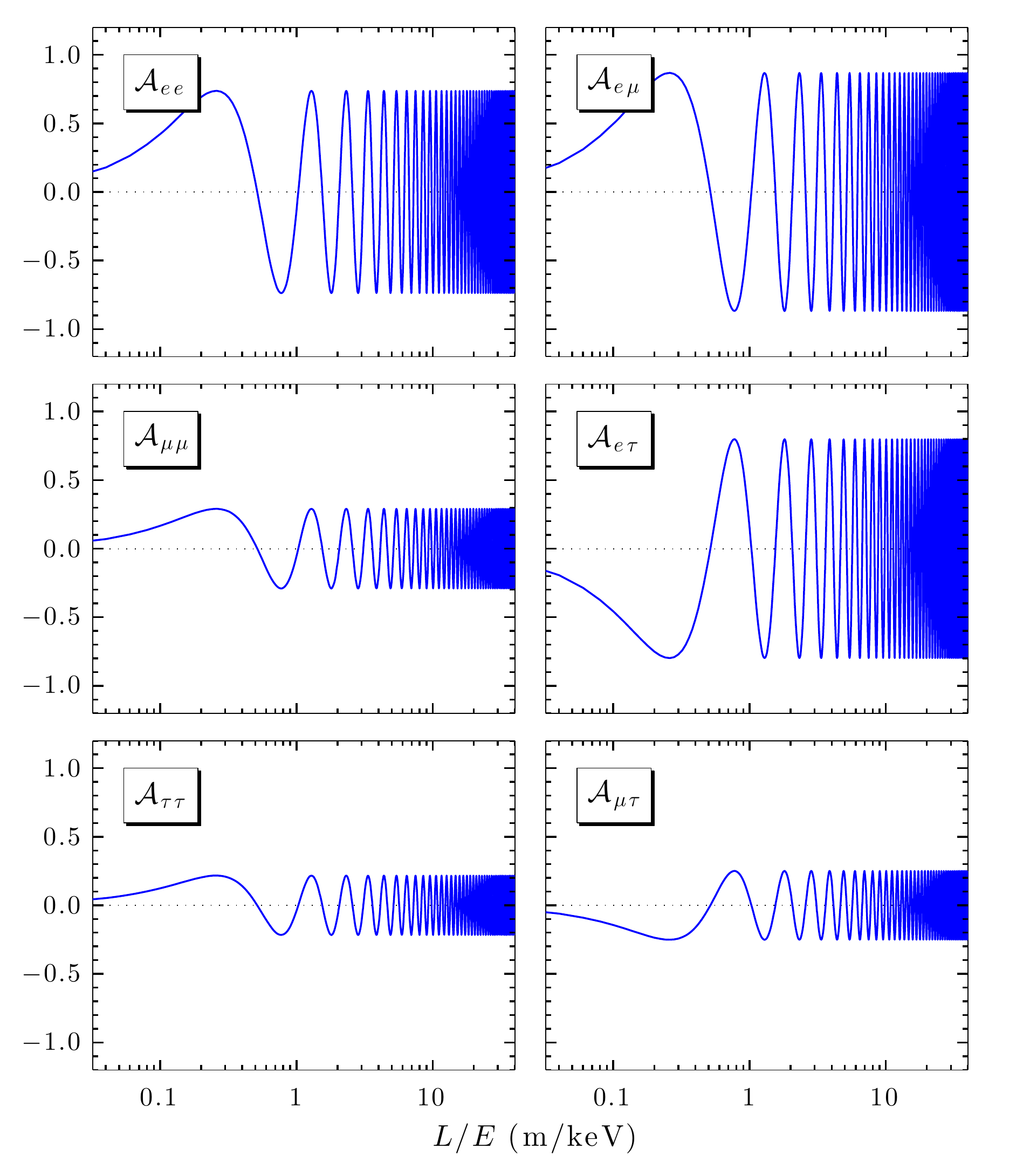}
\end{overpic}
\vspace{-0.3cm} \caption{The CP-violating asymmetries
$\mathcal{A}^{}_{\alpha\beta}$ versus $L/E$ in the normal neutrino
mass hierarchy with $m^{}_1 =0$, $\delta = 0^\circ$ and $\sigma =
45^\circ$.}
\end{figure}
\begin{itemize}
\item     The oscillating region dominated by $\Delta m^2_{31}$
(or $\Delta m^2_{32}$), in which $|\phi^{}_{31}| \sim {\cal O}(1)$
and $\phi^{}_{21} \ll {\cal O}(1)$. In the neglect of the $\Delta
m^2_{21}$-driven contributions, Eq. (12) approximates to
\begin{eqnarray}
\mathcal{A}^{31}_{\alpha\beta} \hspace{-0.2cm} & \simeq &
\hspace{-0.2cm} \frac{\displaystyle 2 m^{}_3 \left(m^{}_1 {\cal
V}^{13}_{\alpha\beta} + m^{}_2 {\cal V}^{23}_{\alpha\beta}\right)
\sin 2\phi^{}_{31}}{\displaystyle \left|\langle
m\rangle^{}_{\alpha\beta} \right|^2 - 4m^{}_3 \left(m^{}_1 {\cal
C}^{13}_{\alpha\beta} + m^{}_2 {\cal C}^{23}_{\alpha\beta}\right)
\sin^2 \phi^{}_{31}} \; ,
\end{eqnarray}
where $\phi^{}_{32} \simeq \phi^{}_{31}$ has been taken into
account.

\item     The oscillating region dominated by $\Delta m^2_{21}$,
in which $\phi^{}_{21} \sim {\cal O}(1)$ and $|\phi^{}_{31}| \gg
{\cal O}(1)$. Hence the $\sin^2\phi^{}_{31}$ and
$\sin^2\phi^{}_{32}$ terms in Eq. (3) oscillate too fast and each of
them averages out to $1/2$, while the $\sin^{} 2\phi^{}_{31}$ and
$\sin^{} 2\phi^{}_{32}$ terms average out to zero. In this case Eq.
(12) approximates to
\begin{eqnarray}
\mathcal{A}^{21}_{\alpha\beta} \hspace{-0.2cm} & \simeq &
\hspace{-0.2cm} \frac{\displaystyle 2m^{}_1 m^{}_2 {\cal
V}^{12}_{\alpha\beta} \sin 2\phi^{}_{21}}{\displaystyle
\left|\langle m\rangle^{}_{\alpha\beta}\right|^2 - 2 m^{}_3
\left(m^{}_1 {\cal C}^{13}_{\alpha\beta} + m^{}_2 {\cal
C}^{23}_{\alpha\beta}\right) - 4 m^{}_1 m^{}_2 {\cal
C}^{12}_{\alpha\beta} \sin^2 \phi^{}_{21}} \; .
\end{eqnarray}
\end{itemize}
Our numerical calculations will be based on the exact formula given
in Eq. (12), but the approximations made in Eqs. (35) and (36) are
helpful to understand the quantitative behaviors of ${\cal
A}^{ij}_{\alpha\beta}$. To reveal the salient features of all the
${\cal A}^{ij}_{\alpha\beta}$, we are going to examine their
dependence on the ratio $L/E$, the three CP-violating phases and the
absolute neutrino mass in Figures 8---16.

\subsection{The dependence of ${\cal A}^{ij}_{\alpha\beta}$ on $L/E$
and $(\delta, \rho, \sigma)$}

Let us consider three special cases for the neutrino mass spectrum,
in which the expressions of ${\cal A}^{ij}_{\alpha\beta}$ can be
more or less simplified, to illustrate their dependence on the ratio
$L/E$ and the phases $\delta$, $\rho$ and $\sigma$.

(A) The normal neutrino mass hierarchy with $m^{}_1 = 0$. In this
case we obtain $m^{}_2 = \sqrt{\Delta m^2_{21}} \simeq 8.66 \times
10^{-3}$ eV and $m^{}_3 = \sqrt{\Delta m^2_{31}} \simeq 4.97 \times
10^{-2}$ eV. Eq. (12) is now simplified to
\begin{eqnarray}
\mathcal{A}^{}_{\alpha\beta} \hspace{-0.2cm} & = & \hspace{-0.2cm}
\frac{\displaystyle 2{\cal V}^{23}_{\alpha\beta} \sin
2\phi^{}_{31}}{\displaystyle \sqrt{\frac{\Delta m^2_{21}} {\Delta
m^2_{31}}} ~{\cal C}^{22}_{\alpha\beta} + \sqrt{\frac{\Delta
m^2_{31}}{\Delta m^2_{21}}} ~{\cal C}^{33}_{\alpha\beta} + 2{\cal
C}^{23}_{\alpha\beta}\cos2\phi_{31}} \; .
\end{eqnarray}
Note that the Majorana phase $\rho$ does not contribute to ${\cal
A}^{}_{\alpha\beta}$ in the $m^{}_1 = 0$ limit \cite{Xing13}. This
point can also be seen in Eq. (37): both ${\cal
C}^{23}_{\alpha\beta}$ and ${\cal V}^{23}_{\alpha\beta}$ do not
contain $\rho$, nor do ${\cal C}^{22}_{\alpha\beta}$ and ${\cal
C}^{33}_{\alpha\beta}$. For simplicity, we choose $(\delta, \sigma)
= (0^\circ, 45^\circ)$, $(90^\circ, 0^\circ)$ and $(90^\circ,
45^\circ)$ to calculate all the six independent ${\cal
A}^{ij}_{\alpha\beta}$, and show their numerical results in Figures
8---10. Some comments are in order.
\begin{figure}[t]
\center
\begin{overpic}[width=12cm]{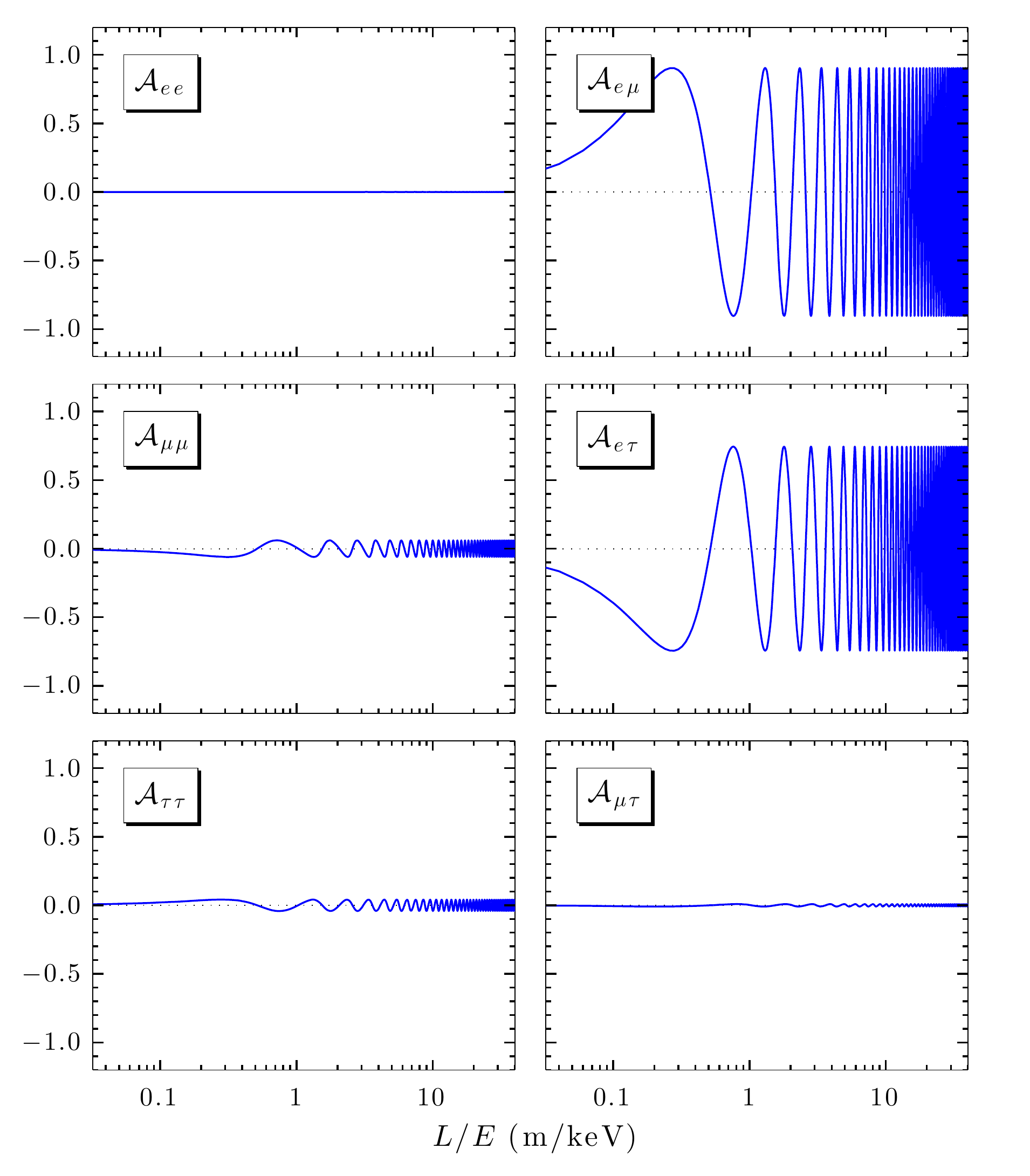}
\end{overpic}
\vspace{-0.3cm} \caption{The CP-violating asymmetries
$\mathcal{A}^{}_{\alpha\beta}$ versus $L/E$ in the normal neutrino
mass hierarchy with $m^{}_1 =0$, $\delta = 90^\circ$ and $\sigma =
0^\circ$.}
\end{figure}
\begin{figure}[t]
\center
\begin{overpic}[width=12cm]{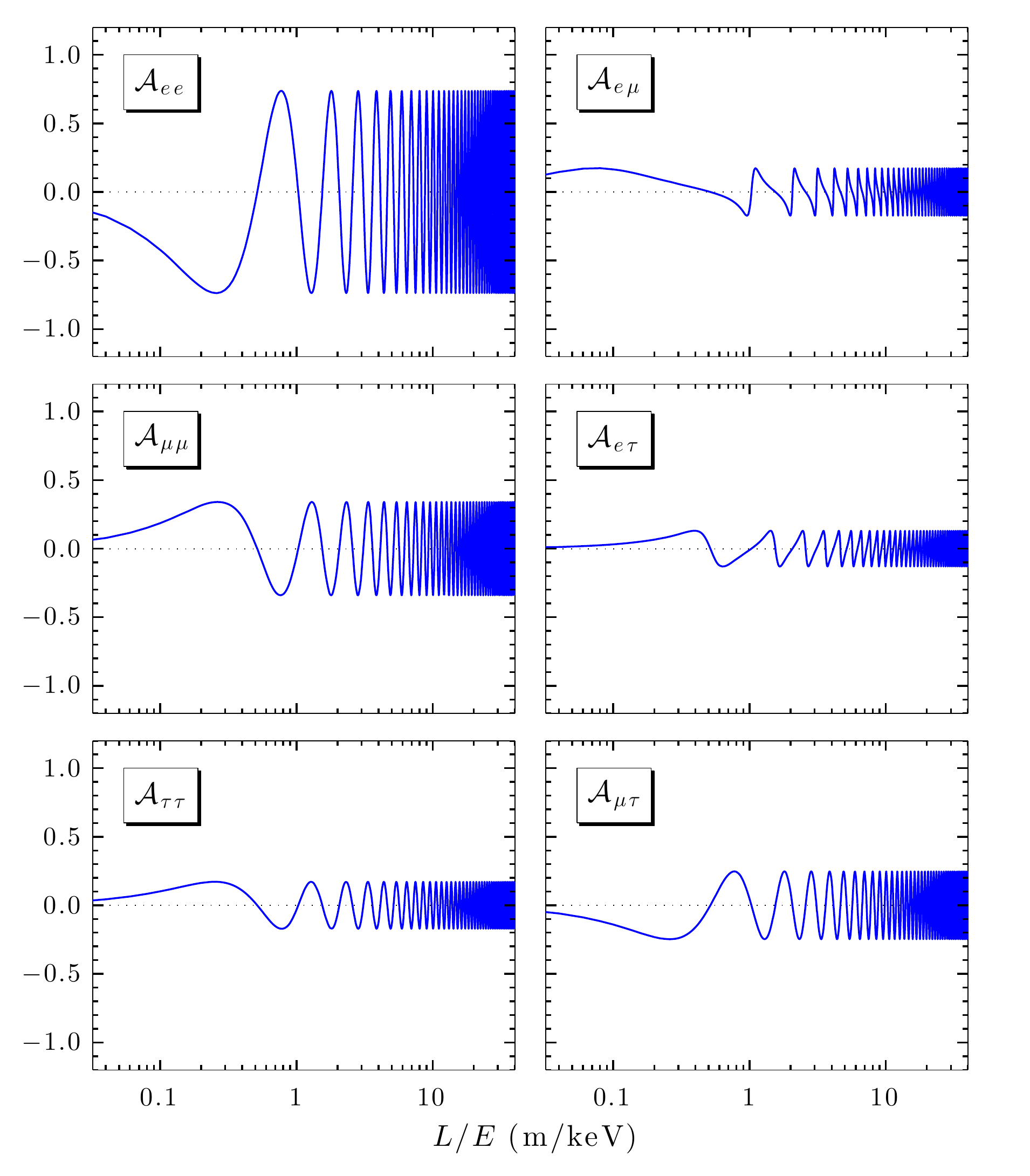}
\end{overpic}
\vspace{-0.3cm} \caption{The CP-violating asymmetries
$\mathcal{A}^{}_{\alpha\beta}$ versus $L/E$ in the normal neutrino
mass hierarchy with $m^{}_1 =0$, $\delta = 90^\circ$ and $\sigma =
45^\circ$.}
\end{figure}
\begin{itemize}
\item     Figure 8 illustrates that significant
CP-violating effects can show up in neutrino-antineutrino
oscillations even though the ``Dirac" phase $\delta$ vanishes. They
arise from the Majorana phase $\sigma$, which has nothing to do with
CP and T violation in normal neutrino-neutrino or
antineutrino-antineutrino oscillations. We have taken $\rho =
45^\circ$ to maximize each CP-violating term in ${\cal
A}^{}_{\alpha\beta}$, where $\rho$ and $\sigma$ enter in the form of
$2\rho$ and $2\sigma$, as one can see from Eqs. (19)---(24)
\footnote{One may redefine $\rho \equiv \rho^\prime/2$ and $\sigma
\equiv \sigma^\prime/2$ in the PMNS matrix $U$, so as to eliminate
the factor 2 from Eqs. (19)---(24).}.

\item     Figure 9 illustrates the nontrivial role of $\delta$ in
generating CP or T violation in neutrino-antineutrino oscillations.
Hence it is intrinsically a Majorana phase. In particular, $\delta =
90^\circ$ (the most favored value to enhance the magnitude of ${\cal
J}$) can lead to large CP-violating asymmetries between $\nu^{}_e
\to \overline{\nu}^{}_\mu$ and $\overline{\nu}^{}_e \to \nu^{}_\mu$
oscillations and between $\nu^{}_e \to \overline{\nu}^{}_\tau$ and
$\overline{\nu}^{}_e \to \nu^{}_\tau$ oscillations. But the other
four CP-violating asymmetries are quite insensitive to $\delta$ in
this case.

\item     A comparison between Figures 9 and 10 tells us again how
important the Majorana phase $\sigma$ is in producing CP and T
violation. The interplay of $\delta$ and $\sigma$ can be either
positive or negative, depending on their explicit values. In order
to determine all the three CP-violating phases, one has to try to
measure the CP-violating effects in as many channels as possible.
Fortunately, not all the channels are strongly suppressed in most
cases, unless $\delta$, $\rho$ and $\sigma$ themselves are too small
or take too special values.
\end{itemize}
When $L/E \gg {\cal O}(1)$ m/keV, all the CP-violating asymmetries
are averaged out to zero in this special normal mass hierarchy.
Hence a measurement of ${\cal A}^{}_{\alpha\beta}$ should better be
done at $L/E \sim {\cal O}(1)$ m/keV. A proper arrangement of $L/E$
may also maximize the signals of CP and T violation.
\begin{figure}[t]
\center
\begin{overpic}[width=12cm]{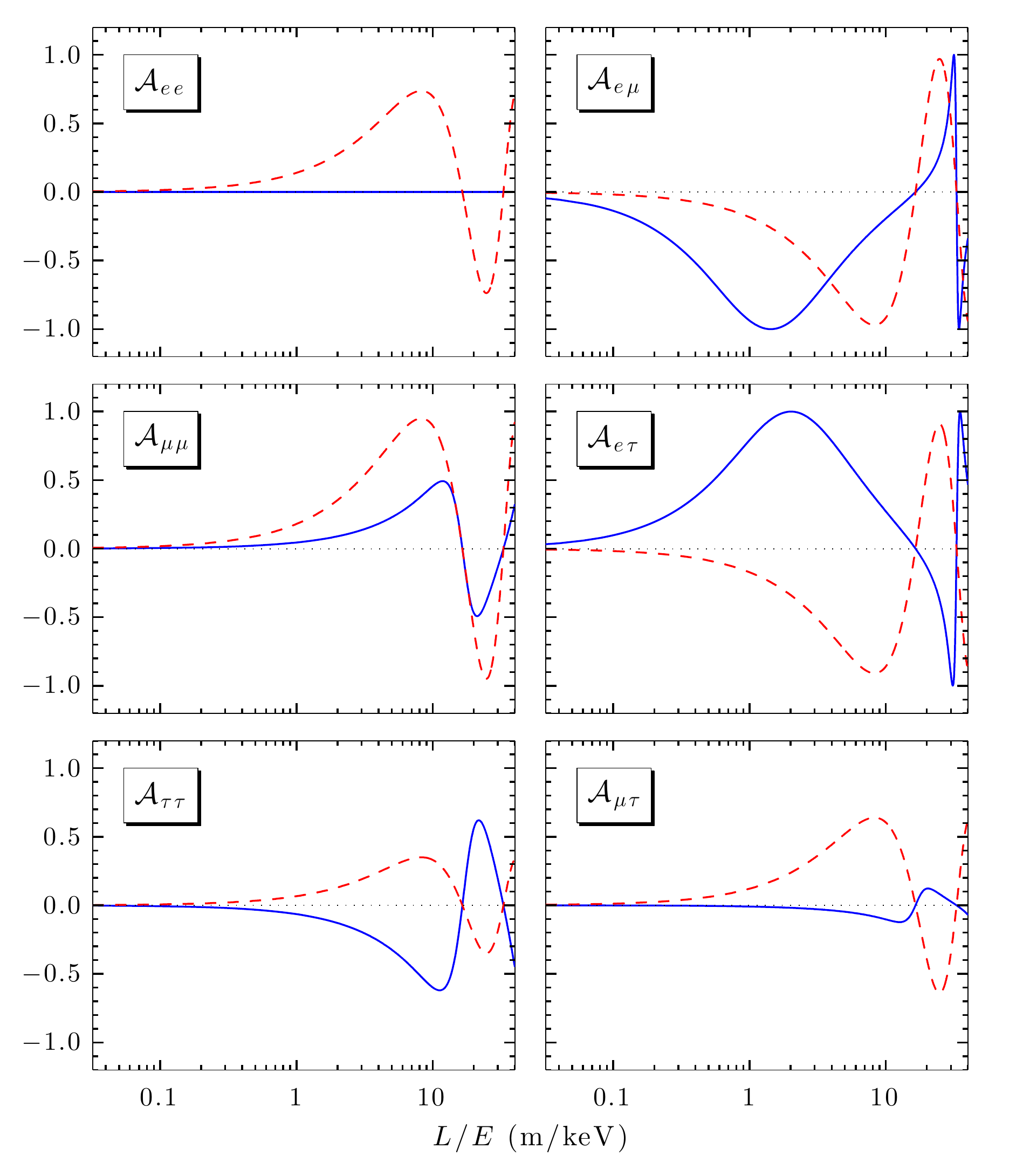}
\end{overpic}
\vspace{-0.3cm} \caption{The CP-violating asymmetries
$\mathcal{A}^{}_{\alpha\beta}$ versus $L/E$ in the inverted neutrino
mass hierarchy with $m^{}_3 =0$: (a) $\delta = 0^\circ$ and
$\rho-\sigma = 45^\circ$ (red dashed lines); (b) $\delta = 90^\circ$
and $\rho-\sigma = 0^\circ$ (blue solid lines).}
\end{figure}

(B) The inverted neutrino mass hierarchy with $m^{}_3 = 0$. In this
case we have $m^{}_2 = \sqrt{-\Delta m^2_{32}} \simeq 4.93 \times
10^{-2}$ eV and $m^{}_1 = \sqrt{-\Delta m^2_{21}-\Delta m^2_{32}}
\simeq 4.85 \times 10^{-2}$ eV. Eq. (12) is then simplified to
\begin{eqnarray}
\mathcal{A}^{}_{\alpha\beta} \hspace{-0.2cm} & = & \hspace{-0.2cm}
\frac{\displaystyle 2{\cal V}^{12}_{\alpha\beta}\sin2\phi^{}_{21}}
{\displaystyle \sqrt{\frac{\Delta m^2_{21} +\Delta m^2_{32}} {\Delta
m^2_{32}}} \ {\cal C}^{11}_{\alpha\beta} + \sqrt{\frac{\Delta
m^2_{32}}{\Delta m^2_{21} + \Delta m^2_{32}}} \ {\cal
C}^{22}_{\alpha\beta}+ 2{\cal
C}^{12}_{\alpha\beta}\cos2\phi^{}_{21}} \; .
\end{eqnarray}
Because of $\Delta m^2_{21} \ll |\Delta m^2_{32}|$, the coefficients
of ${\cal C}^{11}_{\alpha\beta}$ and ${\cal C}^{22}_{\alpha\beta}$
in Eq. (38) are almost equal to one. So the dependence of ${\cal
A}^{}_{\alpha\beta}$ on these two mass-squared differences is rather
weak. Note that all the $\mathcal{A}^{}_{\alpha\beta}$ do not depend
on the absolute values of $\rho$ and $\sigma$ in the $m^{}_3 = 0$
limit, but they depend on $\rho - \sigma$ and $\delta$. To
illustrate, we typically choose $(\delta, \rho-\sigma) = (90^\circ,
0^\circ)$ and $(0^\circ, 45^\circ)$ to calculate the CP-violating
asymmetries $\mathcal{A}^{}_{\alpha\beta}$. The numerical results
are shown in Figure 11. Some discussions are in order.
\begin{figure}[t]
\center
\begin{overpic}[width=12cm]{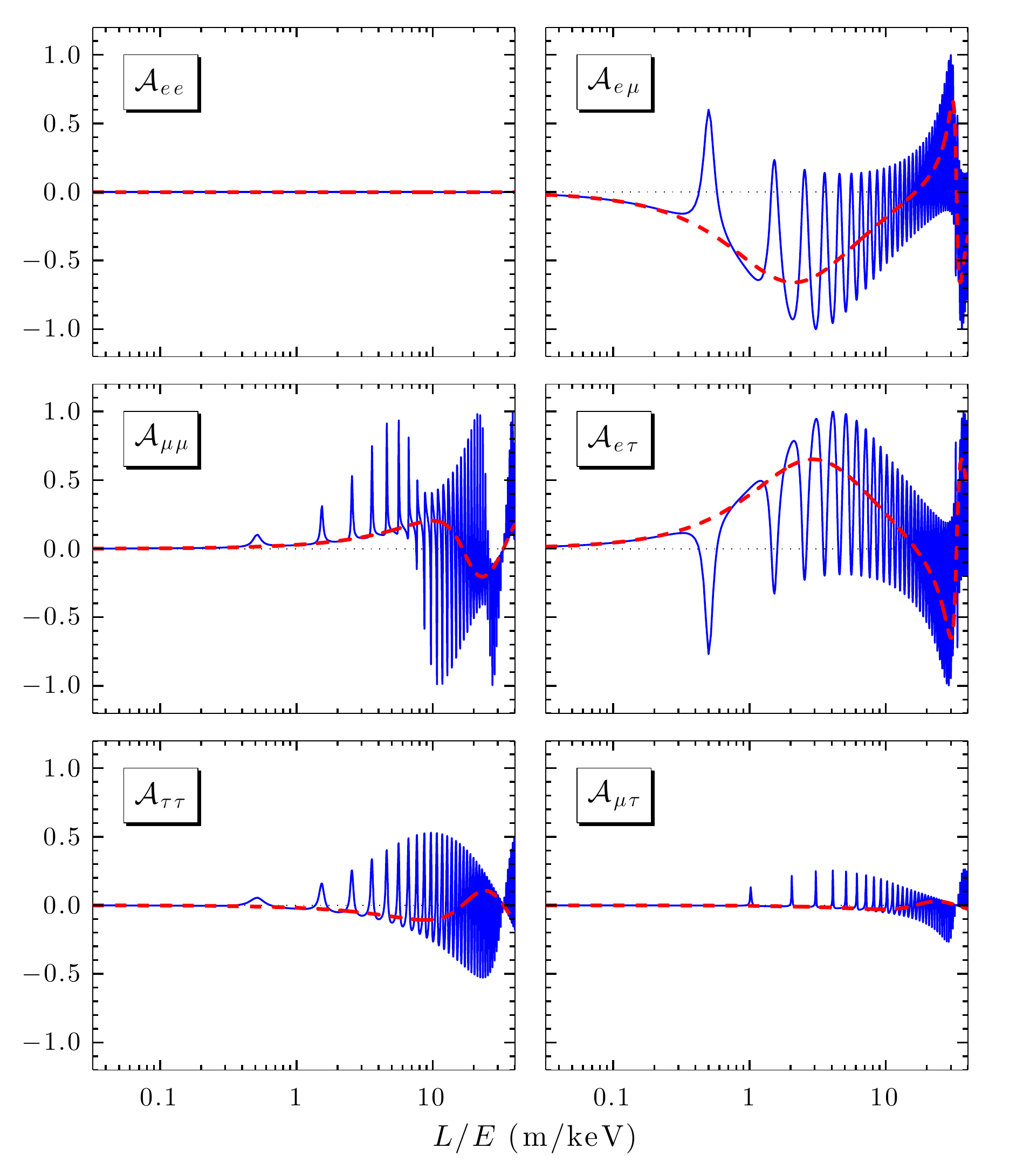}
\end{overpic}
\vspace{-0.3cm}\caption{The CP-violating asymmetries
$\mathcal{A}^{}_{\alpha\beta}$ (blue solid lines) versus $L/E$ in
the nearly degenerate neutrino mass hierarchy with $m^{}_1 \simeq
m^{}_2 \simeq m^{}_3$, $\rho = \sigma = 0^\circ$ and $\delta =
90^\circ$, where the red dashed lines stand for ${\cal
A}^{21}_{\alpha\beta}$ in Eq. (40) with the oscillations driven by
$\Delta m^2_{31}$ and $\Delta m^2_{32}$ being averaged out.}
\end{figure}
\begin{figure}[t]
\center
\begin{overpic}[width=12cm]{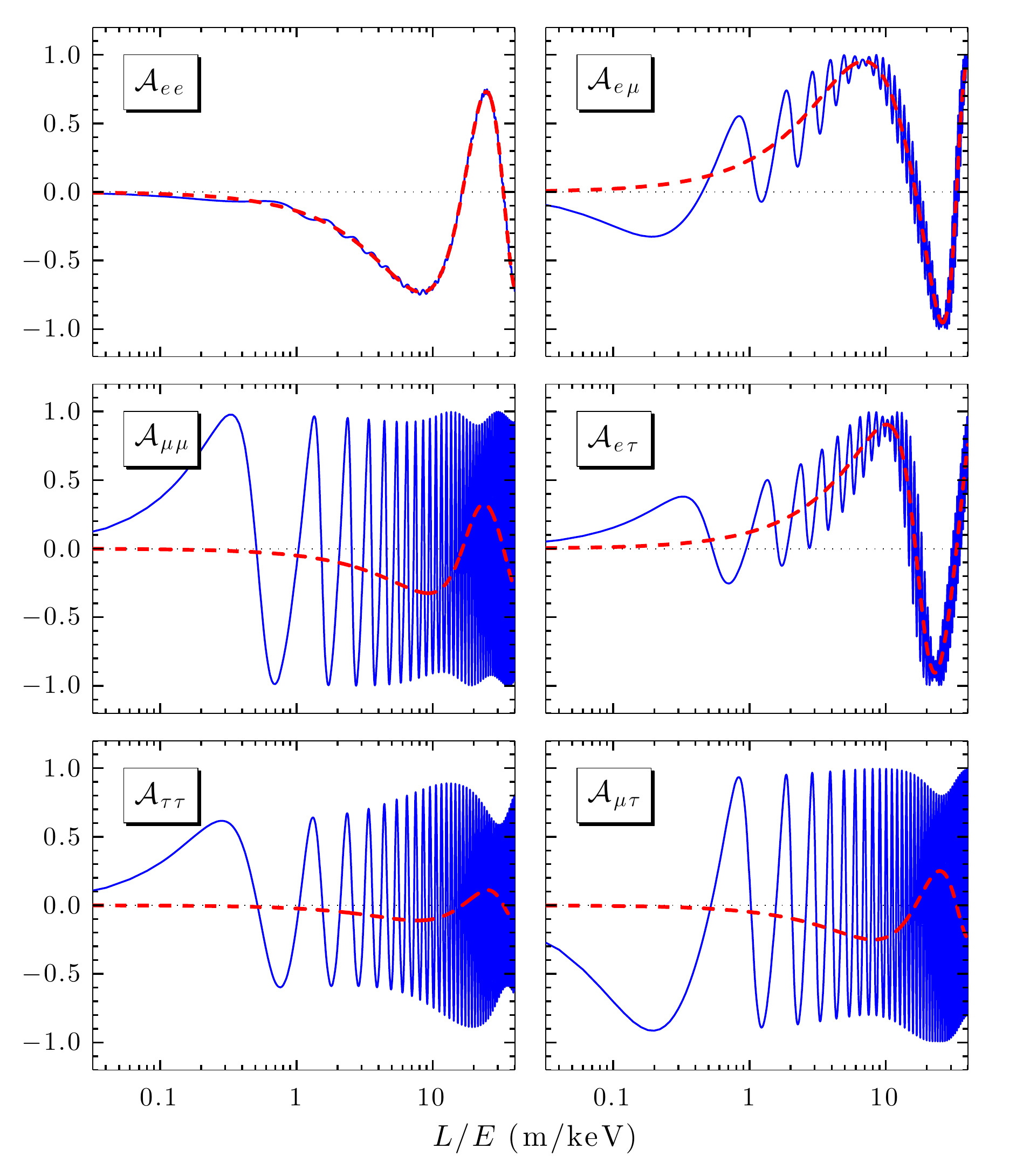}
\end{overpic}
\vspace{-0.3cm}\caption{The CP-violating asymmetries
$\mathcal{A}^{}_{\alpha\beta}$ (blue solid lines) versus $L/E$ in
the nearly degenerate neutrino mass hierarchy with $m^{}_1 \simeq
m^{}_2 \simeq m^{}_3$, $\rho = 0^\circ$, $\sigma = 45^\circ$ and
$\delta = 90^\circ$, where the red dashed lines stand for ${\cal
A}^{21}_{\alpha\beta}$ in Eq. (40) with the oscillations driven by
$\Delta m^2_{31}$ and $\Delta m^2_{32}$ being averaged out.}
\end{figure}
\begin{itemize}
\item     We see again that switching off the ``Dirac" phase
$\delta$ cannot forbid CP and T violation in neutrino-antineutrino
oscillations. Instead, nontrivial values of $\rho-\sigma$ may give
rise to significant CP-violating effects in all the channels under
discussion.

\item     Switching off $\rho-\sigma$ can only lead to ${\cal
A}^{}_{ee} =0$, simply because ${\cal V}^{12}_{ee} =0$ holds in this
case. Thanks to the ``Dirac" phase $\delta$, large CP-violating
asymmetries between $\nu^{}_\alpha \to \overline{\nu}^{}_\beta$ and
$\overline{\nu}^{}_\alpha \to \nu^{}_\beta$ oscillations are
possible to show up.

\item     In either case it is possible to achieve the so-called
``maximal CP violation" (i.e., $|{\cal A}^{}_{\alpha\beta}| =1$).
For example, $|{\cal A}^{}_{e\mu}| \simeq 1$ and $|{\cal
A}^{}_{e\tau}| \simeq 1$ can be obtained for proper values of $L/E$.
Even $|{\cal A}^{}_{\mu\mu}|$ may reach its maximal value at a
suitable point of $L/E$ \cite{Xing13}.
\end{itemize}
In general, both $\delta$ and $\rho-\sigma$ are the sources of CP
and T violation. Since $\delta$ is always associated with
$s^{}_{13}$, its contribution to ${\cal A}^{}_{\alpha\beta}$ is
somewhat suppressed as compared with the contribution from
$\rho-\sigma$. This point can be clearly seen in Eqs. (19)---(24).
Nevertheless, the interplay of $\delta$ and $\rho-\sigma$ sometimes
plays the dominant role in determining the size of ${\cal
A}^{}_{\alpha\beta}$.

(C) The nearly degenerate mass hierarchy with $m^{}_1 \simeq m^{}_2
\simeq m^{}_3$. In this case $m^{}_i \simeq m^{}_j$ can be factored
out and thus canceled on the right-hand side of Eq. (12), leading to
the approximate expressions
\begin{eqnarray}
{\cal A}^{}_{\alpha\beta} \hspace{-0.2cm} & \simeq & \hspace{-0.2cm}
\frac{\displaystyle 2\sum^{}_{i<j} {\cal V}^{ij}_{\alpha\beta} \sin
2\phi^{}_{ji}}{\displaystyle \sum^{}_{i} {\cal C}^{ii}_{\alpha\beta}
+ 2\sum^{}_{i<j} {\cal C}^{ij}_{\alpha\beta} \cos 2\phi^{}_{ji}} \; ,
\end{eqnarray}
which are free from the absolute neutrino masses. In view of Eqs.
(35) and (36), we approximately have
\begin{eqnarray}
\mathcal{A}^{31}_{\alpha\beta} \hspace{-0.2cm} & \simeq &
\hspace{-0.2cm} \frac{\displaystyle 2\left({\cal
V}^{13}_{\alpha\beta} + {\cal V}^{23}_{\alpha\beta}\right) \sin
2\phi^{}_{31}} {\displaystyle \sum^{}_{i}{\cal C}^{ii}_{\alpha\beta}
+2{\cal C}^{12}_{\alpha\beta} + 2\left({\cal C}^{13}_{\alpha\beta}
+{\cal C}^{23}_{\alpha\beta}\right) \cos 2\phi^{}_{31}} \; ,
\nonumber\\
\mathcal{A}^{21}_{\alpha\beta} \hspace{-0.2cm} & \simeq &
\hspace{-0.2cm} \frac{\displaystyle 2{\cal V}^{12}_{\alpha\beta}
\sin 2\phi^{}_{21}} {\displaystyle \sum^{}_{i}{\cal
C}^{ii}_{\alpha\beta} +2{\cal C}^{12}_{\alpha\beta} \cos
2\phi^{}_{21}} \; ,
\end{eqnarray}
corresponding to the oscillating regions dominated by $\Delta
m^2_{31}$ (or $\Delta m^2_{32}$) and $\Delta m^2_{21}$,
respectively. Note that $\mathcal{A}^{31}_{\alpha\beta}$ are
sensitive to all the three CP-violating phases, but only the phase
difference $\rho-\sigma$ and the ``Dirac" phase $\delta$ affect
$\mathcal{A}^{21}_{\alpha\beta}$. For the purpose of illustration,
we typically take $(\rho,\sigma,\delta) = (0^\circ, 0^\circ,
90^\circ)$, $(0^\circ, 45^\circ, 90^\circ)$ and $(45^\circ,
45^\circ, 90^\circ)$ to calculate $\mathcal{A}^{}_{\alpha\beta}$.
The numerical results are given in Figures 12---14
\footnote{We have simply assumed the normal mass hierarchy and input
$m^{}_1 =0.1$ eV in our numerical calculations. We find that the
relevant results are almost the same if the inverted mass hierarchy
with $m^{}_3 =0.1$ eV is taken into account.}.
Some comments and discussions are in order.
\begin{figure}[t]
\center
\begin{overpic}[width=12cm]{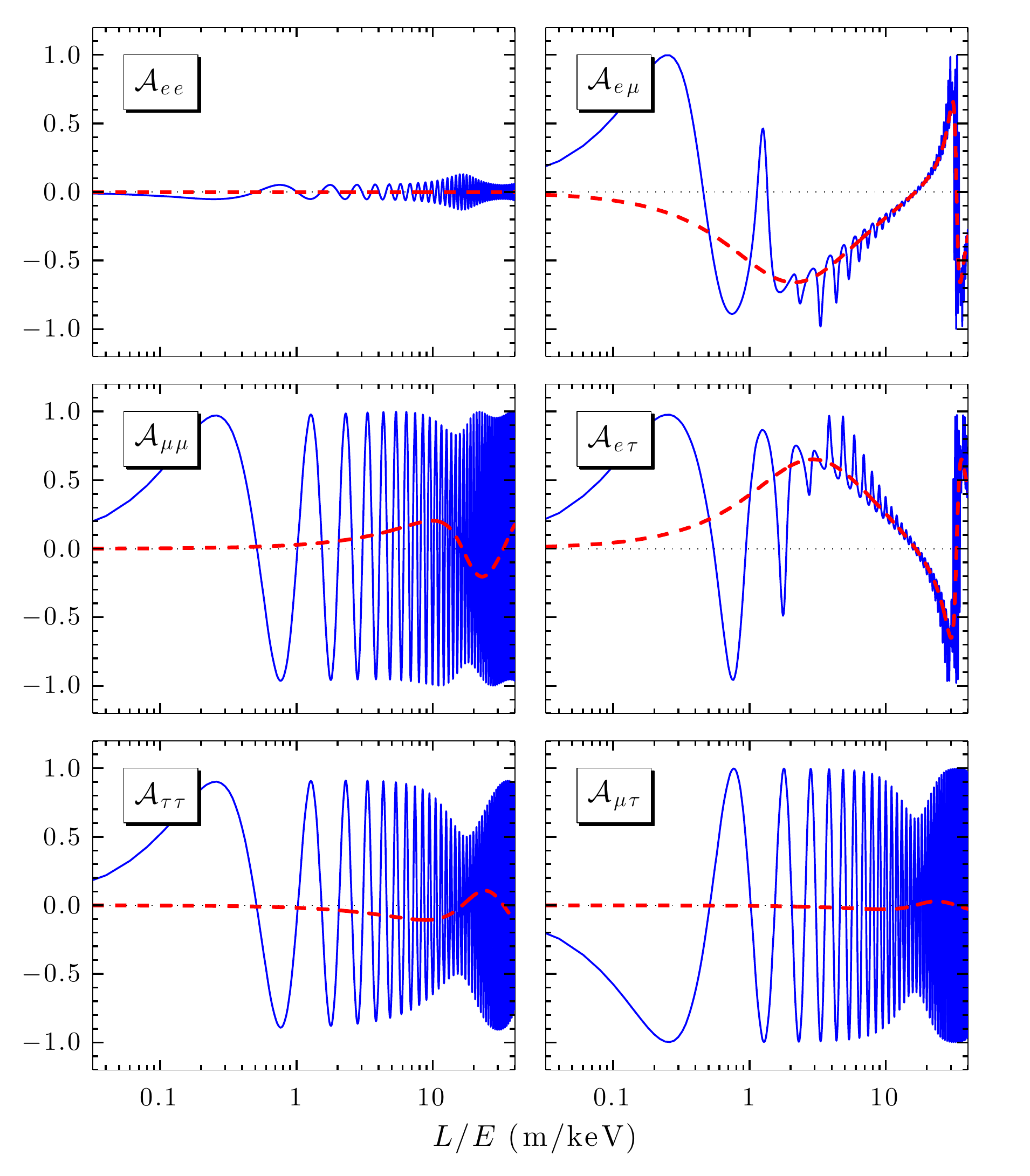}
\end{overpic}
\vspace{-0.3cm}\caption{The CP-violating asymmetries
$\mathcal{A}^{}_{\alpha\beta}$ (blue solid lines) versus $L/E$ in
the nearly degenerate neutrino mass hierarchy with $m^{}_1 \simeq
m^{}_2 \simeq m^{}_3$, $\rho = \sigma = 45^\circ$ and $\delta =
90^\circ$, where the red dashed lines stand for ${\cal
A}^{21}_{\alpha\beta}$ in Eq. (40) with the oscillations driven by
$\Delta m^2_{31}$ and $\Delta m^2_{32}$ being averaged out.}
\end{figure}
\begin{figure}[t]
\center
\begin{overpic}[width=12cm]{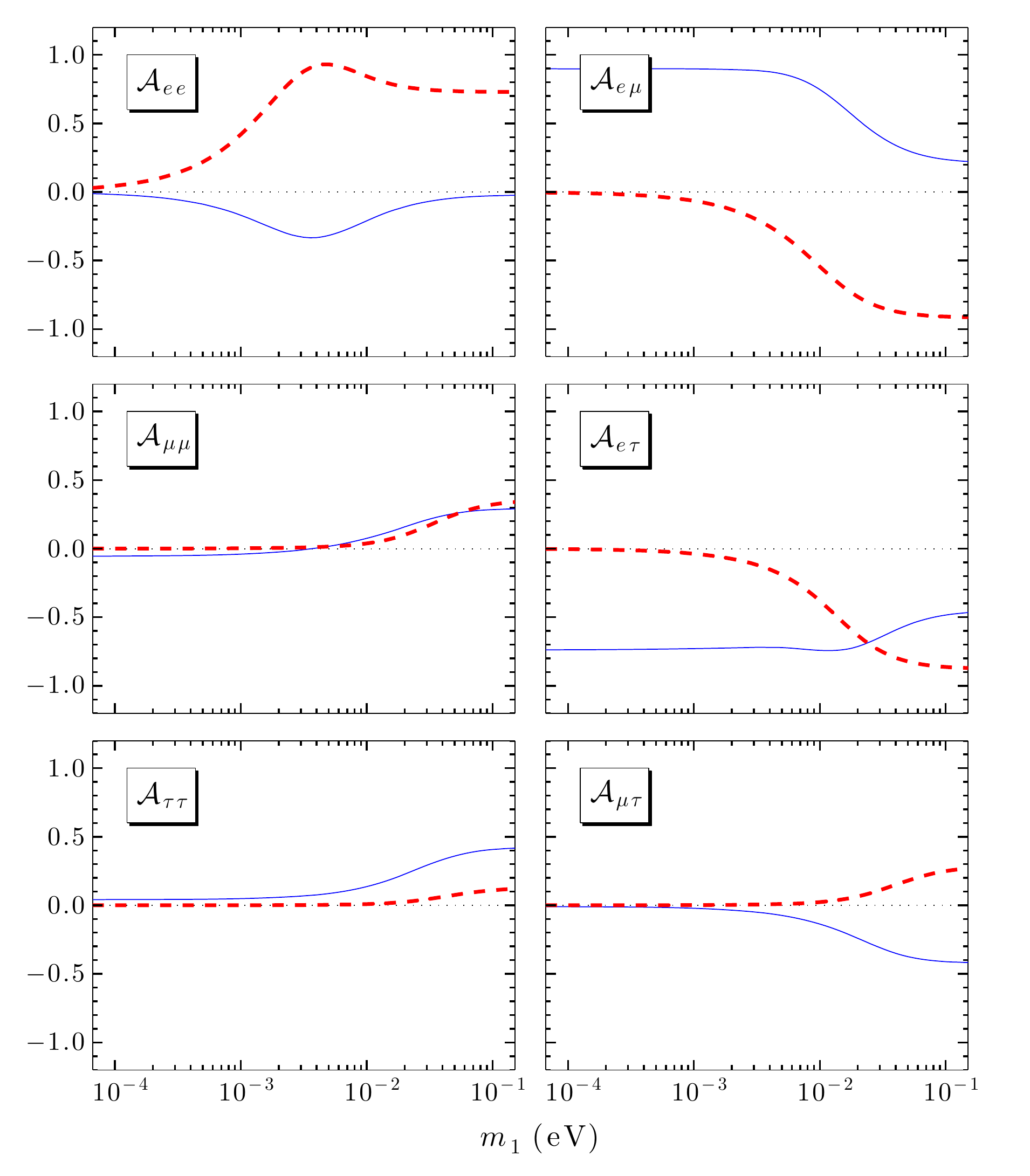}
\end{overpic}
\vspace{-0.3cm}\caption{The CP-violating asymmetries
$\mathcal{A}^{}_{\alpha\beta}$ versus the lightest neutrino mass
$m^{}_1$ in the normal hierarchy with $\rho = 45^\circ$, $\sigma =
0^\circ$ and $\delta = 90^\circ$, where the blue solid lines and red
dashed lines correspond to $L/E\simeq 0.25$ m/keV and 8 m/keV,
respectively.}
\end{figure}
\begin{itemize}
\item     Figure 12 illustrates the CP-violating effects in
neutrino-antineutrino oscillations induced purely by the ``Dirac"
phase $\delta$. We see that ${\cal A}^{}_{ee} = 0$ holds in this
case, simply because the input $\delta = 90^\circ$ is too special to
generate nonvanishing ${\cal V}^{13}_{ee}$ and ${\cal V}^{23}_{ee}$,
as shown in Eq. (19). When $L/E$ is sufficiently large, the $\Delta
m^2_{31}$- and $\Delta m^2_{32}$-dominated terms oscillate too fast
and the observable behaviors of ${\cal A}^{}_{\alpha\beta}$ are
essentially described by ${\cal A}^{21}_{\alpha\beta}$. Once again
we conclude that the CP-violating asymmetries ${\cal A}^{}_{e\mu}$
and ${\cal A}^{}_{e\tau}$ are most sensitive to $\delta$. The same
observation is true for the CP-violating asymmetry between normal
$\nu^{}_e \to \nu^{}_\mu$ (or $\nu^{}_e \to \nu^{}_\tau$) and
$\overline{\nu}^{}_e \to \overline{\nu}^{}_\mu$ (or
$\overline{\nu}^{}_e \to \overline{\nu}^{}_\tau$) oscillations.

\item     Figure 13 illustrates the interplay of
$\sigma$ and $\delta$ in generating CP and T violation in
neutrino-antineutrino oscillations. The suppressed CP-violating
asymmetries in Figure 12 (because of $\rho =\sigma = 0^\circ$) are
now enhanced to a large extent. When both $\rho$ and $\sigma$ are
switched on, as shown in Figure 14, the situation becomes somewhat
more complicated. In either case it is possible to achieve
significant or even maximal CP-violating asymmetries. In the
oscillating region dominated by $\Delta m^2_{31}$ and $\Delta
m^2_{32}$, the first maximum or minimum of ${\cal
A}^{}_{\alpha\beta}$ should be a good place to be detected.

\item     The first maximum or minimum of ${\cal A}^{}_{\alpha\beta}$
in the $\Delta m^2_{31}$-dominated oscillating region roughly occurs
around $L/E \sim 0.25$ m/keV, which corresponds to $\phi^{}_{31}
\sim \pi/4$. In comparison, the first maximum or minimum of ${\cal
A}^{}_{\alpha\beta}$ in the $\Delta m^2_{21}$-dominated oscillating
region may happen around $L/E \sim 8$ m/keV, corresponding to
$\phi^{}_{21}\sim \pi/4$. Of course, these results are more or less
subject to the chosen inputs.
\end{itemize}
The above examples have illustrated the dependence of ${\cal
A}^{}_{\alpha\beta}$ on the ratio $L/E$ and the three CP-violating
phases in three special cases of the neutrino mass spectrum. In the
subsequent subsection we shall examine the sensitivity of ${\cal
A}^{}_{\alpha\beta}$ to the absolute neutrino mass scale in a more
careful way.
\begin{figure}[t]
\center
\begin{overpic}[width=12cm]{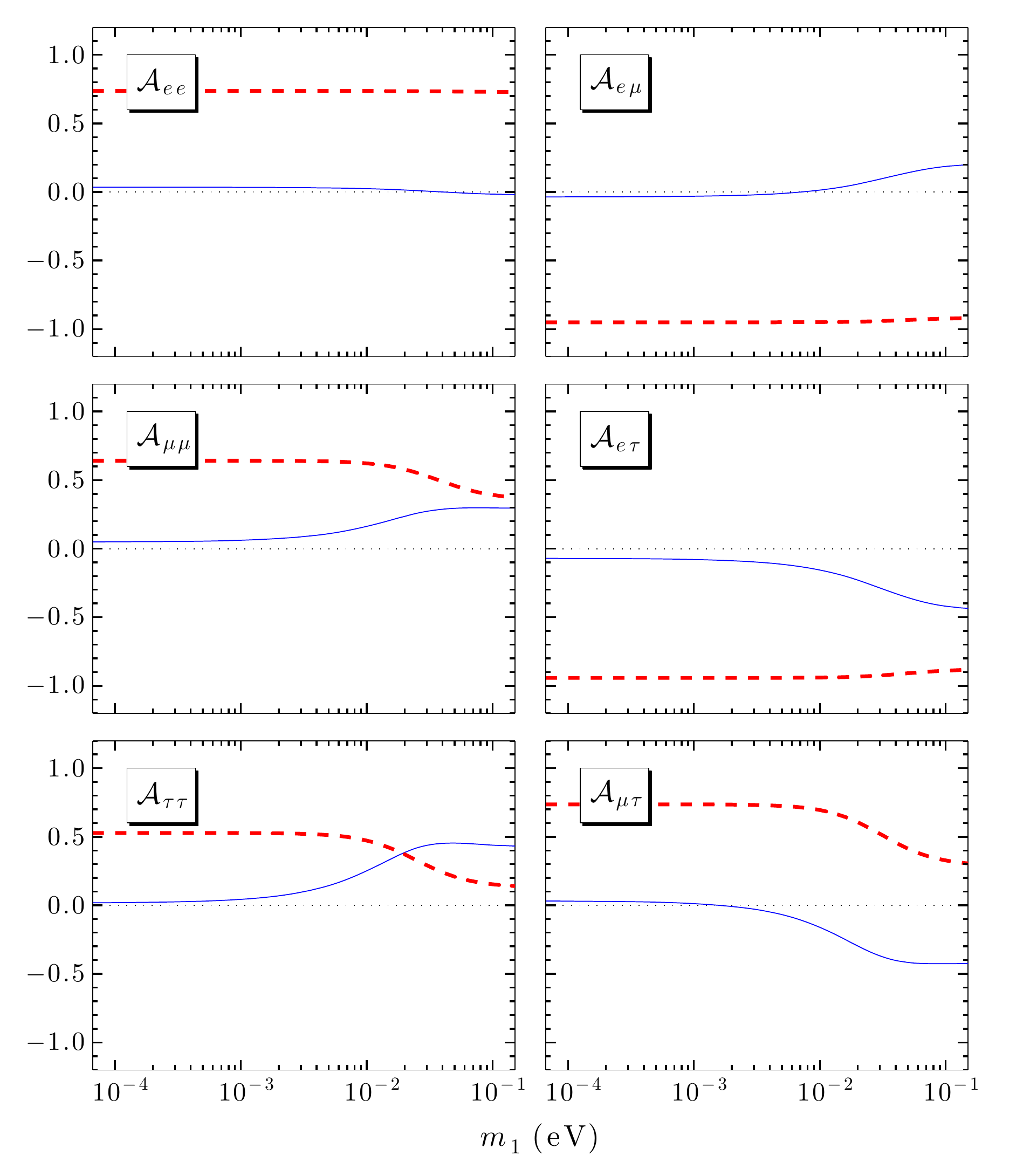}
\end{overpic}
\vspace{-0.3cm}\caption{The CP-violating asymmetries
$\mathcal{A}^{}_{\alpha\beta}$ versus the lightest neutrino mass
$m^{}_3$ in the inverted hierarchy with $\rho = 45^\circ$, $\sigma =
0^\circ$ and $\delta = 90^\circ$, where the blue solid lines and red
dashed lines correspond to $L/E\simeq 0.25$ m/keV and 8 m/keV,
respectively.}
\end{figure}

\subsection{The sensitivity of ${\cal A}^{}_{\alpha\beta}$ to
$m^{}_1$ or $m^{}_3$}

To simplify our numerical calculations, we typically choose
$L/E\simeq 0.25$ m/keV (i.e., $\phi^{}_{31} \simeq \pi/4$) and 8
m/keV (i.e., $\phi^{}_{21} \simeq \pi/4$) which correspond to the
$\Delta m^2_{31}$- and $\Delta m^2_{21}$-dominated oscillating
regions, respectively. We also fix $\rho = 45^\circ$, $\sigma =
0^\circ$ and $\delta = 90^\circ$ to see the changes of ${\cal
A}^{}_{\alpha\beta}$ with the lightest neutrino mass $m^{}_1$
(normal hierarchy) or $m^{}_3$ (inverted hierarchy). The numerical
results are shown in Figures 15 and 16. Some comments are in order.
\begin{itemize}
\item     In Figure 15 the values of $m^{}_1$ change from ${\cal
O}(10^{-4})$ eV to ${\cal O}(10^{-1})$ eV, implying the changes of
the neutrino mass spectrum from $m^{}_1 \ll m^{}_2 \ll m^{}_3$ to
$m^{}_1 \lesssim m^{}_2 \lesssim m^{}_3$. The turning point is
roughly $m^{}_1 \sim \sqrt{\Delta m^2_{31}}$, around which the
sensitivity of ${\cal A}^{}_{\alpha\beta}$ to $m^{}_1$ becomes
stronger. In the chosen parameter space we find that ${\cal
A}^{}_{ee}$, ${\cal A}^{}_{e\mu}$ and ${\cal A}^{}_{e\tau}$ are most
sensitive to $m^{}_1$ at $L/E \simeq 8$ m/keV: the results of these
three CP-violating asymmetries for $m^{}_1 \simeq 0.1$ eV are
significantly different from the ones for $m^{}_1 \simeq 0$.

\item     In Figure 16 the values of $m^{}_3$ change from ${\cal
O}(10^{-4})$ eV to ${\cal O}(10^{-1})$ eV, implying the changes of
the neutrino mass spectrum from $m^{}_3 \ll m^{}_1 \lesssim m^{}_2$
to $m^{}_3 \lesssim m^{}_1 \lesssim m^{}_2$. The turning point is
roughly $m^{}_3 \sim \sqrt{|\Delta m^2_{32}|}$, around which the
sensitivity of ${\cal A}^{}_{\alpha\beta}$ to $m^{}_3$ becomes more
appreciable. But a comparison between Figures 15 and 16 tells us
that the CP-violating asymmetries are in general less sensitive to
the absolute neutrino mass scale in the case of the inverted
hierarchy.

\item     As for the results of ${\cal A}^{}_{\alpha\beta}$,
it does not make much difference whether the nearly degenerate
neutrino mass spectrum is $m^{}_1 \lesssim m^{}_2 \lesssim m^{}_3$
or $m^{}_3 \lesssim m^{}_1 \lesssim m^{}_2$. This point can be
clearly seen in Figures 15 and 16 at $m^{}_1 \simeq m^{}_3 \simeq
0.1$ eV, where the numerical results in these two nearly degenerate
mass spectra approximately match each other.
\end{itemize}
So it is in principle possible to probe the absolute neutrino mass
scale through the study of neutrino-antineutrino oscillation. In
comparison, the normal neutrino-neutrino and
antineutrino-antineutrino oscillations are only sensitive to the
neutrino mass-squared differences.

\section{Summary}

One of the fundamental questions about massive neutrinos is what
their nature is or whether they are the Dirac or Majorana
particles. The absolute neutrino mass scale is so low that it is
extremely difficult to distinguish between the Dirac and Majorana
neutrinos in all the currently available experiments. Today's
techniques have allowed us to push the sensitivity of the
$0\nu\beta\beta$ decay to the level of $|\langle m\rangle^{}_{ee}|
\sim {\cal O}(0.1)$ eV, making it the most feasible way to probe the
Majorana nature of massive neutrinos. The present work is just
motivated by a meaningful question that we have asked ourselves:
what can we proceed to do to determine all the CP-violating phases
in the PMNS matrix $U$ if the massive neutrinos are someday
identified {\it to be} the Majorana particles through a convincing
measurement of the $0\nu\beta\beta$ decay?

In principle, one may determine the Majorana phases of $U$ in
neutrino-antineutrino oscillations
\footnote{Because the Dirac neutrinos do not violate the lepton
number, they cannot undergo neutrino-antineutrino oscillations.
However, it is likely for the Dirac neutrinos to oscillate
between their left-handed and right-handed states in a magnetic
field and in the presence of matter effects \cite{Giunti}. Such
spin-flavor precession processes are beyond the scope of the present
paper and will be further studied elsewhere.}.
In practice, such an experiment might only be feasible in the very
distant future. But we find that a systematic study of CP violation
in neutrino-antineutrino oscillations is still useful, so as to
enrich the phenomenology of Majorana neutrinos. In this work we have
explored the salient features of three-flavor neutrino-antineutrino
oscillations and their CP- and T-violating asymmetries. Six
independent $0\nu\beta\beta$-like mass terms $\langle
m\rangle^{}_{\alpha\beta}$ and nine independent Jalskog-like
parameters ${\cal V}^{ij}_{\alpha\beta}$ have been analyzed in
detail, because they are quite universal and can contribute to the
CP-conserving and CP-violating parts of a number of LNV processes.
We have made a comparison between ${\cal V}^{ij}_{\alpha\beta}$ and
the Jarlskog invariant ${\cal J}$ by switching off the Majorana
phases $\rho$ and $\sigma$, and have demonstrated the Majorana
nature of the ``Dirac" phase $\delta$. As a by-product, the effects
of three CP-violating phases on the LNV decays of doubly- and
singly-charged Higgs bosons have also been reexamined. We have
carried out a comprehensive analysis of the sensitivities of six
possible CP-violating asymmetries ${\cal A}^{}_{\alpha\beta}$ to the
three phase parameters, the neutrino mass spectrum and the ratio of
the neutrino beam energy $E$ to the baseline length $L$. Our
analytical and numerical results provide a complete description of
the distinct roles of Majorana CP-violating phases in
neutrino-antineutrino oscillations and other LNV processes.

Although the particular parametrization of $U$ advocated by the
Particle Data Group \cite{PDG} has been used in this work, one
may always choose a different representation of $U$
which might be more convenient in some aspects of the Majorana
neutrino phenomenology. For instance, the so-called ``symmetrical
parametrization" of the PMNS matrix \cite{SV}
\begin{eqnarray}
K \hspace{-0.2cm} & = & \hspace{-0.2cm} \left( \begin{matrix}
c^{}_{12} c^{}_{13} & s^{}_{12} c^{}_{13} e^{-{\rm i}\varphi^{}_{12}}
& s^{}_{13} e^{-{\rm i} \varphi^{}_{13}} \cr -s^{}_{12} c^{}_{23}
e^{{\rm i} \varphi^{}_{12}} - c^{}_{12} s^{}_{13} s^{}_{23}
e^{-{\rm i} \left(\varphi^{}_{23} - \varphi^{}_{13}\right)} &
c^{}_{12} c^{}_{23} - s^{}_{12} s^{}_{13} s^{}_{23}
e^{-{\rm i} \left(\varphi^{}_{12} + \varphi^{}_{23} - \varphi^{}_{13}
\right)} & c^{}_{13} s^{}_{23} e^{-{\rm i} \varphi^{}_{23}} \cr s^{}_{12}
s^{}_{23} e^{{\rm i} \left(\varphi^{}_{12} + \varphi^{}_{23}\right)}
- c^{}_{12} s^{}_{13} c^{}_{23} e^{{\rm i} \varphi^{}_{13}} &
-c^{}_{12} s^{}_{23} e^{{\rm i} \varphi^{}_{23}} - s^{}_{12} s^{}_{13}
c^{}_{23} e^{-{\rm i} \left(\varphi^{}_{12} - \varphi^{}_{13} \right)}
& c^{}_{13} c^{}_{23} \cr
\end{matrix} \right) \hspace{0.7cm}
\end{eqnarray}
has also been used by some authors to describe neutrino oscillations and LNV processes \cite{RV}. It is easy to establish
the relationship between $U$ in Eq. (1) and $K$ in Eq. (41):
\begin{eqnarray}
U \hspace{-0.2cm} & = & \hspace{-0.2cm} \left(
\begin{matrix} e^{{\rm i} \rho} & 0 & 0 \cr 0 & e^{{\rm i} \sigma}
& 0 \cr 0 & 0 & 1 \cr \end{matrix} \right) K \; ,
\end{eqnarray}
where the three phase parameters of $U$ are related to the three
phase parameters of $K$ as follows:
\begin{eqnarray}
\delta \hspace{-0.2cm} & = & \hspace{-0.2cm} \varphi^{}_{13} -
\varphi^{}_{12} - \varphi^{}_{23} \; , \nonumber \\
\rho \hspace{-0.2cm} & = & \hspace{-0.2cm} \varphi^{}_{12} +
\varphi^{}_{23} \; , \nonumber \\
\sigma \hspace{-0.2cm} & = & \hspace{-0.2cm} \varphi^{}_{23} \; .
\end{eqnarray}
Therefore, it is straightforward to reexpress ${\cal C}^{ij}_{\alpha\beta}$
and ${\cal V}^{ij}_{\alpha\beta}$ in terms of the angle and phase
parameters of $K$ simply with the help of Eq. (43). Given three light
or heavy {\it sterile} neutrinos, it is also straightforward to extend Eq. (41)
to a full parametrization of the $6\times 6$ neutrino mixing matrix \cite{Xing}.

The Schechter-Valle (Black Box) theorem \cite{SV2}
has told us that an observation of the $0\nu\beta\beta$ decay
points to the Majorana nature of massive neutrinos, but such a
LNV process may be dominated either by a tree-level Majorana neutrino
mass term or by other possible new physics which is essentially
unrelated to the neutrino masses. The radiative
mass term induced by the Black Box (loop) diagram itself is
extremely small in most cases, although this is not always true
\cite{Lindner}. Hence one has to be careful when relating the
rate of the $0\nu\beta\beta$ decay fully to the neutrino masses.
In this work we have assumed the existence of a tree-level Majorana
mass term dominating the Black Box diagram, leading to
$\langle m\rangle^{}_{ee}$ which has a direct relation to the
rate of the $0\nu\beta\beta$ decay. The same observation is
expected to be true for all the LNV processes which depend on the
effective Majorana mass terms $\langle m\rangle^{}_{\alpha\beta}$.
Of course, the situation will change if other types of LNV physics
exist \cite{Lindner}.

While it is still a dream to fully determine the flavor dynamics of
Majorana neutrinos, including their CP-violating phases, one should
not be too pessimistic. The reason is simply that the history of
neutrino physics has been full of surprises in making the impossible
possible, but one has to be patient.

\vspace{0.3cm}
\begin{flushleft}
{\Large\bf Acknowledgement}
\end{flushleft}

We would like to thank Y.F. Li for helpful discussions. This work
was supported in part by the National Natural Science Foundation of
China under Grant No. 11135009.

\newpage

\appendix
\section{Explicit expressions of ${\cal C}_{\alpha\beta}^{ij}$}

Given the standard parametrization of the PMNS matrix $U$ in Eq.
(1), one may explicitly write out all the CP-conserving quantities
${\cal C}^{ij}_{\alpha\beta}$ defined in Eq. (5). Such formulas are
expected to be useful to understand the behaviors of
neutrino-antineutrino oscillations and make reasonable analytical
approximations for their oscillation probabilities and CP-violating
asymmetries.

First of all, we have $C_{\alpha\alpha}^{ii} = |U^{}_{\alpha i}|^4$
(for $\alpha = e, \mu, \tau$ and $i = 1, 2, 3$). The explicit
expressions of these nine quantities are
\begin{eqnarray}
{\cal C}_{{ee}}^{11} \hspace{-0.2cm} & = & \hspace{-0.2cm}
c_{12}^4c_{13}^4  \; ,
\nonumber\\
{\cal C}_{{ee}}^{22} \hspace{-0.2cm} & = & \hspace{-0.2cm}
s_{12}^4c_{13}^4  \; ,
\nonumber\\
{\cal C}_{{ee}}^{33} \hspace{-0.2cm} & = & \hspace{-0.2cm} s_{13}^4
\; ;
\nonumber\\
{\cal C}_{\mu \mu }^{11} \hspace{-0.2cm} & = & \hspace{-0.2cm}
\left(s_{12}^2c_{23}^2 +2c_{12}^{} s_{12}^{}s_{13}^{} c_{23}^{}
s_{23}^{}\cos\delta + c_{12}^2 s_{13}^2 s_{23}^2\right)^2 \; ,
\nonumber\\
{\cal C}_{\mu \mu }^{22} \hspace{-0.2cm} & = & \hspace{-0.2cm}
\left(c_{12}^2 c_{23}^2-2  c_{12}^{} s_{12}^{}s_{13}^{}
c_{23}^{}s_{23}^{} \cos\delta+s_{12}^2 s_{13}^2 s_{23}^2\right)^2 \;
,
\nonumber\\
{\cal C}_{\mu \mu }^{33} \hspace{-0.2cm} & = & \hspace{-0.2cm}
c_{13}^4 s_{23}^4 \; ;
\nonumber\\
{\cal C}_{\tau \tau }^{11} \hspace{-0.2cm} & = & \hspace{-0.2cm}
\left(s_{12}^2 s_{23}^2-2 c_{12}^{}
s_{12}^{}s_{13}^{}c_{23}^{}s_{23}^{}\cos\delta +c_{12}^2s_{13}^2
c_{23}^2 \right)^2 \; ,
\nonumber\\
{\cal C}_{\tau \tau }^{22} \hspace{-0.2cm} & = & \hspace{-0.2cm}
\left(c_{12}^2 s_{23}^2+ 2 c_{12}^{}
s_{12}^{}s_{13}^{}c_{23}^{}s_{23} ^{}
\cos\delta+s_{12}^2s_{13}^2c_{23}^2\right)^2 \; ,
\nonumber\\
{\cal C}_{\tau \tau }^{33} \hspace{-0.2cm} & = & \hspace{-0.2cm}
c_{13}^4 c_{23}^4 \; .
\end{eqnarray}
Because ${\cal C}^{ii}_{\alpha\beta} = \sqrt{{\cal
C}^{ii}_{\alpha\alpha}{\cal C}^{ii}_{\beta\beta}}$ holds, it is
straightforward to write out the expressions of ${\cal
C}^{ii}_{\alpha\beta}$ (for $\alpha \neq \beta$) with the help of
Eq. (44). The following sum rule is also valid:
\begin{eqnarray}
\sum_\alpha \sum_{\beta}{\cal C}^{ii}_{\alpha\beta} \hspace{-0.2cm}
& = & \hspace{-0.2cm} \sum_\alpha \left|U^{}_{\alpha i}\right|^2 = 1
\; .
\end{eqnarray}
We see that all the ${\cal C}^{ii}_{\alpha\beta}$ are independent of
the Majorana phases $\rho$ and $\sigma$.

Next, we calculate ${\cal C}^{ij}_{\alpha\alpha}$ and ${\cal
C}^{ij}_{\alpha\beta}$ in terms of the flavor mixing parameters of
the PMNS matrix $U$ given in Eq. (1). The results are
\begin{eqnarray}
{\cal C}_{ee}^{12} \hspace{-0.2cm} & = & \hspace{-0.2cm} c_{12}^2
s_{12}^2 c_{13}^4 \cos 2\left(\rho - \sigma\right) \; ,
\nonumber \\
{\cal C}_{ee}^{13} \hspace{-0.2cm} & = & \hspace{-0.2cm} c_{12}^2
c_{13}^2 s_{13}^2 \cos 2\left(\delta + \rho\right) \; ,
\nonumber \\
{\cal C}_{ee}^{23} \hspace{-0.2cm} & = & \hspace{-0.2cm} s_{12}^2
c_{13}^2 s_{13}^2 \cos 2\left(\delta + \sigma\right) \; ;\nonumber\\
{\cal C}_{\mu \mu }^{12} \hspace{-0.2cm} & = & \hspace{-0.2cm}
c_{12}^2 s_{12}^2 \left(c_{23}^4 - 4 s_{13}^2 c_{23}^2 s_{23}^2 +
s_{13}^4 s_{23}^4\right) \cos 2\left(\rho -\sigma\right)
\nonumber \\
\hspace{-0.2cm} & & \hspace{-0.2cm} + 2c_{12}^{} s_{12}^{} s_{13}^{}
c_{23}^{} s_{23} ^{} \left(c_{23}^2 - s_{13}^2 s_{23}^2\right)
\left[c_{12}^2 \cos \left(2 \rho -2 \sigma +\delta\right) - s_{12}^2
\cos \left(2 \rho - 2\sigma -\delta\right) \right]
\nonumber \\
\hspace{-0.2cm} & & \hspace{-0.2cm} + s_{13}^2 c_{23}^2 s_{23}^2
\left[c_{12}^4 \cos 2\left(\rho -\sigma + \delta\right) + s_{12}^4
\cos 2\left(\rho - \sigma - \delta\right) \right] \; ,
\nonumber \\
{\cal C}_{\mu \mu }^{13} \hspace{-0.2cm} & = & \hspace{-0.2cm}
c_{13}^2 s_{23}^2 \left[ s_{12}^2 c_{23}^2 \cos 2\rho + 2 c_{12}^{}
s_{12}^{} s_{13}^{} c_{23}^{} s_{23}^{} \cos \left(\delta +2
\rho\right) + c_{12}^2 s_{13}^2 s_{23}^2 \cos 2\left(\delta +
\rho\right) \right] \; ,
\nonumber \\
{\cal C}_{\mu \mu }^{23} \hspace{-0.2cm} & = & \hspace{-0.2cm}
c_{13}^2 s_{23}^2 \left[ c_{12}^2 c_{23}^2 \cos 2\sigma - 2
c_{12}^{} s_{12}^{} s_{13}^{} c_{23}^{} s_{23}^{} \cos \left(\delta
+ 2\sigma\right) + s_{12}^2 s_{13}^2 s_{23}^2 \cos 2\left(\delta +
\sigma\right) \right] \; ;
\nonumber\\
{\cal C}_{\tau \tau }^{12} \hspace{-0.2cm} & = & \hspace{-0.2cm}
c_{12}^2 s_{12}^2 \left(s_{23}^4 - 4 s_{13}^2c_{23}^2 s_{23}^2 +
s_{13}^4c_{23}^4 \right) \cos 2\left(\rho - \sigma\right)
\nonumber \\
\hspace{-0.2cm} & & \hspace{-0.2cm} - 2c_{12}^{} s_{12}^{} s_{13}^{}
c_{23}^{} s_{23}^{} \left(s_{23}^2 - s_{13}^2 c_{23}^2\right)
\left[c_{12}^2 \cos \left(2 \rho - 2\sigma + \delta\right) -
s_{12}^2 \cos \left(2 \rho - 2\sigma - \delta\right) \right]
\nonumber \\
\hspace{-0.2cm} & & \hspace{-0.2cm} + s_{13}^2 c_{23}^2 s_{23}^2
\left[c_{12}^4 \cos 2\left(\rho - \sigma + \delta\right) + s_{12}^4
\cos 2\left(\rho - \sigma - \delta\right) \right] \; ,
\nonumber \\
{\cal C}_{\tau \tau }^{13} \hspace{-0.2cm} & = & \hspace{-0.2cm}
c_{13}^2 c_{23}^2 \left[ s_{12}^2 s_{23}^2 \cos 2\rho - 2 c^{}_{12}
s^{}_{12} s^{}_{13} c^{}_{23} s^{}_{23} \cos \left(\delta +
2\rho\right) + c_{12}^2 s_{13}^2 c_{23}^2 \cos 2\left(\delta +
\rho\right) \right] \; ,
\nonumber \\
{\cal C}_{\tau \tau }^{23} \hspace{-0.2cm} & = & \hspace{-0.2cm}
c_{13}^2 c_{23}^2 \left[ c_{12}^2 s_{23}^2 \cos 2\sigma + 2
c^{}_{12} s^{}_{12} s^{}_{13} c^{}_{23} s^{}_{23} \cos \left(\delta
+ 2\sigma\right) + s_{12}^2 s_{13}^2 c_{23}^2 \cos 2\left(\delta +
\sigma\right) \right] \; ;
\end{eqnarray}
and
\begin{eqnarray}
{\cal C}_{e\mu}^{12} \hspace{-0.2cm} & = & \hspace{-0.2cm} -c_{12}^2
s_{12}^2 c_{13}^2 \left(c_{23}^2 - s_{13}^2 s_{23}^2\right) \cos
2\left(\rho - \sigma\right)
\nonumber \\
\hspace{-0.2cm} & & \hspace{-0.2cm} - c^{}_{12} s^{}_{12} c_{13}^2
s^{}_{13} c^{}_{23} s^{}_{23} \left[c_{12}^2 \cos \left(2\rho -
2\sigma + \delta\right) - s_{12}^2 \cos \left(2\rho - 2\sigma -
\delta\right) \right] \; ,
\nonumber \\
{\cal C}_{e\mu}^{13} \hspace{-0.2cm} & = & \hspace{-0.2cm} -
c^{}_{12} c_{13}^2 s^{}_{13} s^{}_{23} \left[s^{}_{12} c^{}_{23}
\cos \left(\delta + 2\rho\right) - c^{}_{12} s^{}_{13} s^{}_{23}
\cos 2\left(\delta + \rho\right) \right] \; ,
\nonumber \\
{\cal C}_{e\mu}^{23} \hspace{-0.2cm} & = & \hspace{-0.2cm}
+s^{}_{12} c_{13}^2 s^{}_{13} s^{}_{23} \left[c^{}_{12} c^{}_{23}
\cos \left(\delta + 2\sigma\right) - s^{}_{12} s^{}_{13} s^{}_{23}
\cos 2\left(\delta + \sigma\right) \right] \; ;
\nonumber\\
{\cal C}_{e\tau}^{12} \hspace{-0.2cm} & = & \hspace{-0.2cm} c_{12}^2
s_{12}^2 c_{13}^2 \left(c_{23}^2 s_{13}^2 - s_{23}^2\right) \cos
2\left(\rho - \sigma\right)
\nonumber \\
\hspace{-0.2cm} & & \hspace{-0.2cm} + c^{}_{12} s^{}_{12} c_{13}^2
s^{}_{13} c^{}_{23} s^{}_{23} \left[c_{12}^2 \cos \left(2\rho -
2\sigma + \delta\right) - s_{12}^2 \cos \left(2\rho - 2\sigma -
\delta\right) \right] \; ,
\nonumber \\
{\cal C}_{e\tau}^{13} \hspace{-0.2cm} & = & \hspace{-0.2cm}
+c^{}_{12} c_{13}^2 s^{}_{13} c^{}_{23} \left[s^{}_{12} s^{}_{23}
\cos \left(\delta + 2\rho\right) - c^{}_{12} s^{}_{13} c^{}_{23}
\cos 2\left(\delta + \rho\right) \right] \; ,
\nonumber \\
{\cal C}_{e\tau}^{23} \hspace{-0.2cm} & = & \hspace{-0.2cm}
-s^{}_{12} c_{13}^2 s^{}_{13} c^{}_{23} \left[c^{}_{12} s^{}_{23}
\cos \left(\delta + 2\sigma\right) + s^{}_{12} s^{}_{13} c^{}_{23}
\cos 2\left(\delta + \sigma\right) \right] \; ;
\nonumber\\
{\cal C}_{\mu \tau }^{12} \hspace{-0.2cm} & = & \hspace{-0.2cm}
-c_{12}^2 s_{12}^2 \left[c_{23}^4 s_{13}^2 - \left(1 +
s_{13}^2\right)^2 c_{23}^2 s_{23}^2 + s_{13}^2 s_{23}^4\right] \cos
2\left(\rho - \sigma\right)
\nonumber \\
\hspace{-0.2cm} & & \hspace{-0.2cm} - c^{}_{12} s^{}_{12} s^{}_{13}
c^{}_{23} s^{}_{23} \left(1 + s_{13}^2\right) \left(c_{23}^2 -
s_{23}^2\right) \left[c_{12}^2 \cos \left(2\rho - 2\sigma +
\delta\right) - s_{12}^2 \cos \left(2\rho - 2\sigma - \delta\right)
\right]
\nonumber \\
\hspace{-0.2cm} & & \hspace{-0.2cm} -s_{13}^2 c_{23}^2 s_{23}^2
\left[c_{12}^4 \cos 2\left(\rho - \sigma + \delta\right) + s_{12}^4
\cos 2\left(\rho - \sigma - \delta\right)\right] \; ,
\nonumber \\
{\cal C}_{\mu \tau }^{13} \hspace{-0.2cm} & = & \hspace{-0.2cm}
c_{13}^2 c^{}_{23} s^{}_{23} \left[- s_{12}^2 c^{}_{23} s^{}_{23}
\cos 2\rho + c^{}_{12} s^{}_{12} s^{}_{13} \left(c_{23}^2 -
s_{23}^2\right) \cos \left(\delta + 2\rho\right) + c_{12}^2 s_{13}^2
c^{}_{23} s^{}_{23} \cos 2\left(\delta + \rho\right)\right] \; ,
\nonumber \\
{\cal C}_{\mu \tau }^{23} \hspace{-0.2cm} & = & \hspace{-0.2cm}
c_{13}^2 c^{}_{23} s^{}_{23} \left[- c_{12}^2c^{}_{23} s^{}_{23}
\cos 2\sigma - c^{}_{12} s^{}_{12} s^{}_{13} \left(c_{23}^2 -
s_{23}^2\right) \cos \left(\delta + 2\sigma\right) + s_{12}^2
s_{13}^2 c^{}_{23} s^{}_{23} \cos 2\left(\delta +
\sigma\right)\right] \; . \hspace{0.8cm}
\end{eqnarray}
By definition, ${\cal C}^{ij}_{\alpha\beta} = {\cal
C}^{ij}_{\beta\alpha} = {\cal C}^{ji}_{\alpha\beta} = {\cal
C}^{ji}_{\beta\alpha}$ holds. Then Eq. (8) allows us to establish
the following relations between the results of ${\cal
C}^{ij}_{\alpha\alpha}$ in Eq. (46) and those of ${\cal
C}^{ij}_{\alpha\beta}$ in Eq. (47):
\begin{eqnarray}
{\cal C}^{ij}_{e\mu} \hspace{-0.2cm} & = & \hspace{-0.2cm}
\frac{1}{2}\left({\cal C}^{ij}_{\tau\tau} - {\cal C}^{ij}_{ee} -
{\cal C}^{ij}_{\mu\mu}\right) \; ,
\nonumber \\
{\cal C}^{ij}_{\mu\tau} \hspace{-0.2cm} & = & \hspace{-0.2cm}
\frac{1}{2}\left({\cal C}^{ij}_{ee} - {\cal C}^{ij}_{\mu\mu} - {\cal
C}^{ij}_{\tau\tau}\right) \; ,
\nonumber \\
{\cal C}^{ij}_{\tau e} \hspace{-0.2cm} & = & \hspace{-0.2cm}
\frac{1}{2}\left({\cal C}^{ij}_{\mu\mu} - {\cal C}^{ij}_{ee} - {\cal
C}^{ij}_{\tau\tau}\right) \; .
\end{eqnarray}
In view of the smallness of $\theta^{}_{13}$, one may make some
analytical approximations for the above results by neglecting the
terms proportional to $s^2_{13}$.

\section{Analytical approximations of $\langle m\rangle_{\alpha\beta}^{}$}

The exact expressions of $\langle m \rangle^{}_{\alpha\beta}$
have been given in Eq. (27). To understand Figure 4 in a better way,
here we make some analytical approximations
for all the $|\langle m\rangle_{\alpha\beta}^{}|$ in four special but
interesting cases.

Case (A): the lightest neutrino mass $m^{}_1$ satisfies $m^2_1 \ll
\Delta m^2_{21}$. In this case, $m^{}_2 \simeq \sqrt{\Delta
m^2_{21}}$ and $m^{}_3 \simeq \sqrt{\Delta m^2_{31}}$ hold. One may
therefore neglect those terms proportional to $m^{}_1$ in Eq. (27).
In view of the smallness of $\theta^{}_{13}$, we approximately have
\begin{eqnarray}
\langle m\rangle_{ee}^{} \hspace{-0.2cm} & \simeq & \hspace{-0.2cm}
+\sqrt{\Delta m_{21}^{2}} \ s_{12}^2 e^{2 {\rm i}\sigma }
+ \sqrt{\Delta m_{31}^{2}} \ s_{13}^2e^{-2 {\rm i}\delta } \; ,
\nonumber\\
\langle m\rangle_{\mu\mu}^{} \hspace{-0.2cm} & \simeq & \hspace{-0.2cm}
+\sqrt{\Delta m_{21}^{2}} \ c_{12}^2 c_{23}^2 e^{2 {\rm i}\sigma }
+ \sqrt{\Delta m_{31}^{2}} \ s_{23}^2 \; ,
\nonumber\\
\langle m\rangle_{\tau\tau}^{} \hspace{-0.2cm} & \simeq & \hspace{-0.2cm}
+\sqrt{\Delta m_{21}^{2}} \ c_{12}^2 s_{23}^2 e^{2 {\rm i}\sigma }
+ \sqrt{\Delta m_{31}^{2}} \ c_{23}^2 \; ,
\nonumber\\
\langle m\rangle_{e\mu}^{} \hspace{-0.2cm} & \simeq &
\hspace{-0.2cm} +\sqrt{\Delta m_{21}^{2}} \ c_{12}^{} s_{12}^{}
c_{23}^{} e^{2 {\rm i}\sigma } +\sqrt{\Delta m_{31}^{2}} \ s_{13}^{}
s_{23}^{}e^{- {\rm i} \delta} \; ,
\nonumber\\
\langle m\rangle_{e\tau}^{} \hspace{-0.2cm} & \simeq &
\hspace{-0.2cm} -\sqrt{\Delta m_{21}^{2}} \ c_{12}^{} s_{12}^{}
s_{23}^{} e^{2 {\rm i}\sigma } +\sqrt{\Delta m_{31}^{2}} \ s_{13}^{}
c_{23}^{} e^{- {\rm i} \delta } \; ,
\nonumber\\
\langle m\rangle_{\mu\tau}^{} \hspace{-0.2cm} & \simeq & \hspace{-0.2cm}
-\sqrt{\Delta m_{21}^{2}} \ c_{12}^2 c_{23}^{}
s_{23}^{} e^{2 {\rm i} \sigma } + \sqrt{\Delta m_{31}^{2}} \ c_{23}^{}
s_{23}^{} \; .
\end{eqnarray}
The lower and upper bounds of $|\langle m\rangle|^{}_{\alpha\beta}$
turn out to be
\begin{eqnarray}
\sqrt{\Delta m_{21}^{2}} \ s_{12}^2 - \sqrt{\Delta m_{31}^{2}} \ s_{13}^2
\lesssim \hspace{-0.2cm} & \left|\langle m\rangle_{ee}^{}\right|
& \hspace{-0.2cm} \lesssim \sqrt{\Delta m_{21}^{2}} \ s_{12}^2
+ \sqrt{\Delta m_{31}^{2}} \ s_{13}^2 \; ,
\nonumber\\
\sqrt{\Delta m_{31}^{2}} \ s_{23}^2 - \sqrt{\Delta m_{21}^{2}} \
c_{12}^2 c_{23}^2 \lesssim \hspace{-0.2cm} & \left|\langle
m\rangle_{\mu\mu}^{} \right| & \hspace{-0.2cm} \lesssim \sqrt{\Delta
m_{31}^{2}} \ s_{23}^2 + \sqrt{\Delta m_{21}^{2}} \ c_{12}^2
c_{23}^2 \; ,
\nonumber\\
\sqrt{\Delta m_{31}^{2}} \ c_{23}^2 - \sqrt{\Delta m_{21}^{2}} \
c_{12}^2 s_{23}^2 \lesssim \hspace{-0.2cm} & \left|\langle
m\rangle_{\tau\tau}^{}\right| & \hspace{-0.2cm} \lesssim
\sqrt{\Delta m_{31}^{2}} \ c_{23}^2 + \sqrt{\Delta m_{21}^{2}} \
c_{12}^2 s_{23}^2 \; ,
\nonumber\\
\sqrt{\Delta m_{31}^{2}} \ s_{13}^{} s_{23}^{} - \sqrt{\Delta
m_{21}^{2}} \ c_{12}^{} s_{12}^{} c_{23}^{} \lesssim \hspace{-0.2cm}
& \left|\langle m\rangle_{e\mu}^{}\right| & \hspace{-0.2cm} \lesssim
\sqrt{\Delta m_{31}^{2}} \ s_{13}^{} s_{23}^{} + \sqrt{\Delta
m_{21}^{2}} \ c_{12}^{} s_{12}^{} c_{23}^{} \; ,
\nonumber\\
\sqrt{\Delta m_{31}^{2}} \ s_{13}^{} c_{23}^{} - \sqrt{\Delta
m_{21}^{2}} \ c_{12}^{} s_{12}^{} s_{23}^{} \lesssim \hspace{-0.2cm}
& \left|\langle m\rangle_{e\tau}^{}\right| & \hspace{-0.2cm}
\lesssim \sqrt{\Delta m_{31}^{2}} \ s_{13}^{} c_{23}^{} +
\sqrt{\Delta m_{21}^{2}} \ c_{12}^{} s_{12}^{} s_{23}^{} \; ,
\nonumber\\
\sqrt{\Delta m_{31}^{2}} \ c_{23}^{} s_{23}^{} - \sqrt{\Delta
m_{21}^{2}} \ c_{12}^2 c_{23}^{} s_{23}^{} \lesssim \hspace{-0.2cm}
& \left|\langle m\rangle_{\mu\tau}^{}\right| & \hspace{-0.2cm}
\lesssim \sqrt{\Delta m_{31}^{2}} \ c_{23}^{} s_{23}^{} +
\sqrt{\Delta m_{21}^{2}} \ c_{12}^2 c_{23}^{} s_{23}^{} \; .
\end{eqnarray}
We see that the dominant terms of $|\langle m\rangle_{\mu\mu}^{}|$,
$|\langle m\rangle_{\tau\tau}^{}|$ and $|\langle
m\rangle_{\mu\tau}^{}|$ are associated with $\sqrt{\Delta
m_{31}^{2}}$, and they receive some small corrections from the terms
proportional to $\sqrt{\Delta m_{21}^{2}}$. In comparison, the
magnitude of $|\langle m\rangle_{ee}^{}|$ is much smaller because
its $\sqrt{\Delta m_{31}^{2}}$ term is suppressed by $s^2_{13}$. The
dominant terms of $|\langle m\rangle_{e\mu}^{}|$ and $|\langle
m\rangle_{e\tau}^{}|$ are associated with $\sqrt{\Delta m_{31}^{2}}
\ s^{}_{13}$, and thus they are somewhat less suppressed.

Case (B): the lightest neutrino mass $m^{}_3$ satisfies $m^2_3 \ll
\Delta m^2_{21}$. In this case, $m^{}_1 \simeq m^{}_2 \simeq
\sqrt{-\Delta m^2_{32}}$ holds. We are allowed to neglect those
terms proportional to $m^{}_3$ in Eq. (27). Given the smallness of
$\theta^{}_{13}$, we approximately obtain
\begin{eqnarray}
\langle m\rangle_{ee}^{} \hspace{-0.2cm} & \simeq & \hspace{-0.2cm}
+\sqrt{-\Delta m_{32}^{2}} \left( c_{12}^2 e^{2 {\rm i} \rho }
+ s_{12}^2 e^{2 {\rm i} \sigma } \right) \; ,
\nonumber\\
\langle m\rangle_{\mu\mu}^{} \hspace{-0.2cm} & \simeq & \hspace{-0.2cm}
+\sqrt{-\Delta m_{32}^{2}} \ c_{23}^2 \left( s_{12}^2 e^{2 {\rm i} \rho }
+ c_{12}^2 e^{2 {\rm i} \sigma } \right) \; ,
\nonumber\\
\langle m\rangle_{\tau\tau}^{} \hspace{-0.2cm} & \simeq & \hspace{-0.2cm}
+\sqrt{-\Delta m_{32}^{2}} \ s_{23}^2 \left( c_{12}^2 e^{2 {\rm i} \sigma }
-s_{12}^2 e^{2 {\rm i} \rho } \right) \; ,
\nonumber\\
\langle m\rangle_{e\mu}^{} \hspace{-0.2cm} & \simeq & \hspace{-0.2cm}
+\sqrt{-\Delta m_{32}^{2}} \ c_{12}^{} s_{12}^{} c_{23}^{}
\left( e^{2 {\rm i} \sigma }  - e^{2 {\rm i} \rho }\right) \; ,
\nonumber\\
\langle m\rangle_{e\tau}^{} \hspace{-0.2cm} & \simeq & \hspace{-0.2cm}
+\sqrt{-\Delta m_{32}^{2}} \ c_{12}^{} s_{12}^{} s_{23}^{}
\left( e^{2 {\rm i} \rho } -  e^{2 {\rm i} \sigma } \right) \; ,
\nonumber\\
\langle m\rangle_{\mu\tau}^{} \hspace{-0.2cm} & \simeq & \hspace{-0.2cm}
-\sqrt{-\Delta m_{32}^{2}} \ c_{23}^{}s_{23}^{} \left(s_{12}^2  e^{2
{\rm i} \rho } + c_{12}^2 e^{2 {\rm i} \sigma } \right) \; .
\end{eqnarray}
Then the lower and upper bounds of $|\langle
m\rangle^{}_{\alpha\beta}|$ are approximately given by
\begin{eqnarray}
\sqrt{-\Delta m_{32}^{2}} \ \cos2\theta_{12}^{} \lesssim
\hspace{-0.2cm} & \left|\langle m\rangle_{ee}^{}\right| &
\hspace{-0.2cm} \lesssim \sqrt{-\Delta m_{32}^{2}} \; ,
\nonumber\\
\sqrt{-\Delta m_{32}^{2}} \ c_{23}^2 \cos2\theta_{12}^{} \lesssim
\hspace{-0.2cm} & \left|\langle m\rangle_{\mu\mu}^{}\right| &
\hspace{-0.2cm} \lesssim \sqrt{-\Delta m_{32}^{2}} \ c_{23}^2 \; ,
\nonumber\\
\sqrt{-\Delta m_{32}^{2}} \ s_{23}^2 \cos2\theta_{12}^{} \lesssim
\hspace{-0.2cm} & \left|\langle m\rangle_{\tau\tau}^{}\right| &
\hspace{-0.2cm} \lesssim \sqrt{-\Delta m_{32}^{2}} \ s_{23}^2 \; ,
\nonumber\\
0 \lesssim \hspace{-0.2cm} & \left|\langle m\rangle_{e\mu}^{}\right|
& \hspace{-0.2cm} \lesssim \sqrt{-\Delta m_{32}^{2}} \ \sin
2\theta^{}_{12} c_{23}^{} \; ,
\nonumber\\
0 \lesssim \hspace{-0.2cm} & \left|\langle
m\rangle_{e\tau}^{}\right| & \hspace{-0.2cm} \lesssim \sqrt{-\Delta
m_{32}^{2}} \ \sin 2\theta^{}_{12} s_{23}^{} \; ,
\nonumber\\
\sqrt{-\Delta m_{32}^{2}} \ c_{23}^{}s_{23}^{} \cos2\theta_{12}^{}
\lesssim \hspace{-0.2cm} & \left|\langle
m\rangle_{\mu\tau}^{}\right| & \hspace{-0.2cm} \lesssim
\sqrt{-\Delta m_{32}^{2}} \ c_{23}^{} s_{23}^{} \; .
\end{eqnarray}
We see that the lower bounds of $|\langle m\rangle_{e\mu}^{}|$ and
$|\langle m\rangle_{e\tau}^{}|$ are zero, while the allowed ranges
of the other four effective mass terms are quite narrow and at the
level of $\sqrt{-\Delta m_{32}^{2}}~$. Given $\rho \simeq \sigma$,
even the texture zeros $\langle m\rangle_{e\mu}^{} \simeq \langle
m\rangle_{e\tau}^{} \simeq 0$ can be achieved. A systematic analysis
of such two-zero textures of the Majorana neutrino mass matrix
$M^{}_\nu$ has been done in the literature \cite{TWO}.

Case (C): the lightest neutrino mass $m^{}_1$ satisfies $m^2_1 \gg
\Delta m^2_{31}$. This case corresponds to a nearly degenerate mass
hierarchy: $m^{}_1 \simeq m^{}_2$ and $m^{}_3 \simeq m^{}_1 + \Delta
m^2_{31}/(2m^{}_1)$. Given the smallness of $\theta^{}_{13}$, the
expressions of $\langle m\rangle^{}_{\alpha\beta}$ in Eq. (27)
approximate to
\begin{eqnarray}
\langle m\rangle_{ee}^{} \hspace{-0.2cm} & \simeq & \hspace{-0.2cm}
m^{}_1 \left(c_{12}^2 e^{2 {\rm i} \rho} + s_{12}^2 e^{2 {\rm i}
\sigma } \right) \; ,
\nonumber\\
\langle m\rangle_{\mu\mu}^{} \hspace{-0.2cm} & \simeq &
\hspace{-0.2cm} m^{}_1 \left(s_{23}^2 + s_{12}^2 c_{23}^2 e^{2 {\rm
i} \rho } + c_{12}^2 c_{23}^2 e^{2 {\rm i} \sigma } \right) +
\frac{\Delta m^2_{31}}{2m^{}_1} s_{23}^2 \; ,
\nonumber\\
\langle m\rangle_{\tau\tau}^{} \hspace{-0.2cm} & \simeq &
\hspace{-0.2cm} m^{}_1 \left( c_{23}^2 + c_{12}^2 s_{23}^2 e^{2 {\rm
i} \sigma } + s_{12}^2 s_{23}^2 e^{2 {\rm i} \rho } \right) +
\frac{\Delta m^2_{31}}{2m^{}_1} c_{23}^2 \; ,
\nonumber\\
\langle m\rangle_{e\mu}^{} \hspace{-0.2cm} & \simeq & \hspace{-0.2cm}
m^{}_1 c^{}_{12} s^{}_{12} c^{}_{23} \left( e^{2 {\rm i} \sigma }
- e^{2 {\rm i} \rho } \right) \; ,
\nonumber\\
\langle m\rangle_{e\tau}^{} \hspace{-0.2cm} & \simeq &
\hspace{-0.2cm} m^{}_1 c^{}_{12} s^{}_{12} s^{}_{23} \left( e^{2
{\rm i} \rho } - e^{2 {\rm i} \sigma } \right) \; ,
\nonumber\\
\langle m\rangle_{\mu\tau}^{} \hspace{-0.2cm} & \simeq &
\hspace{-0.2cm} m^{}_1 c^{}_{23} s^{}_{23} \left( 1 - c_{12}^2 e^{2
{\rm i} \sigma } - s_{12}^2 e^{2 {\rm i} \rho } \right) +
\frac{\Delta m^2_{31}}{2m^{}_1} c_{23}^{} s_{23}^{} \; .
\end{eqnarray}
The lower and upper bounds of $|\langle m\rangle^{}_{\alpha\beta}|$
turn out to be
\begin{eqnarray}
m^{}_1 \cos2\theta_{12}^{} \lesssim \hspace{-0.2cm} & \left|\langle
m\rangle_{ee}^{}\right| & \hspace{-0.2cm} \lesssim m^{}_1 \; ,
\nonumber\\
0 \lesssim \hspace{-0.2cm} &
\left|\langle m\rangle_{\mu\mu}^{}\right| & \hspace{-0.2cm} \lesssim
m^{}_1 \; ,
\nonumber\\
m^{}_1 \cos2\theta_{23}^{} + \frac{\Delta m^2_{31}}{2m^{}_1}
c_{23}^2 \lesssim \hspace{-0.2cm} & \left|\langle
m\rangle_{\tau\tau}^{}\right| & \hspace{-0.2cm} \lesssim m^{}_1 \; ,
\nonumber\\
0 \lesssim \hspace{-0.2cm} & \left|\langle m\rangle_{e\mu}^{}\right|
& \hspace{-0.2cm} \lesssim m^{}_1 \sin 2\theta^{}_{12} c^{}_{23} \;
,
\nonumber\\
0 \lesssim \hspace{-0.2cm} & \left|\langle
m\rangle_{e\tau}^{}\right| & \hspace{-0.2cm} \lesssim m^{}_1 \sin
2\theta^{}_{12} s^{}_{23} \; ,
\nonumber\\
\frac{\Delta m^2_{31}}{4m^{}_1} \sin 2\theta^{}_{23} \lesssim
\hspace{-0.2cm} & \left|\langle m\rangle_{\mu\tau}^{}\right| &
\hspace{-0.2cm} \lesssim m^{}_1 \sin 2\theta^{}_{23} \; .
\end{eqnarray}
Note that the lower bounds of $|\langle m\rangle_{\mu\mu}^{}|$ and
$|\langle m\rangle_{\tau\tau}^{}|$ obtained above hold only in the
$\theta^{}_{23}<45^\circ$ case, consistent with the numerical
calculations done in section 3.2. If $\theta^{}_{23}>45^\circ$ is
supported by the future experimental data, then the lower bounds of
$|\langle m\rangle_{\mu\mu}^{}|$ and $|\langle
m\rangle_{\tau\tau}^{}|$ in Eq. (54) should be exchanged.

Case (D): the lightest neutrino mass $m^{}_3$ satisfies $m^2_3 \gg
|\Delta m^2_{32}|$. This case also corresponds to a nearly
degenerate mass hierarchy: $m^{}_1 \simeq m^{}_2$ and $m^{}_3 \simeq
m^{}_1 + \Delta m^2_{32}/(2m^{}_1)$, where $\Delta m^2_{32} <0$. The
approximate expressions of $\langle m\rangle^{}_{\alpha\beta}$ can
similarly be obtained as in Eq. (53) by replacing $\Delta m^2_{31}$
with $\Delta m^2_{32}$. Then we arrive at
\begin{eqnarray}
m^{}_1 \cos2\theta_{12}^{} \lesssim \hspace{-0.2cm} & \left|\langle
m\rangle_{ee}^{}\right| & \hspace{-0.2cm} \lesssim m^{}_1 \; ,
\nonumber\\
0 \lesssim \hspace{-0.2cm} &
\left|\langle m\rangle_{\mu\mu}^{}\right| & \hspace{-0.2cm}
\lesssim m^{}_1 \; ,
\nonumber\\
m^{}_1 \cos2\theta_{23}^{} + \frac{\Delta m^2_{32}}{2m^{}_1}
c_{23}^2 \lesssim \hspace{-0.2cm} &
\left|\langle m\rangle_{\tau\tau}^{}\right| & \hspace{-0.2cm}
\lesssim m^{}_1 \; ,
\nonumber\\
0 \lesssim \hspace{-0.2cm} &
\left|\langle m\rangle_{e\mu}^{}\right| & \hspace{-0.2cm}
\lesssim m^{}_1 \sin 2\theta^{}_{12} c^{}_{23} \; ,
\nonumber\\
0 \lesssim \hspace{-0.2cm} &
\left|\langle m\rangle_{e\tau}^{}\right| & \hspace{-0.2cm}
\lesssim m^{}_1 \sin 2\theta^{}_{12} s^{}_{23} \; ,
\nonumber\\
0 \lesssim \hspace{-0.2cm} &
\left|\langle m\rangle_{\mu\tau}^{}\right| & \hspace{-0.2cm}
\lesssim m^{}_1 \sin 2\theta^{}_{23} \; .
\end{eqnarray}
It is worth pointing out that the lower bound of $|\langle
m\rangle^{}_{\alpha\beta}\rangle|$ given in Eq. (51) or Eq. (52)
should be zero if $\theta^{}_{23}$ is finally found to lie in the
second quadrant and makes $m^{}_1 \cos 2\theta^{}_{23} + \Delta
m^2_{31} c^2_{23}/(2 m^{}_1)$ or $m^{}_1 \cos 2\theta^{}_{23} +
\Delta m^2_{32} c^2_{23}/(2 m^{}_1)$ negative.

\end{document}